\input harvmac
\input amssym
\let\includefigures=\iftrue
\newfam\black
\input rotate
\input epsf
\noblackbox
\includefigures
\message{If you do not have epsf.tex (to include figures),} \message{change the
option at the top of the tex file.}
\def\figin{\epsfcheck\figin}\def\figins{\epsfcheck\figins}
\def\epsfcheck{\ifx\epsfbox\UnDeFiNeD
\message{(NO epsf.tex, FIGURES WILL BE IGNORED)}
\gdef\figin##1{\vskip2in}\gdef\figins##1{\hskip.5in}
\else\message{(FIGURES WILL BE INCLUDED)}%
\gdef\figin##1{##1}\gdef\figins##1{##1}\fi}
\def\DefWarn#1{}

\def\equation{\eqn}
\def\figinsert{\goodbreak\midinsert}
\def\ifig#1#2#3{\DefWarn#1\xdef#1{fig.~\the\figno}
\writedef{#1\leftbracket fig.\noexpand~\the\figno}%
\figinsert\figin{\centerline{#3}}\medskip\centerline{\vbox{\baselineskip12pt
\advance\hsize by -1truein\noindent\footnotefont{\bf Fig.~\the\figno:} #2}}
\bigskip\endinsert\global\advance\figno by1}
\else
\def\ifig#1#2#3{\xdef#1{fig.~\the\figno}
\writedef{#1\leftbracket fig.\noexpand~\the\figno}%
\global\advance\figno by1} \fi

\def\Title#1#2{\nopagenumbers\abstractfont\hsize=\hstitle\rightline{#1}%
\vskip 0.5in\centerline{\titlefont#2}\abstractfont\vskip .4in\pageno=0}


\lref\Dst{M. Douglas, ``The Statistics of String/M-theory Vacua,''
JHEP {\bf 0305} (2003) 046,
hep-th/0303194.}

\lref\AD{S. Ashok and M. Douglas, ``Counting Flux Vacua,''
JHEP {\bf 0401} (2004) 060,
hep-th/0307049.}

\lref\DD{F. Denef and M. Douglas, ``Distributions of Flux Vacua,''
JHEP {\bf 0405} (2004) 072,
hep-th/0404116.}

\lref\mrdsusy{M. Douglas, ``Statistical analysis of the
supersymmetry breaking scale,'' hep-th/0405279.}

\lref\DGT{M. Dine, E. Gorbatov and S. Thomas, ``Low Energy
Supersymmetry from the Landscape,'' hep-th/0407043\semi
M. Dine, ``Supersymmetry, naturalness and the landscape,'' hep-th/0410201.}

\lref\lennysusy{L. Susskind, ``Supersymmetry breaking in the
anthropic landscape,'' hep-th/0405189.}
\lref\KT{ A.~Klemm and S.~Theisen,
``Considerations of One-Modulus Calabi-Yau Compactifications:
Picard-Fuchs Equations, K\"ahler Potentials and
Mirror Maps,'' Nucl.Phys. {\bf B389} (1993) 153, hep-th/9205041.}

\lref\cdgp{P. Candelas, X. De la Ossa, P. Green and L. Parkes,
``A pair of Calabi-Yau manifolds as an exactly soluble superconformal
theory,'' Nucl. Phys. {\bf B359} (1991) 21.}

\lref\GKP{S. Giddings, S. Kachru, and J. Polchinski, ``Hierarchies
from Fluxes in String Compactifications,'' Phys. Rev. {\bf D66}
(2002) 106006, hep-th/0105097.}

\lref\noscale{E. Cremmer, S. Ferrara, C. Kounnas and D.V. Nanopoulos, ``Naturally
vanishing cosmological constant in N=1 supergravity,'' Phys. Lett. {\bf B133} (1983) 61\semi
J. Ellis, A.B. Lahanas, D.V. Nanopoulos and K. Tamvakis, ``No-scale supersymmetric
standard model,'' Phys. Lett. {\bf B134} (1984) 429.}

\lref\nelson{A. Nelson and N. Seiberg, ``R-symmetry breaking versus supersymmetry
breaking,'' Nucl. Phys. {\bf B416} (1994) 46, hep-ph/9309299.}

\lref\IINK{
S. Gurrieri, J. Louis, A. Micu and D. Waldram, ``Mirror Symmetry
in Generalized Calabi-Yau Compactifications,'' Nucl. Phys. {\bf B654}
(2003) 61, hep-th/0211102\semi
S. Kachru, M. Schulz, P. Tripathy and S. Trivedi, ``New Supersymmetric
String Compactifications,'' JHEP {\bf 0303} (2003) 061, hep-th/0211182.}

\lref\CDFKM{
P.~Candelas, X.~De La Ossa, A.~Font, S.~Katz and D.~R.~Morrison,
``Mirror symmetry for two parameter models. I,''
Nucl.\ Phys.\ B {\bf 416} (1994) 481, hep-th/9308083.}

\lref\dghs{M.\ Hindry and J.\ Silverman, {\it Diophantine Geometry, an
introduction}, Springer Verlag, New York 2000.}

\lref\BT{J.D. Brown and C. Teitelboim, ``Dynamical Neutralization of the
Cosmological Constant,'' Phys. Lett. {\bf B195} (1987) 177\semi
J.D. Brown and C. Teitelboim, ``Neutralization of the Cosmological Constant
by Membrane Creation,'' Nucl. Phys. {\bf B297} (1988) 787.}

\lref\resultantreference{reference on resultants}

\lref\Tschinkel{Tschinkel reference}

\lref\fmsw{J.~L.~Feng, J.~March-Russell, S.~Sethi and F.~Wilczek,
``Saltatory relaxation of the cosmological constant,''
Nucl.\ Phys.\ B {\bf 602}, 307 (2001),
{\tt hep-th/0005276}.
}

\lref\CHSW{P. Candelas, G. Horowitz, A. Strominger and E. Witten,
``Vacuum Configurations for Superstrings,'' Nucl. Phys. {\bf B258} (1985) 46.}

\lref\StromW{
A.~Strominger and E.~Witten,
``New Manifolds For Superstring Compactification,''
Commun.\ Math.\ Phys.\  {\bf 101}, 341 (1985).
}

\lref\unp{S. Kachru, unpublished; E. Silverstein, unpublished; L. Susskind, unpublished.}

\lref\ddf{F. Denef, M. Douglas and B. Florea, ``Building a better
racetrack,'' JHEP {\bf 0406} (2004) 034, hep-th/0404257.}

\lref\BergKatz{P. Berglund and S. Katz, ``Mirror symmetry for hypersurfaces
in weighted projective space and topological couplings,'' Nucl. Phys.
{\bf B420} (1994) 289, hep-th/9311014.}

\lref\KST{S. Kachru,~M. Schulz and S. P. Trivedi,
``Moduli Stabilization from Fluxes in a Simple IIB Orientifold,''
JHEP {\bf 0310} (2003) 007, hep-th/0201028.}
\lref\Freypol{A. Frey
and J. Polchinski, ``N=3 Warped Compactifications,'' Phys. Rev.
{\bf D65} (2002) 126009, hep-th/0201029.}

\lref\GP{B. Greene and R. Plesser, ``Duality in Calabi-Yau Moduli Space,''
Nucl. Phys. {\bf B338} (1990) 15.}

\lref\BP{R. Bousso and J. Polchinski, ``Quantization of Four-Form Fluxes and
Dynamical Neutralization of the Cosmological Constant,'' JHEP {\bf 0006} (2000) 006,
hep-th/0004134.}

\lref\counter{E. Silverstein, ``Counter-intuition and scalar masses,''
hep-th/0407202.}

\lref\conlon{J. Conlon and F. Quevedo, ``On the explicit construction
and statistics of Calabi-Yau flux vacua,'' hep-th/0409215.}

\lref\kumar{J. Kumar and J.D. Wells, ``Landscape cartography: A coarse
survey of gauge group rank and stabilization of the proton,''
hep-th/0409218.}

\lref\acharya{B.S. Acharya, ``A moduli fixing mechanism in M-theory,''
hep-th/0212294\semi
B. de Carlos, A. Lukas and S. Morris, ``Non-perturbative vacua for
M-theory on $G_2$ manifolds,'' hep-th/0409255.}

\lref\DG{O. DeWolfe and S. Giddings, ``Scales and Hierarchies in
Warped Compactifications and Brane Worlds,'' Phys. Rev. {\bf D67} (2003)
066008, hep-th/0208123.}

\lref\font{A. Font, ``Periods and Duality Symmetries in Calabi-Yau
Compactifications,'' Nucl. Phys. {\bf B391} (1993) 358, hep-th/9203084.}

\lref\Font{A. Font, ``$Z_N$ Orientifolds with Flux,'' hep-th/0410206.}

\lref\morrison{D. Morrison, ``Picard-Fuchs equations and mirror maps
for hypersurfaces,'' hep-th/9111025.}

\lref\TT{P. Tripathy and S.P. Trivedi, ``Compactification with flux
on $K3$ and Tori,'' JHEP {\bf 0303} (2003) 028, hep-th/0301139.}

\lref\cdgp{P. Candelas, X. de la Ossa, P. Green and L. Parkes, ``A Pair of
Calabi-Yau Manifolds as an Exactly Soluble Superconformal Theory,''
Nucl. Phys. {\bf B359} (1991) 21.}

\lref\TT{P. Tripathy and S. P. Trivedi, ``Compactification with Flux
on K3 and Tori,''  JHEP {\bf 0303} (2003) 028, hep-th/0301139.}

\lref\GKT{A. Giryavets, S. Kachru and P. Tripathy, ``On the Taxonomy
of Flux Vacua,'' JHEP {\bf 0408} (2004) 002, hep-th/0404243.}

\lref\GKTT{A. Giryavets, S. Kachru, P. Tripathy and S. Trivedi,
``Flux Compactifications on Calabi-Yau Threefolds,''
JHEP {\bf 0404} (2004) 003, hep-th/0312104.}

\lref\KKLT{S. Kachru, R. Kallosh, A. Linde and S.P. Trivedi, ``de Sitter Vacua
in String Theory,'' Phys. Rev. {\bf D68} (2003) 046005, hep-th/0301240.
}

\lref\split{N. Arkani-Hamed and S. Dimopoulos, ``Supersymmetric
Unification without Low Energy Supersymmetry and Signatures
for Fine-tuning at the LHC,'' hep-th/0405159\semi
G.F. Giudice and A. Romanino, ``Split Supersymmetry,'' hep-ph/0406088\semi
N. Arkani-Hamed, S. Dimopoulos, G.F. Giudice and A. Romanino, ``Aspects
of split supersymmetry,'' hep-ph/0409232;
I. Antoniadis and S. Dimopoulos, ``Splitting supersymmetry in
string theory,'' hep-th/0411032.
}

\lref\feng{J. Feng, J. March-Russell, S. Sethi and F. Wilczek,
``Saltatory Relaxation of the Cosmological Constant,''
Nucl. Phys. {\bf B602} (2001) 307, hep-th/0005276.}

\lref\Herman{H. Verlinde, ``Holography and Compactification,''
Nucl. Phys. {\bf B580} (2000) 264, hep-th/9906182\semi
C. Chan, P. Paul and H. Verlinde, ``A Note on Warped
String Compactification,'' Nucl. Phys. {\bf B581} (2000) 156,
hep-th/0003236.}

\lref\saltman{
A. Saltman and E. Silverstein, ``The Scaling of the No-Scale Potential
and de Sitter Model Building,'' hep-th/0402135.}

\lref\OtherIIB{
V. Balasubramanian and P. Berglund, ``Stringy corrections to
K\"ahler potentials, SUSY breaking, and the cosmological constant
problem,'' hep-th/0408054\semi
C.P. Burgess, R. Kallosh and F. Quevedo, ``De Sitter Vacua from Supersymmetric
D-terms,'' JHEP {\bf 0310} (2003) 056, hep-th/0309187\semi
C. Escoda, M. Gomez-Reino and F. Quevedo, ``Saltatory de Sitter String
Vacua,'' JHEP {\bf 0311} (2003) 065, hep-th/0307160\semi
A. Frey, M. Lippert and B. Williams, ``The Fall of Stringy
de Sitter,'' Phys. Rev. {\bf D68} (2003) 046008, hep-th/0305018.
}

\lref\KPV{
S. Kachru, J. Pearson and H. Verlinde, ``Brane/Flux Annihilation and
the String Dual of a Non-supersymmetric Field Theory,'' JHEP
{\bf 0206} (2002) 021, hep-th/0112197.}

\lref\DKV{
O. DeWolfe, S. Kachru and H. Verlinde, ``The giant inflaton,''
JHEP {\bf 0405} (2004) 017, hep-th/0403123.}

\lref\nonkahler{For interesting work in this direction with further
references, see:
G.L. Cardoso, G. Curio, G. Dall'Agata, D. L\"ust, P. Manousselis
and G. Zoupanos, ``Nonkahler String Backgrounds and their Five
Torsion Classes,'' Nucl. Phys. {\bf B652} (2003) 5, hep-th/0211118\semi
K. Becker, M. Becker, K. Dasgupta and P. Green, ``Compactifications
of Heterotic Theory on NonK\"ahler Complex Manifolds I,''
JHEP {\bf 0304} (2003) 007, hep-th/0301161\semi
K. Becker, M. Becker, K. Dasgupta, P. Green,
E. Sharpe, ``Compactifications of Heterotic Strings on NonK\"ahler
Complex Manifolds II," Nucl. Phys. {\bf B678} (2004) 19, hep-th/0310058\semi
J. Gauntlett, D. Martelli and D. Waldram, ``Superstrings with
Intrinsic Torsion,'' Phys. Rev. {\bf D69} (2004) 086002, hep-th/0302158.
}
\lref\ps{J. Polchinski and M. Strassler, ``The String Dual of a
Confining Four-Dimensional Gauge Theory,'' hep-th/0003136.}

\lref\RS{L. Randall and R. Sundrum, `` A Large Mass Hierarchy from
a Small Extra Dimension,'' Phys. Rev. Lett. {\bf 83} (1999) 3370,
hep-th/9905221.}

\lref\GV{S. Gukov and C. Vafa, ``Rational Conformal Field Theories and
Complex Multiplication,'' hep-th/0203213.}

\lref\Borcea{C. Borcea, ``Calabi-Yau Threefolds and Complex Multiplication,''
in {\it Essays on Mirror Manifolds}, S.T. Yau ed., International Press, 1992.}

\lref\GVW{S. Gukov, C. Vafa and E. Witten, ``CFTs from Calabi-Yau Fourfolds,''
Nucl. Phys. {\bf B584} (2000) 69, hep-th/9906070\semi
T.R. Taylor and C. Vafa, ``RR flux on Calabi-Yau and partial supersymmetry
breaking,'' Phys. Lett. {\bf B474} (2000) 130, hep-th/9912152\semi
P. Mayr, ``On Supersymmetry Breaking in String Theory and its Realization
in Brane Worlds,'' Nucl. Phys. {\bf B593} (2001) 99, hep-th/0003198.}

\lref\olded{E. Witten, ``Some properties of $O(32)$ Superstrings,'' Phys. Lett.
{\bf B149} (1984) 351.}

\lref\MSS{A. Maloney, E. Silverstein and A. Strominger, ``De Sitter Space
in Noncritical String Theory,'' hep-th/0205316.}

\lref\lerche{W. Lerche, ``Special Geometry and Mirror Symmetry for Open
String Backgrounds with ${\cal N}=1$ Supersymmetry,'' hep-th/0312326.}

\lref\Heterotic{
S. Gurrieri, A. Lukas and A. Micu, ``Heterotic on Half-flat,''
hep-th/0408121\semi
K. Becker, M. Becker, K. Dasgupta, P. Green and E. Sharpe, ``Compactifications
of Heterotic Strings on Non-Kahler complex manifolds II,'' hep-th/0310058\semi
M. Becker, G. Curio and A. Krause, ``De Sitter Vacua from
Heterotic M Theory,'' Nucl.Phys. {\bf B693} (2004) 223, hep-th/0403027\semi
R. Brustein and S.P. de Alwis, ``Moduli Potentials in String
Compactifications with Fluxes: Mapping the Discretuum,''
Phys.Rev. {\bf D69} (2004) 126006, hep-th/0402088\semi
S. Gukov, S. Kachru, X. Liu and L. McAllister,
``Heterotic Moduli Stabilization with Fractional Chern-Simons Invariants,''
Phys.Rev. {\bf D69} (2004) 086008, hep-th/0310159\semi
E. Buchbinder and B. Ovrut, ``Vacuum Stability in Heterotic M-theory,''
Phys.Rev. {\bf D69} (2004) 086010, hep-th/0310112.}

\lref\svw{S. Sethi, C. Vafa and E. Witten, ``Constraints on Low-Dimensional
String Compactifications,'' Nucl. Phys. {\bf B480} (1996) 213, hep-th/9606122.}

\lref\oldflux{
A. Strominger, ``Superstrings with Torsion,'' Nucl. Phys. {\bf B274} (1986) 253\semi
J. Polchinski and A. Strominger, ``New vacua for
type II string theory,'' Phys. Lett. {\bf B388} (1996) 736,
hep-th/9510227\semi K. Becker and M. Becker, ``M-theory on eight
manifolds,'' Nucl. Phys. {\bf B477} (1996) 155,
hep-th/9605053\semi J.Michelson, ``Compactifications of type IIB
strings to four-dimensions with non-trivial classical potential,''
Nucl. Phys. {\bf B495} (1997) 127, hep-th/9610151\semi
B. Greene,
K. Schalm and G. Shiu, ``Warped compactifications in M and F theory,''
Nucl. Phys. {\bf B584} (2000) 480, hep-th/0004103\semi
M.~Grana and J.~Polchinski,
``Supersymmetric three-form flux perturbations on AdS(5),''
Phys.\ Rev.\ D {\bf 63}, 026001 (2001),
hep-th/0009211\semi
G. Curio, A. Klemm, D. L\"ust and S. Theisen, ``On the Vacuum Structure of
Type II String Compactifications on Calabi-Yau Spaces with H-Fluxes,''
Nucl. Phys. {\bf B609} (2001) 3, hep-th/0012213\semi
K. Becker and M. Becker, ``Supersymmetry Breaking, M Theory and
Fluxes,'' JHEP {\bf 010} (2001) 038, hep-th/0107044\semi
M. Haack and J. Louis, ``M theory compactified on Calabi-Yau
fourfolds with background flux,'' Phys. Lett. {\bf B507} (2001)
296, hep-th/0103068\semi J. Louis and A. Micu, ``Type II theories
compactified on Calabi-Yau threefolds in the presence of
background fluxes,'' Nucl.Phys. {\bf B635} (2002) 395, hep-th/0202168.}

\lref\drs{K. Dasgupta, G. Rajesh and S. Sethi, ``M-theory, orientifolds
and G-flux,'' JHEP {\bf 9908} (1999) 023, hep-th/9908088.}

\lref\bbhl{K. Becker, M. Becker, M. Haack and J. Louis, ``Supersymmetry breaking
and $\alpha^\prime$ corrections to flux induced potentials,'' JHEP {\bf 060} (2002) 0206,
hep-th/0204254.}

\lref\cvetic{
K. Behrndt and M. Cvetic, ``General ${\cal N}=1$ supersymmetric
flux vacua of (massive) type IIA string theory,'' hep-th/0403049\semi
K. Behrndt and M. Cvetic, ``General ${\cal N}=1$ supersymmetric
fluxes in massive type IIA string theory,'' hep-th/0407263\semi
S. Kachru and A. Kashani-Poor, to appear.}

\lref\shiu{
F. Marchesano and G. Shiu, ``Building MSSM flux vacua,'' hep-th/0409132\semi
F. Marchesano and G. Shiu, ``MSSM vacua from
flux compactifications,'' hep-th/0408059}

\lref\cveticliu{
M. Cvetic and T. Liu, ``Three-family supersymmetric standard
models, flux compactifications, and moduli stabilization,''
hep-th/0409032.}

\lref\lust{D. L\"ust, S. Reffert and S. Stieberger, ``MSSM with
soft SUSY breaking terms from D7-branes with fluxes,'' hep-th/0410074\semi
D. L\"ust, S. Reffert and S. Stieberger, ``Flux induced soft supersymmetry
breaking in chiral type IIB orientifolds with D3/D7 branes,'' hep-th/0406092.}

\lref\Klemm{A. Klemm, B. Lian, S.S. Roan and S.T. Yau, ``Calabi-Yau
Fourfolds for M-theory and F-theory Compactifications,''
Nucl.Phys. {\bf B518} (1998) 515, hep-th/9701023.}

\lref\ferrara{R. D'Auria, S. Ferrara and S. Vaula, ``${\cal N}=4$
gauged supergravity and a IIB orientifold with fluxes,'' New J.
Phys. {\bf 4} (2002) 71, hep-th/0206241\semi S. Ferrara and M.
Porrati, ``${\cal N}=1$ no-scale supergravity from IIB
orientifolds,'' Phys. Lett. {\bf B545} (2002) 411,
hep-th/0207135\semi R. D'Auria, S. Ferrara, M. Lledo and S. Vaula,
``No-scale ${\cal N}=4$ supergravity coupled to Yang-Mills: the
scalar potential and super-Higgs effect,'' Phys. Lett. {\bf B557}
(2003) 278, hep-th/0211027\semi R. D'Auria, S. Ferrara, F.
Gargiulo, M. Trigiante and S. Vaula, ``${\cal N}=4$ supergravity
Lagrangian for type IIB on $T^6/Z_2$ orientifold in presence of
fluxes and D3 branes,'' JHEP {\bf 0306} (2003) 045,
hep-th/0303049\semi L. Andrianopoli, S. Ferrara and M. Trigiante,
``Fluxes, supersymmetry breaking and gauged supergravity,''
hep-th/0307139\semi B. de Wit, Henning Samtleben and M. Trigiante,
``Maximal Supergravity from IIB Flux Compactifications,''
hep-th/0311224.}

\lref\sen{A. Sen, ``Orientifold limit of F-theory vacua,'' Phys. Rev.
{\bf D55} (1997) 7345, hep-th/9702165.}

\lref\softsusy{
L.E. Ibanez, ``The fluxed MSSM,'' hep-ph/0408064\semi
P.G. Camara, L.E. Ibanez and A.M. Uranga, ``Flux-induced SUSY
breaking soft terms on D7-D3 brane systems,''
hep-th/0408036\semi
P.G. Camara, L.E. Ibanez and A.M Uranga, ``Flux Induced
SUSY Breaking Soft Terms, hep-th/0311241\semi
M. Grana, T. Grimm, H. Jockers and J. Louis, ``Soft Supersymmetry
Breaking in Calabi-Yau Orientifolds with D-branes and Fluxes,''
hep-th/0312232\semi
A. Lawrence and J. McGreevy, ``Local String Models of Soft Supersymmetry
Breaking,'' hep-th/0401034.}

\lref\canfinite{
P.~Candelas, X.~de la Ossa and F.~Rodriguez-Villegas,
``Calabi-Yau manifolds over finite fields. I,''
arXiv:hep-th/0012233.
}

\lref\canfinitetwo{
P.~Candelas, X.~de la Ossa and F.~Rodriguez Villegas,
``Calabi-Yau manifolds over finite fields. II,''
arXiv:hep-th/0402133.
}

\lref\susskind{L. Susskind, ``The anthropic landscape of string theory,''
hep-th/0302219.}

\lref\KS{I. Klebanov and M. Strassler,
``Supergravity and a Confining Gauge Theory: Duality Cascades and $\chi$SB
Resolution of Naked Singularities,'' JHEP {\bf 008} (2000) 052, hep-th/0007191.}

\lref\uranga{ J. Cascales, M. Garcia de Moral, F. Quevedo and A.
Uranga, ``Realistic D-brane Models on Warped Throats: Fluxes,
Hierarchies and Moduli Stabilization,'' hep-th/0312051\semi J.
Cascales and A. Uranga, ``Chiral 4d String Vacua with D-branes and
Moduli Stabilization,'' JHEP {\bf 0402} (2004) 031, hep-th/0311250\semi
J. Cascales and A. Uranga, ``Chiral 4d ${\cal N}=1$ String Vacua with D-Branes and
NSNS and RR Fluxes,'' JHEP {\bf 0305} (2003) 011,
hep-th/0303024\semi
R. Blumenhagen, D. L\"ust and T. Taylor, ``Moduli Stabilization in Chiral
Type IIB Orientifold Models with Fluxes,'' Nucl. Phys. {\bf B663} (2003) 219,
hep-th/0303016.
}

\lref\BDG{T. Banks, M. Dine and E. Gorbatov,
``Is there a string theory landscape?,'' hep-th/0309170.}
\lref\Susskind{L. Susskind, ``The Anthropic Landscape of String Theory,''
hep-th/0302219.}

\lref\Cremmer{
E.~Cremmer, S.~Ferrara, L.~Girardello and A.~Van Proeyen,
``Yang-Mills Theories With Local Supersymmetry: Lagrangian, Transformation
Laws And Superhiggs Effect,''
Nucl.\ Phys.\ B {\bf 212}, 413 (1983).
}

\lref\Berg{
M.~Berg, M.~Haack and B.~Kors,
``An orientifold with fluxes and branes via T-duality,''
Nucl.\ Phys.\ B {\bf 669}, 3 (2003)
[arXiv:hep-th/0305183].
}

\lref\mooreaa{
G.~W.~Moore,
``Arithmetic and attractors,'' hep-th/9807087.
}

\lref\mooreleshouches{
G.~W.~Moore,``Les Houches lectures on strings and arithmetic,'' hep-th/0401049.
}

\lref\DDtwo{
F.~Denef and M.~R.~Douglas,
``Distributions of nonsupersymmetric flux vacua,''
arXiv:hep-th/0411183.
}

\def\det{{\rm det}}


\def\IL{\relax{\rm I\kern-.18em L}}
\def\IH{\relax{\rm I\kern-.18em H}}
\def\IR{\relax{\rm I\kern-.18em R}}
\def\IC{\relax\hbox{$\inbar\kern-.3em{\rm C}$}}
\def\IZ{\relax\ifmmode\mathchoice
{\hbox{\cmss Z\kern-.4em Z}}{\hbox{\cmss Z\kern-.4em Z}}
{\lower.9pt\hbox{\cmsss Z\kern-.4em Z}} {\lower1.2pt\hbox{\cmsss Z\kern-.4em
Z}}\else{\cmss Z\kern-.4em Z}\fi}


\def\det{{\rm det}}


\def\IZ{\relax\ifmmode\mathchoice
{\hbox{\cmss Z\kern-.4em Z}}{\hbox{\cmss Z\kern-.4em Z}}
{\lower.9pt\hbox{\cmsss Z\kern-.4em Z}} {\lower1.2pt\hbox{\cmsss Z\kern-.4em
Z}}\else{\cmss Z\kern-.4em Z}\fi}

\chardef\tempcat=\the\catcode`\@ \catcode`\@=11
\def\cyracc{\def\u##1{\if \i##1\accent"24 i%
    \else \accent"24 ##1\fi }}
\newfam\cyrfam


\def\bar{\overline}

\def\rt2{\sqrt{2}}
\def\irt2{{1\over\sqrt{2}}}

\def\hat{\widehat}
\def\slashchar#1{\setbox0=\hbox{$#1$}           
   \dimen0=\wd0                                 
   \setbox1=\hbox{/} \dimen1=\wd1               
   \ifdim\dimen0>\dimen1                        
      \rlap{\hbox to \dimen0{\hfil/\hfil}}      
      #1                                        
   \else                                        
      \rlap{\hbox to \dimen1{\hfil$#1$\hfil}}   
      /                                         
   \fi}
\writedefs

\newbox\tmpbox\setbox\tmpbox\hbox{\abstractfont }%
\Title{\vbox{\baselineskip12pt\hbox{\rightline{hep-th/0411061}}
\hbox{\rightline{MIT-CTP-3545, PUPT-2142}} 
\hbox{\rightline{SU-ITP-04/40, SLAC-PUB-10801}} 
}}
 {\vbox{ {\centerline{Enumerating Flux Vacua with Enhanced Symmetries}}}}
\centerline{Oliver DeWolfe$^a$\footnote{$^1$}{\tt odewolfe@princeton.edu $^2$giryav@stanford.edu},
Alexander Giryavets$^{b2}$\footnote{*}{On leave from Steklov Mathematical Institute, Moscow, Russia},
Shamit Kachru$^b$\footnote{$^3$}{\tt skachru@stanford.edu  $^4$
wati\ {\rm at}\ mit.edu} and
Washington Taylor$^{c4}$}
\bigskip\smallskip
\centerline{$^{a}$ Department of Physics}
\centerline{Princeton University}
\centerline{Princeton, NJ 08544, U.S.A.}
\smallskip
\centerline{$^{b}$ Department of Physics and SLAC}
\centerline{Stanford University} \centerline{Stanford, CA 94305/94309, U.S.A.}
\smallskip
\centerline{$^{c}$ Center for Theoretical Physics}
\centerline{MIT, Bldg. 6}
\centerline{Cambridge, MA 02139, U.S.A.}
\bigskip\bigskip\smallskip
\noindent
We study properties of flux vacua in type IIB string theory in several
simple but illustrative models.  We initiate the study of the relative
frequencies of vacua with vanishing superpotential $W=0$ and with
certain discrete symmetries.  For the models we investigate we also
compute the overall rate of growth of the number of vacua as a
function of the D3-brane charge associated to the fluxes, and the
distribution of vacua on the moduli space. The latter two questions
can also be addressed by the statistical theory developed by Ashok,
Denef and Douglas, and our results are in good agreement with their
predictions.  Analysis of the first two questions requires methods which
are more number-theoretic in nature.  We develop some elementary
techniques of this type, which are based on arithmetic properties of
the periods of the compactification geometry at the points in moduli
space where the flux vacua are located.
\bigskip\medskip
\Date{November 2004}

\newsec{Introduction}

Flux vacua of type IIB string theory have been under detailed study
for some time
now (see
\refs{\oldflux,\GVW,\drs,\GKP,\KST,\Freypol,\DG,\TT,\GKTT,\ferrara,\Berg,\Font}
and references
therein).  Much of their interest derives from the fact that,
in the presence of nontrivial RR and NS three-form
fluxes $F_3$ and $H_3$, the IIB theory on a Calabi-Yau orientifold
manifests a computable potential for the complex structure
moduli and the dilaton.
This potential generically has isolated critical points for
these moduli, and such vacua provide good starting points for
the inclusion of further effects that can
stabilize the remaining geometric moduli \refs{\KKLT,\ddf,\saltman,\OtherIIB}.
Similar mechanisms have been proposed in the other corners
of the M-theory parameter space including the 11d limit \acharya,
the heterotic theory \Heterotic, the IIA theory \cvetic, and
supercritical string theory \MSS.
One interesting feature of the flux potentials is that they
give rise to a very large ``discretuum'' of vacua \refs{\BT,\BP,\fmsw}.
For this reason, it is useful (and probably necessary) to find
statistical descriptions which can summarize the features of
large classes of vacua at once.
Such a statistical theory, which is useful for at least the
simplest questions about the IIB Calabi-Yau flux vacua (questions
such as ``how many vacua are there?'' and ``where are they located
on the moduli space?''),
was proposed in \refs{\Dst,\AD}.  It has been further refined, applied and tested
in \refs{\DD,\GKT,\mrdsusy,\conlon,\kumar,\DDtwo}.

In this paper, we continue this statistical study of IIB flux vacua.
In addition to addressing the questions of vacuum degeneracy and
distribution on the moduli space in some simple examples, we also
initiate the study of two more detailed questions which may be
relevant for models of low-energy particle physics.  The first
question is, what fraction of the IIB flux vacua are
``supersymmetric'' in the sense of \GKP, i.e. have $W=0$ in the
leading approximation?\foot{This should not be confused with the
notion of supersymmetry generally used in \AD\ and \DD, where the
K\"ahler moduli are ignored and all of the no-scale vacua are called
supersymmetric.  The latter language is appropriate in many cases
because, after proper inclusion of both K\"ahler moduli and
nonperturbative effects, many no-scale vacua do indeed give rise to
supersymmetric AdS vacua \KKLT.}  This question is important because
in recent discussions of the typical scale of supersymmetry breaking
in the landscape \refs{\lennysusy,\mrdsusy,\DGT,\counter}, it has been
argued that the fraction of $W=0$ vacua may decide the issue of
whether the typical SUSY breaking scale is high or low.  A very small
fraction of $W=0$ vacua would favor high scale SUSY breaking.  This
observation has motivated interesting new testable models of weak
scale physics \split.  On the other hand, as discussed in \DGT, even a
modest fraction of $W=0$ vacua could lead (after further effects) to a
greater number of models with low SUSY breaking scale.  We will remain
neutral about the question of preferred scale, and simply present
results about the relative fraction of $W=0$ vacua in some tractable
examples.

The second question we address is, how frequently do discrete symmetries
arise in flux vacua?  Such symmetries are required in many models of
weak scale physics, to, {\it e.g.}, prevent a catastrophic rate of proton decay.
While our simple models (which do not even include a standard model
sector) are by no means realistic laboratories in which to face
these detailed issues, they are a good place to begin understanding
how frequently discrete symmetries may arise ``accidentally'' in
the landscape.  One can envision a systematic improvement of
the realism of examples, by working through the same kinds of
calculations in the pseudo-realistic models presented in
{\it e.g.} \refs{\uranga,\shiu,\cveticliu,\lust,\softsusy}.

Throughout the paper we use a combination of techniques to analyze the
various models of interest.  For some questions involving enumeration
of the total number of vacua we use a continuous approximation to the
fluxes, along the lines of the techniques developed in \refs{\AD,\DD}.
For overconstrained systems like the set of $W = 0$ vacua, these
techniques are not directly applicable, and the discrete nature of the
fluxes must be taken into account.  We develop several simple methods
for dealing with such problems, which are more essentially
number-theoretic in nature.  Our approach to dealing with these
problems relies on the algebraic structure of the periods of the
compactification manifold at the points in moduli space associated
with flux vacua.  In all cases we consider, these periods lie in a
field extension of finite degree over the rationals ${\bf Q}$, so that we
can reduce the problems of interest to a finite set of algebraic
equations over the integers.  In most cases we have also used
numerical methods to check the analytic predictions.
We summarize our results in the final section.

Our paper is organized as follows: we begin in \S2\ by presenting some
material which will be generally useful in understanding our analysis.
In \S3-\S5, we answer some of the questions of interest in a
particular compactification, with the compactifications arranged in
order of increasing complexity. In \S3, we study vacua in a rigid
Calabi-Yau.  In \S4, we study vacua on the torus $T^6$.  We focus
attention in particular on the subclass of vacua where the torus has
the symmetric form $(T^2)^3$, where each $T^2$ has the same modular
parameter $\tau$.  In \S5, we describe our results (at and near a
special locus in moduli space) for the four Calabi-Yau hypersurfaces
with a single complex structure parameter.  Finally, we close in \S6
with a summary and a discussion of the implications of
these results and promising directions for future investigation.

\newsec{Taxonomy of IIB flux vacua}

The addition of fluxes can lift the moduli of string theory compactifications, leaving isolated vacua at various locations around what was the moduli space.  We are interested in the ``taxonomy" of these vacua -- their general distribution as well as the incidence of vacua with certain special properties.

In this section, we review the ``imaginary self-dual" flux
vacua in type IIB string theory that we will study in the rest of the paper.
After describing the fixing of the dilaton and complex structure moduli, we
briefly discuss the K\"ahler moduli and our philosophy towards their treatment.
We then describe the modular symmetries that must be
fixed in order to count inequivalent vacua, before
reviewing the statistical methods applied to the problem of
enumerating vacua by Douglas and collaborators, which primarily approximate the fluxes as continuous.
We introduce some basic ideas from number theory and discuss how these
ideas will help in
our analysis of  counting problems which are not amenable to a
straightforward continuous approximation.
Finally we motivate and describe the two types of ``special" vacua we will
be interested in counting: those with vanishing
superpotential $W=0$, and those preserving discrete symmetries in the low-energy theory.

\subsec{Basic equations}

As the properties of flux superpotentials on Calabi-Yau orientifolds
in type IIB string theory have been reviewed  many
times, we will be brief.  Our conventions are those of \GKT.

We consider a Calabi-Yau threefold $M$ with $h_{2,1}$ complex
structure deformations, and choose a symplectic basis $\{A^a, B_b \}$
for the $b_3 = 2 h_{2,1} + 2$ three-cycles, $a,b = 1, \ldots, h_{2,1}
+ 1$, with dual cohomology elements $\alpha_a$, $\beta^b$ such that:
\eqn\cyclebasis{
\int_{A^a} \alpha_b = \delta^a_b \,, \quad \quad \int_{B_b} \beta^a =
- \delta_b^a \,, \quad \quad \int_{M} \alpha_a \wedge \beta^b =
\delta_a^b.
} Fixing a normalization for the unique holomorphic three-form
$\Omega$, we assemble the periods $z^a \equiv \int_{A^a} \Omega$,
${\cal G}_b \equiv \int_{B_b} \Omega$ into a $b_3$-vector $\Pi(z) \equiv
({\cal G}_b, z^a)$.  The $z^a$ are taken as projective coordinates on
the complex structure moduli space, with ${\cal G}_b = \partial_b{\cal
G}(z)$.  The K\"ahler potential ${\cal K}$ for the $z^a$ as well as
the axio-dilaton $\phi \equiv C_0 + i e^{-\varphi}$ is
\eqn\kahler{
{\cal K} = -\log (i \int_{M} \Omega \wedge \overline\Omega) - \log (-i (\phi - \bar\phi))
= - \log(-i \Pi^\dagger \cdot \Sigma \cdot \Pi) - \log(-i(\phi - \bar\phi)) \,,
}
where $\Sigma$ is the symplectic matrix $\Sigma \equiv \pmatrix{\;0\;\;\; 1 \cr -1\;\; 0}$.
The axio-dilaton and complex structure moduli take values in the moduli space ${\cal M}$; a correct global description of the moduli space requires that we identify points in ${\cal M}$ related by modular symmetries, as we describe in \S2.2.

We now consider nonzero fluxes of the RR and NSNS 3-form field
strengths $F_3$ and $H_3$ over these three cycles, defining the
integer-valued $b_3$-vectors $f$ and $h$ via \eqn\fluxes{ F_{3} =
- (2 \pi)^2 \alpha'(f_a\, \alpha^a + f_{a +h_{2,1}+1}\, \beta_a)
\,, \quad H_{3} = - (2 \pi)^2 \alpha'(h_a \, \alpha^a + h_{a
+h_{2,1}+1} \,\beta_a) \,. } The fluxes induce a superpotential
for the complex structure moduli as well as the axio-dilaton \GVW:
\eqn\gvw{ W = \int_{M} G_3 \wedge \Omega(z) = (2\pi)^2\alpha'\,(f - \phi h) \cdot
\Pi(z) \,, } where $G_3 \equiv F_3 - \phi H_3$.

We will be interested exclusively in vacua satisfying the F-flatness conditions:
\eqn\origfflat{
D_\phi W = D_a W = 0 \,,
}
where $D_a W \equiv \partial_a W + W \partial_a {\cal K}$, and we have allowed $a$ to run only over $h_{2,1}$ inhomogeneous coordinates.
This is alternately
\eqn\fflat{
(f - \bar\phi h) \cdot \Pi(z) = (f - \phi h) \cdot (\partial_a \Pi + \Pi \partial_a {\cal K}) = 0 \,.
}
These conditions force the complex structure to align such that  the $(3,0)$ and $(1,2)$ parts of the fluxes vanish, leaving the fluxes ``imaginary self-dual," $*_6 G_3 = i G_3$.

The fluxes also induce a contribution to the total D3-brane charge
\eqn\nflux{
N_{\rm flux} = {1 \over (2 \pi)^4 (\alpha')^2} \int_{M}  F_3 \wedge H_3  = f \cdot \Sigma \cdot h \,.
}
In the rest of the paper, we will set $(2\pi)^2 \alpha^\prime = 1$ for convenience.
For vacua satisfying \origfflat, the physical dilaton condition Im $\phi > 0$ implies $N_{\rm flux} > 0$.
As the total charge on a compact manifold must vanish, sources of negative D3-charge must be present as well. For a given IIB orientifold compactification, a fixed amount of negative charge is induced by the orientifolds, leading to an effective bound on $N_{\rm flux}$:
\eqn\fluxbound{
N_{\rm flux} \leq L \,,
}
where for instance in a IIB orientifold arising as a limit of a
fourfold compactification of F-theory \sen, $L$ can be computed from
the Euler character of the fourfold \svw.  Although the number of imaginary-self dual flux vacua is infinite, the set satisfying \fluxbound\ for fixed $L$ is in general finite, as we shall discuss in \S2.3.  In this paper, we do not restrict ourselves to particular values of $L$ associated with a specific orientifold, but instead following \AD\ we take $L$ to be an arbitrary positive integer, and focus on developing tools for estimating numbers of vacua with certain properties, as a function of the fixed parameter $L$, in the context of this simpler mathematical model.

In a more realistic model in which one applies the methods developed
here to a particular compactification of potential phenomenological
interest, one would need to impose a value of $L$ associated with
a particular orientifold of the Calabi-Yau.  The charge difference $L
- N_{\rm flux}$ can be made up by mobile D3-branes; note that
antibranes, another potential source of negative charge, violate the
imaginary self-dual structure. Their inclusion in this class of models
(which is only consistent with finding a tadpole-free configuration
after one has stabilized K\"ahler modes) does not greatly extend the
class of fluxes which must be considered, since the number which can
be included without inducing a perturbative instability is small
compared to $L$ \refs{\KPV, \DKV}.

\medskip
\noindent
{\it Inclusion of K\"ahler Moduli}

In general, K\"ahler moduli are also present in these systems, and they do not acquire a potential from the flux superpotential \gvw.
In the leading approximation, however, they modify the problem in a very simple
way.  The supergravity potential takes the form
\eqn\sugrapot{V = e^{{\cal K}_{tot}}\Big( \sum_{i,j}
D_{i}W g^{i \bar j}D_{\bar j} \overline W
- 3 |W|^2 \Big) \,,}
where $i,j$ run over the complex structure moduli, the dilaton, and
the K\"ahler moduli, and ${\cal K}_{tot}$ includes the full K\"ahler potential for all these fields.    As described in \GKP, the K\"ahler contribution to the
scalar potential in these models can be shown to be
\eqn\cancel{g^{\alpha\bar\beta}D_{\alpha}W D_{\bar \beta}\overline W  =  3|W|^2 \,,} where $\alpha$, $\beta$ run over K\"ahler moduli only.
The cancellation between the contribution \cancel\ and the $-3 |W|^2$ in
\sugrapot\
exemplifies (at leading order in $g_s$ and $\alpha^\prime$)
the famous ``no-scale supergravity'' structure \noscale.  As an example, models with a single volume modulus $\rho$ (whose imaginary part controls the Calabi-Yau volume and whose real part is an RR axion arising from reduction of the $C_4$ field) have a K\"ahler potential contribution
\eqn\kahpot{{\cal K}(\rho,\bar\rho) = -3 ~{\rm log}(-i(\rho - \bar\rho)) \,,}
which can be obtained from dimensional reduction along
the lines of \olded;  given the $\rho$-independence of \gvw, it can easily be seen that \cancel\ follows.

In a no-scale supergravity background,  the potential becomes positive semidefinite, with vacua only arising at $V=0$.  To obtain a vacuum at $V=0$, one must solve the
dilaton and complex structure F-flatness conditions \origfflat.  If one has $D_{\phi}W$ or $D_{a}W \neq 0$, the $e^{{\cal K}_{tot}}$ factor in \sugrapot\ causes
the (positive) potential to fall rapidly to zero as Im $\rho \to \infty$, leading to
rapid decompactification.

On the other hand, in general one {\it cannot} solve $D_\alpha W = 0$.  The equations $D_{a} W = D_{\phi}W = 0$ provide $h^{2,1}({M}) + 1$  equations in $h^{2,1}({M}) + 1$ unknowns.  Imposing $D_\alpha W = 0$  requires $W=0$ as well by \cancel, but since $W$ depends only on $\phi$ and the $z^a$, this overdetermines the constraints on these fields, with no solution in general.
This means that for general choices of the fluxes, supersymmetry is broken by the auxiliary field in the $\rho$ multiplet, while the tree-level vacuum energy vanishes due to the no-scale cancellation \cancel.  Vacua with $W=0$
may, however, exist for special fluxes; we discuss these in \S2.5\ and count such
vacua in various examples in later sections of the paper.

It is critical to note that the special no-scale structure \cancel\ is not expected to be preserved by quantum corrections.  For instance, the first $\alpha^\prime$  correction to the K\"ahler potential \bbhl\ ruins the cancellation \cancel.   As described in \KKLT, generic corrections to ${\cal K}$ and $W$ break the no-scale structure, for example due to a non-perturbatively generated
superpotential for the K\"ahler fields, permitting nontrivial solutions to all the equations $D_i W =0$.
Therefore, one should expect that under many circumstances, fluxes which yield
only non-supersymmetric no-scale vacua with $D_{\alpha}W \neq 0$, yield
supersymmetric AdS vacua with $D_{\alpha}W = 0$ but $W \neq 0$ after inclusion of further corrections. These corrections are, however, in general difficult to compute exactly.

Due to this difficulty, we follow the lead of Douglas and
collaborators and neglect the K\"ahler moduli entirely, focusing
exclusively on the dilaton and complex structure moduli from now on.
Their philosophy is that the distribution of vacua for the dilaton and
complex structure moduli alone is more representative of the character
of the ultimately frozen K\"ahler moduli then the inevitably broken
no-scale structure \cancel\ would be.  For the questions we are trying
to address, concerning the statistics of discrete symmetries which act
only on the complex structure and dilaton, or the frequency of vacua
with $W=0$ at tree level, the detailed dynamics of the K\"ahler moduli
would be irrelevant in any case.

This approach should be a reasonable first step in learning about the
space of real IIB string vacua, as in many cases further
corrections will yield AdS or dS vacua starting from generic no-scale vacua,
along the lines of \refs{\KKLT,\saltman,\ddf}.

\subsec{The modular group and vacua}

Obtaining proper statistics of flux vacua requires not overcounting multiple copies of the same vacuum that are produced by symmetry transformations.  We review these symmetries here.

In the absence of fluxes, a symmetry group ${\cal G}= SL(2,Z)_\phi \times \Gamma$ acts on the moduli space ${\cal M}$, where $SL(2,Z)_\phi$ is the S-duality of type IIB string theory and $\Gamma$ is the modular group of the complex structure moduli space.  Points on ${\cal M}$ related by ${\cal G}$ are considered equivalent, and a fundamental domain for the moduli space arises from dividing out by ${\cal G}$.

For the vacua we consider, the fluxes are affected by ${\cal G}$ as well.
$SL(2,Z)_\phi$ acts in the ordinary way: given an $SL(2,Z)$ matrix $\pmatrix{a \;\;\;b \cr c \;\;\; d}$ we have
\eqn\sltwoz{
\phi \rightarrow {a \, \phi + b \over c\,  \phi + d}   \,, \quad \quad \pmatrix{f \cr h} \rightarrow \pmatrix{a \;\;\; b \cr c \;\;\; d} \pmatrix{f \cr h} \,.
} Under this transformation $(f - \phi h) \rightarrow (f - \phi h)/(c
\phi + d)$, and hence solutions of \fflat\ are carried into other
solutions, while $N_{\rm flux}$ \nflux\ is preserved.  The action of
$SL(2,Z)$ generates a K\"ahler transformation on $W$ \gvw\ and ${\cal K}$ \kahler:
\eqn\kahlertrans{
W \rightarrow \Lambda W \,, \quad \quad {\cal K} \rightarrow {\cal K} - \log \Lambda - \log \bar\Lambda \,,
}
with in this case $\Lambda = 1 /(c \phi + d)$.  K\"ahler transformations \kahlertrans\ are by definition symmetries of ${\cal N} = 1$ supergravity; the potential $V \equiv e^{\cal K} ( |DW|^2 - 3 |W|^2)$ is manifestly invariant, while an appropriate transformation for other fields, including fermions, is determined by $\Lambda$ (see, for example, \Cremmer).  We shall find later that preserved discrete symmetries in various low-energy theories also take the form of K\"ahler transformations.

Since $SL(2,Z)$ is of infinite order, we immediately see that the
action of this group on any given vacuum will produce an infinite
number of copies.  To usefully count inequivalent vacua, one must mod
out by this redundancy.  One may do this either by restricting $\phi$
to a fundamental domain or by placing a restriction on the fluxes; we
will find both methods to be useful.  It is worth noting that the
matrix $-1$ is in $SL(2,Z)$, although its action on $\phi$ is trivial;
hence vacua with the signs of all fluxes flipped are gauge-equivalent.

The precise nature of the complex structure modular group $\Gamma$ will vary from case to case.  The general form for our examples is that under the transformation of the complex structure moduli $z^a \to {z'}^a$, the periods change as
\eqn\periodmod{
\Pi(z^a) \rightarrow \Lambda(z^a)  \, M \cdot \Pi(z^a) \,,
}
where $\Lambda$ is a function of the moduli and $M$ is a constant $Sp(b_3, Z)$ matrix.
 \periodmod\ then induces a K\"ahler transformation \kahlertrans, and is consequently
 a symmetry, provided the fluxes transform as
\eqn\modfluxtrans{
f \to f \cdot M^{-1} \,, \quad \quad h \to h \cdot M^{-1} \,,
}
which thanks to the symplectic condition have the same value of $N_{\rm flux}$ as $(f,h)$, and which lead to solutions of \fflat\ with moduli $\phi, {z'}^a$.  Again to count only inequivalent vacua, one must mod out by these images under the modular group.  We shall have more to say about fixing $\Gamma$ for the various cases as they arise.

\subsec{Enumerating vacua}

Here we discuss the volumes in flux-space occupied by sets of
imaginary self-dual vacua given fixed restrictions from the modular
group and the D3-brane charge bound $L$, and describe how one can
estimate the total growth of vacua with $L$.  We additionally review
the theory of Ashok, Denef and Douglas \refs{\AD,\DD}, which takes a
statistical approach to describing the set of vacua for this class of
compactifications.

The set of integral fluxes  $f_i, h_i$ form a lattice ${\bf Z}^{2b_3}
\subset {\bf R}^{2b_3}$.  Because the F-flatness conditions \origfflat\
are linear in these fluxes, the set of possible continuous values for the fluxes solving these equations and giving rise to dilaton and complex structure moduli in any fixed finite size region in the moduli space ${\cal M}$ forms a cone $C$ in ${\bf R}^{2b_3}$; when the restriction on the moduli space is to a fundamental domain of ${\cal G}$, we call the cone $C_F$.  The cone $C$ intersects the unit sphere $S^{2b_3-1}$ in a region whose boundary is determined by the constraints imposed on the moduli.  Because the number of F-flatness conditions is equal to the number of moduli, these equations generically have a solution, so that the cone $C$ is in general a $2b_3$-dimensional
subspace.

We saw in \S2.1\ that in a given compactification, tadpole constraints
require one to impose the condition $N_{\rm flux} \leq L$ on the
fluxes for some fixed $L$ determined by the global properties of the
model.  Because $N_{\rm flux}$ becomes a positive definite quadratic
form on the fluxes for solutions to the F-flatness conditions
\origfflat, the upper bound on $N_{\rm flux}$ has the effect of
cutting the cone $C$ along a
hyperboloid $H$ in ${\bf R}^{2b_3}$; the set of discrete fluxes giving
rise to moduli in the appropriate region then corresponds to the set
of integral points in the cone $C$ which also lie inside the bounding
hyperboloid $H$.  Generically, the resulting set of vacua is finite.

In particular, if a compact region of finite volume in ${\cal M}$ is
chosen, then the truncation by $H$ of the cone $C$ is a compact region in
${\bf R}^{2b_3}$ with finite volume.  This volume scales as
$\sqrt{L}^{2b_3}$ as $L$ is increased, so that for sufficiently large
$L$ the number of integral points in the volume scales as:
\eqn\crude{N_{\rm vacua}(N_{\rm flux} \leq L) \sim L^{b_3} \,,}
where we have assumed that $L$ is larger than other quantities in the problem, notably that $L \gg b_3$.  Obtaining formulae like \crude\ for more specific subclasses of vacua will be one of our primary goals; in the future we will simply write $N_{\rm vacua}(L)$ to mean a set of vacua with $N_{\rm flux} \leq L$.

Subtleties may arise in this counting, and one of our goals in this
paper is to investigate examples to help understand under what
circumstances the estimate \crude\ is valid.  As an example of a case
where this asymptotic formula may break down, the boundary of the cone
$C_F$ in continuous flux space associated with a fundamental domain
may intersect the codimension one region $Z\subset S^{ 2b_3 -1}$ on
which the quadratic form $N_{\rm flux}$ vanishes.  In this case the
cone truncated by $H$ is no longer compact.  As discussed in \AD, in
this situation the number of lattice points in the truncated cone may
not scale according to \crude.  As $L$ is increased, there are lattice
points in the truncated cone whose projection onto $S^{2b_3 -1}$
approaches $Z$; the fluxes at these lattice points are generally of
order $L$ rather than of order $\sqrt{L}$.  We will encounter an
example of this behavior for the ``special symmetric torus'' in \S4.2,
where restricting to a subset of the allowed fluxes gives rise to an
extra factor of ${\rm log}~ L$ in the number of vacua beyond that
which would be indicated from the counting \crude.

At this point, we can see why one needs to develop a statistical
theory to gain some understanding of this class of vacua.  Simple
orientifold models with $L \sim 10^3$ and $b_3 \sim 100$ can be
obtained by starting with the Calabi-Yau fourfolds in appendix B.4 of
\Klemm.  In such models the naive estimate \crude\ gives rise to $\sim
10^{300}$ vacua.  To gain any credible understanding of the detailed
properties of such a collection, one needs to have a statistical
description.

The idea that discrete fluxes would lead to a large number of closely
spaced discrete vacua was developed in  \refs{\BT,\BP,\fmsw}.
A precise refinement of the rough counting summarized above, giving
rise to such a statistical description, has been proposed in
\refs{\AD,\DD}.  The theory of these authors actually provides much more information than \crude, however.
In addition to giving the scaling of the number
of vacua with $L$, they also provide a formula for the
distribution of the vacua on the moduli space, under the fundamental assumption that $L$ is sufficiently large that the fluxes can be treated as continuous parameters.  The most transparent formula is that for an ``index" where vacua satisfying \origfflat\ are
counted weighted by sgn$(\det\, D^2W)$:
\eqn\fullres{N_{\rm vacua}(L) \sim L^{b_3} ~\int_{\cal M} {\det(-R - \omega)} \,,}
where $\omega$ and $R$ denote the K\"ahler and curvature 2-forms on
${\cal M}$, respectively.  This formula  suggests a smooth
discretization of flux vacua across the moduli space.
For fixed $L$, however, a sufficiently small
${\cal M'}$ will begin to probe the ``fractal" distribution of vacua
coming from the true integral nature of the fluxes, invalidating
\fullres.

Analogous (but more complicated) formulae to \fullres\ are derived in \refs{\AD,\DD} for the (unweighted) number of vacua satisfying \origfflat, as well as more general solutions to the equations $\partial_{a} V = \partial_{\phi} V = 0$ not necessarily satisfying \origfflat.\foot{These authors count as ``supersymmetric" any solutions to \origfflat, and term these more general vacua ``non-supersymmetric".  Although the more general vacua would have tadpoles in the no-scale approximation after inclusion of K\"ahler moduli, one can play off these tadpoles against further effects to obtain stable, nonsupersymmetric solutions
\saltman.  Due to confusion over whether ``supersymmetric" means only solving \origfflat\ or implies $W=0$ as well, we avoid the term altogether.}  Numerical results consistent with the formula \fullres\ have been
obtained in \refs{\GKT,\conlon}, and indications so far are that it is correct within its range of validity.

Our goal will be to compute the numbers and distributions of vacua
without making the primary assumption that leads directly to the formula
\fullres: neglecting the discreteness of the fluxes.
In part, this will be to allow meaningful independent comparison with
\fullres.  But this is also important because the more detailed
questions we are interested in --- including the relative frequency of
vacua with $W=0$, and the relative frequency of vacua with discrete
symmetries --- involve overconstrained systems of equations, giving new
mathematical questions which cannot be answered by the statistical
theory.  Here, we briefly summarize some useful ways of thinking that
emerged in our analysis of these models, and may be more generally
useful.  We then describe the two relevant classes of vacua in more
detail in subsequent subsections.

\subsec{Algebraic classification of vacua}

One approach to counting flux vacua is to choose a set of allowed
integral fluxes and then to try to solve the equations
\fflat\ for each combination of fluxes.  In some very simple
cases these equations can be solved exactly, as we demonstrate below
for the symmetric torus, but this is in general a difficult task.  In
simple Calabi-Yau models, for example, the periods are given by
transcendental functions, and one is
often reduced to approximation techniques as in \refs{\GKT,\conlon}.
While we will also use such techniques here, they cannot easily be
used to establish the existence of vacua with $W$ ${\it precisely}$
vanishing, or vacua at a given point in moduli space where a discrete
symmetry could be restored.

Instead of specifying fluxes and then searching for solutions, one can
also choose to specify a point in ${\cal M}$ and then search for
integral fluxes which yield solutions precisely at this point.  If the
points in ${\cal M}$ associated with allowed vacua (``allowed'' as defined
by the interest of the questioner) can be identified
to lie in some particular discrete family, we can then in principle
sum over the allowed vacuum values in ${\cal M}$.  We can sometimes do
this exactly, to arrive at an exact formula for the number of vacua as
a discrete sum.  Sometimes we can proceed by finding generic features
of the allowed values of ${\cal M}$ which affect the scaling of the
number of vacua, and which allow us to estimate the total number of
vacua in terms of the asymptotic number of points in ${\cal M}$ with
those features.  We now consider this in more detail.

\bigskip
\noindent
{\it 2.4.1. Vacuum counting at a point in moduli space: field extensions}
\medskip

In all cases we examine in this paper, it turns out that the allowed
points in moduli space where vacua arise are points where (with an
appropriate normalization on $\Omega$) the period vector $\Pi(z)$ and
its first covariant derivatives have components taking values in
a finite extension of the field ${\bf Q}$ of rational numbers.  For
example, one family of extensions we encounter on the torus are the
imaginary quadratic number fields ${\bf Q}[\sqrt{-D}]$ with $D > 0$.
On Calabi-Yau manifolds, we have periods which lie in one of the
cyclotomic fields ${\cal F}({\bf Q})_k$, which for integer $k$ is the
extension of the rationals by powers of the $k$th root of unity
$\alpha$.  The algebraic structure of such points in Calabi-Yau moduli
spaces has also played an important role in the works
\refs{\Borcea,\mooreaa,\GV,\mooreleshouches,\canfinite,\canfinitetwo}.

It seems that
the most important characteristic of a vacuum associated with a point
$p\in{\cal M}$ is the degree ${\cal D}_p$ of the field extension over ${\bf Q}$
associated with the periods at that point.  The degree of the
extension ${\bf Q}[\sqrt{-D}]$ is ${\cal D} = 2$, while the degree of the extension
${\cal F}_k$ is given by ${\cal D} = \phi(k)$, where $\phi(k)$ is the Euler totient function,\foot{$\phi(k)$ should not be confused with the axio-dilaton $\phi$.} the number of positive integers $< k$ which
are relatively prime to $k$.  It can also be computed as
\eqn\Euler{\phi(k) = k \;  \prod_i \left( 1 - {1\over P_i} \right) }
where $P_i$ are the prime factors of $k$, with each prime factor appearing
once in the product.

The degree of the field extension at a point provides a simple estimate of how
many flux vacua should arise there, for a fixed bound $N_{\rm flux} \leq L$ on the fluxes.   The point is that, when a single equation in \fflat\ fixes a modulus given the fluxes, it imposes precisely one condition on the moduli, since they take complex values.  If, however, the moduli are fixed and one tries to satisfy the equation by constraining the integer-valued fluxes, the number of constraints that arise is larger, and is given by the degree ${\cal D}$ of the extension.

At a point in moduli space where the field extension ${\cal F}$ over ${\bf Q}$ is of
degree ${\cal D}$, one can view the equations \fflat\ (or their generalization including $W=0$) as a system of linear equations which the integral fluxes must satisfy, with coefficients lying in the field  ${\cal F}$. Note that because these equations are also linear in $\phi$, any vacuum satisfying these equations with periods (and derivatives) in ${\cal F}$ must also have $\phi \in{\cal F}$. A general element $t$ of ${\cal F}$  can be written in an expansion
\equation\expansion{
t =~\sum_{i=1}^{\cal D} t_i \sigma_i \,,} where $t_i \in {\bf Q}$ and
$\sigma_i$ form a basis of elements of ${\cal F}$ which are linearly
independent over ${\bf Q}$.  For example, for the field extension
${\bf Q} [\sqrt{-D}]$, $D > 0$, we can use $\sigma_1 = 1, \sigma_2 = i \sqrt{D}$,
while for the cyclotomic fields ${\cal F}_k$, we can take
$\sigma_i=\alpha^{n_i -1}$, with $\alpha$ a $k$th root of unity and $n_i$
relatively prime to $k$.  We therefore find that solving \fflat\ at a
point in ${\cal M}$ with periods in a field extension of degree ${\cal
D}$ can be thought of as solving a system of $\eta \equiv {\cal D} \,
(b_{2,1}(M)+1)$ equations in the $2b_3(M)$ integral components of the
flux vectors $f, h$.

This analysis leads directly to a prediction for the generic
$L$ scaling of the number of solutions associated with any given point
in moduli space.
Since the quadratic constraint $N_{\rm flux} \leq L$ implies
$f_i, h_i \leq {\cal O} (\sqrt{L})$, we expect
\eqn\nonsusy{N_{\rm vacua}(L) \sim L^{\kappa/2}, ~\kappa \equiv 2b_3 - \eta\,,}
as the scaling of solutions to \fflat\ at such a point in ${\cal
M}$, and
\eqn\susy{N_{\rm vacua}(L; W =0) \sim L^{\gamma/2}, ~\gamma \equiv 2b_3 - \eta - {\cal D} = \kappa - {\cal D} \,,}
as the scaling of solutions with $W=0$.
These arguments are only clearly justified
for generic ($W=0$) vacua when $\kappa > 0$
($\gamma > 0$).  This indicates that they will be most useful
for extensions of very low degree ${\cal D}$, though they may
apply more generally.

In our examples later in the paper, we find some situations in which these formulae apply and immediately give a correct estimate of the number of solutions associated with a fixed point in moduli space.  In many concrete examples, however, subtleties arise.  First of all, it is not always true that the equations \fflat\ are linearly independent; we will find for the example of Landau-Ginzburg vacua of $M_8$ in \S5 that the two equations in \fflat\ are degenerate when the expansion \expansion\ of the dilaton is restricted.  Secondly, in many cases the system is overdetermined, generically constraining the flux vectors $f, h$ to vanish, such as the examples of Landau-Ginzburg vacua $M_5$ and $M_{10}$, as well as $W=0$ vacua for $M_8$, in \S5.  Thirdly, in some cases one of the constraints on fluxes following from \fflat\ may require, {\it e.g.}, $N_{\rm flux}=0$, in which case there can be no nontrivial ISD solution, regardless of the counting of degrees of freedom;
this occurs for $W=0$ vacua in the rigid CY model of \S3.  It is possible other subtleties could arise in other examples.

\bigskip
\noindent
{\it 2.4.2. Vacuum counting over a region in moduli space: heights}
\medskip

The analysis we have just described gives an order-of-magnitude estimate for the number
of flux vacua associated with a particular point in moduli space.  In
most cases, we are interested not just in a single point in moduli
space, but in the ensemble of points admitting vacuum solutions.
Thus, we want to sum the contribution to the total number of vacua
over a large set of points in ${\cal M}$.  In some cases we want to
sum over all points in moduli space admitting vacua.  In other cases,
such as on one-parameter Calabi-Yau manifolds, we may wish to fix the
complex structure moduli and sum over all possible values of the
axio-dilaton $\phi$.
In any of these situations, we require a more refined estimate of the
number of vacua than \nonsusy,\ \susy.  In particular, for two points
$p_1, p_2 \in{\cal M}$ we need some measure of the relative number of
vacua associated with these two points.  If these points are
associated with vacua having periods with the same degree of field
extension over ${\bf Q}$, then even if the estimate \nonsusy\ holds,
we need to know the relative scale of the overall coefficient
multiplying $L^{\kappa/2}$.

A very useful tool in developing a more precise estimate for the
number of vacua at some point $p \in {\cal M}$ is the notion of {\sl
height}, which is a fundamental tool in the modern approach to
Diophantine geometry (i.e., the study of  solutions of systems of
equations over the integers using algebraic geometry---see for
example \dghs.  For another discussion of heights in a somewhat
related physical context, see \mooreaa).
The basic concept of a height is very simple.  It is
really just a systematic way of developing an order-of-magnitude
estimate for problems involving integers and rational numbers.  For
example, the height of a rational number $p/q$ in reduced form (so
that $p, q$ have no common divisors) is just the maximum of $| p |, |
q |$.  More generally, if we have a $n$-plet of rational numbers
$p_1/q_1, \ldots, p_n/q_n$, we can think of this as a point
\equation\projective{
x =\left( 1, {p_1\over q_1}, {p_2\over q_2}, \ldots,
{p_n\over q_n} \right) \in {\bf P}^{n} ({\bf Q}) \,,
}
in the $n$-dimensional projective space over ${\bf Q}$.  The height of
a point \projective\ in ${\bf P}^n ({\bf Q})$ is given by taking a
representative
of the point  where all denominators are cleared and common divisors
are removed,
\equation\cleared{
x \sim \left(m_0,m_1, \ldots, m_n \right), \; \; {\rm gcd} (m_0, \ldots,
m_n) = 1\,,
}
and then defining the height of $x$ to be
\equation\height{
 H (x) = {\rm max} (| m_0 |, | m_1 |, \ldots, | m_n |) \,.
}
This notion of height is extremely useful in the mathematical study of
Diophantine equations because it systematizes the notion of
order-of-magnitude in a way which is natural to number-theoretic
problems.  While this may seem rather arcane, we will only use this
notion in a fairly straightforward physicist's sense, in order to
compute the scaling behavior of simple systems of equations.

The basic idea is the following.  Imagine we have a linear equation
\equation\linear{
A_0 +
{p_1\over q_1} A_1 +
{p_2\over q_2} A_2 +
\ldots
{p_n\over q_n} A_n = 0 \,,
}
for fixed values of the rational coefficients $p_i/q_i$,
and we wish to estimate the number of solutions of this equation for
integers $A_i$ with $|A_i |  \leq N$, for large $N$ (that is, the height of the solution $(A_0, \ldots A_n)$ is bounded at $N$).  As above, the
scaling is clearly generically as $N^n$, but we are also interested in
the overall coefficient.  We can understand this by considering the height of the {\it coefficients}, treating them as a point \projective.

Clearing denominators as above, the solutions to the equation \linear\ with rational coefficients are clearly the same as the solutions of the equation with integral
coefficients
\equation\linearintegral{
m_0 A_0 + m_1 A_1 + \cdots m_n A_n \,.
}
It is easy to see that generically the number of solutions of this
linear equation goes as $N^n/H (x)$ where the height $H (x)$ of the coefficients is
defined through \height. (For example, assume that $m_0$ is the
largest of the integers $m_i$.  Then the remaining sum over $m_iA_i, i
\neq 0$ is of order $H N$ or less, and generically $1/H$ of the possible
values of that sum will be divisible by $m_0$; thus, we expect of order
$N^n/H$ solutions.)

We can generalize the notion of height slightly to include the field
extensions over ${\bf Q}$ discussed above.  If we have, for example, a
modulus $\phi \in {\cal F}$, where ${\cal F}$ is a field extension of
degree ${\cal D}$ over ${\bf Q}$, we can as in \expansion\ consider
$\phi$ to be a ${\cal D}$-plet of rational numbers.  Considering these
numbers to define coordinates in rational projective space, we can
define the height $H (\phi)$ as above.  Similarly, we can define the
height of a polynomial
\equation\polynomialheight{
a_nx^n + \cdots a_1 x + a_0 \in {\bf Q}[x] \,,
}
with rational coefficients $a_i = p_i/q_i$ to be the height of the
corresponding point $(a_0, \ldots, a_n) \in  {\bf P}^n ({\bf Q})$.  We
will use this generalization of the notion of height in particular in
the example discussed in Section \S4.3.

One helpful property of heights is that they behave well
under multiplication.  If $\alpha, \beta \in{\cal F}$ have heights $H
(\alpha), H (\beta)$ then generically
\equation\heightproduct{
H (\alpha \beta) \sim H (\alpha) H (\beta) \,.
}
Similarly, if polynomials $f (x), g (x) \in {\bf Q}[x]$ have heights
$H (f), H (g)$ then generically $H (fg) \sim H (f)H (g)$.
This statement can be made more precise (see for example \dghs), but we
will simply treat it as an order-of-magnitude estimate.

We will also find it very useful to estimate the number $N (n, H)$ of points $x \in {\bf
P}^n ({\bf Q})$ of height $ \leq H$ as $H^{n +1}$.  Since generically $n +1$
integers of order $H$ have no common divisor, this estimate follows
directly from \cleared.  A more precise calculation \dghs\ shows that
\equation\nasymptotic{
N (n, H) ={2^n\over \zeta (n +1)} H^{n +1} +{\cal O} (H^n) \,,
}
where the subleading term goes as $H ~{\rm log} ~H$ for $n = 1$.
The factor of $1/\zeta ( n +1) = \prod_p (1-1/ p^{n + 1})$ where $p$ runs over all prime numbers 
serves to remove the  fraction of integers with common divisors---for example, $1/2^{n +1}$ of
all $ (n +1)$-plets of integers share the divisor 2, 1/$3^{n +1}$ of all
$(n +1)$-plets of integers share the divisor 3, etc.

We can now describe a general estimate for the number of vacua where
we sum over points in the moduli space for the axio-dilaton and/or the
complex structure moduli.  Assume, for example, that we fix the
complex structure moduli $z$ so that the periods take values in a
particular field extension ${\cal F}_z$ of degree ${\cal D}$ over ${\bf
Q}$, and we consider all values of the axio-dilaton lying in ${\cal
F}_z$.  For simplicity, let us take the periods to be of order 1.  For
each possible value of the axio-dilaton, we have as above $\eta ={\cal
D} (b_{2, 1} +1)$ linear equations which must be satisfied.  The total
number of fluxes is $2b_3$, with each flux of order $\sqrt{L}$.  If
the axio-dilaton has height $H$, then each linear equation is solved by
a fraction of order $1/(H \sqrt{L})$ of the possible fluxes.
Generically then, the number of vacua at some
fixed $\phi$ of height $H (\phi)$
is given by
\equation\nvphi{
N_{\rm vacua} (L; \phi, z) \sim{L^{b_3-\eta/2}\over H (\phi)^\eta }\,.
}
One subtlety that will be important to us is the following.  For each
equation of the form $f \cdot \hat\Pi = \phi h \cdot \hat\Pi$, where
$\hat\Pi$ is the period vector or one of its covariant derivatives
projected onto a basis element of ${\cal F}_z$, generically $h \cdot
\hat\Pi$ will be nonzero, and the height of $\phi$ will constrain the
possible solutions for $f$ by $1/H$, leading to the counting in
\nvphi.  However, it is also possible for both sides of the equation
to vanish separately; this is supressed by a factor $1/\sqrt{L}$, but
enhanced by $H$, since the height of $\phi$ provides no constraint.
In most cases the generic solutions will dominate, but 
for $H > \sqrt{L}$ solutions with $f \cdot \hat{\Pi} = h \cdot
\hat{\Pi} = 0$ may contribute to, or even dominate the generic
solutions enumerated by \nvphi.
As we discuss in the next subsection, $W=0$ vacua involve
precisely these special solutions for the $D_\phi W =0$ equation.

We can also sum \nvphi\ over all values of the dilaton.
Since there are on the order of $H^{\cal D}$ allowed values for the
axio-dilaton of height $H$, the total number of vacua for all $\phi$ would be
\equation\nvt{
N_{\rm vacua}(L; z) \sim \sum_{H = 1}^{H_{\rm max}}  H^{{\cal D} -\eta}
L^{b_3-\eta/2} \,,
}
where the upper bound $H_{\rm max}$ is the maximum dilaton height consistent with our constraint on the fluxes.  (In all cases we are studying in this paper, either the degree of the extension is 2, in which case $H_{\rm max} = L$, or the sum
converges so rapidly that the precise upper bound is irrelevant.)

Summing over various complex structure moduli $z$ is more complicated,
particularly when we wish to sum over a range of $z$ involving
different field extensions.  Understanding the proper techniques for
performing this type of sum is an important question for the future.

These kinds of arguments can be generalized, for example to solving
for vacua with $W=0$, which we do in the next subsection.  If the
preceding discussion seems a bit abstract, the reader may find it
useful to go through the example discussed in section \S3.3, which
gives a concrete example of how to use this kind of analysis.  We will
use the considerations described in this subsection as one tool for
estimating the number of vacua in many of the simple models we study.

In many cases subtleties arise that must be taken into account.  In
particular, the formulae \nvphi, \nvt\ were derived under the
assumption that the coefficients in the linear equations satisfied by
the fluxes are not correlated in any way.  In many cases this is not a
valid assumption---for example the periods may satisfy algebraic
relations which need to be incorporated in a correct analysis of the
space of solutions.  Furthermore, in some cases logarithmic factors
arise which correct \nvt.  
Thus, these formulae are not applicable in all models.
Even when these precise formulae are not correct, however, the
considerations we have used in deriving them provide a useful paradigm
for framing the enumeration problems of interest.  By understanding
precisely how these formulae are violated in specific models, we can
often develop a more concrete analysis giving precise answers.

\subsec{Vacua with $W = 0$}

For several reasons, it is of interest to count vacua with
$W=0$ in the no-scale approximation.  While we will reduce this to a clean
mathematics problem in the models and approximations we use, it may be useful
to provide the underlying physical motivation(s) to study these vacua.
We will give two motivations, and then describe why the problem cannot be answered by the statistical theory of \S2.3.

\medskip
\noindent
1) By standard nonrenormalization theorems, one expects that the condition $W = 0$ persists to all orders in perturbation theory, with possible corrections only being generated by non-perturbative effects in the $\alpha^\prime$ or $g_s$ expansion.
Such effects are exponentially small in parameters (like bare gauge couplings)
that one may expect to have a uniform distribution. Hence, the distribution of $W$ values, in any vacua which have $W_{\rm tree} = 0$, will quite plausibly satisfy
\eqn\expdist{N_{\rm vacua}(|W|^2/M_P^6 \leq \epsilon) \sim -{1 \over {{\rm log} (\epsilon)}}\,,}
for given $\epsilon$.  This in particular makes small values of $W$
natural.  Given the formula \sugrapot\ for
the vacuum energy in ${\cal N}=1$ supergravity, one sees that any SUSY-breaking F-terms in a vacuum with small cosmological constant $\Lambda$ need to satisfy
\eqn\need{\Lambda \ll M_P^4 \to {|F|^2 \over M_P^4} \sim {|W|^2 \over M_P^6}~.}
In a scenario where the entire $W \neq 0$ is generated by nonperturbative dynamics,
it is very plausible that the same holds true for any SUSY breaking F expectation
values.
In this case, one could also expect the $F_{a}$ and $F_{\phi}$ VEVs to be distributed
according to a formula analogous to \expdist.

It was argued in \DGT\ that vacua with the distribution \expdist\ for
$W$ and the SUSY breaking F-terms could give a significant or even
dominant contribution to the total number of vacua with $\Lambda \ll
M_P^4$, as long as the conditions leading to \expdist\ (namely, $W=0$
at tree-level, together with suitable nonperturbative effects) are not
too rare.  Since this is contrary to the intuition coming from other
arguments
\refs{\BP,\lennysusy,\mrdsusy,\counter} which point to high-scale SUSY breaking as being preferred in a landscape picture, finding the fraction of $W=0$ vacua is rather well motivated.

\medskip
\noindent
2) It follows from the work of \nelson\ that there is a deep
connection between the existence of R-symmetries and the possibility
of spontaneously breaking supersymmetry in the true ground state of a
supersymmetric field theory.  Assuming the superpotential is a ${\it
generic}$ function of the fields consistent with all symmetries, the
existence of an R-symmetry is a necessary condition for SUSY breaking,
and a spontaneously broken R-symmetry is a sufficient condition.
Since unbroken R-symmetries imply $W = 0$, the study of $W = 0$ vacua
which have $W$ generated by nonperturbative effects in accord with
\expdist, is closely related to the problem of studying R-symmetries in string models
\DGT.
We note, however, that it is well known that the superpotential is
often ${\it not}$ a generic function of the fields consistent with all
symmetries in supersymmetric field theories, so the correspondence is
not precise.

\medskip

Having motivated their study, consider now the equations describing a
$W=0$ vacuum.  As we have described, imposing $W=0$ as well as the
F-flatness conditions $D_\phi W = D_a W = 0$ imposes $h^{2,1} + 2$
conditions on the $h^{2,1} + 1$ moduli; in general, this system cannot
be solved.  We also have, however, $2b_3(M)$ discrete variables at our
disposal -- the choices of $f_i$ and $h_i$.  At least at special loci
in moduli space in special models, the freedom obtained by varying the
integer flux choices $\it does$ suffice to give solutions to the
system \refs{\drs,\KST,\Freypol,\TT,\GKTT}.

Hence the existence of $W=0$ vacua will depend crucially on the choice
of fluxes.  For {\it continuous} fluxes, the system would in fact not
be overdetermined -- one could view the fluxes as additional fields,
and roughly speaking solve the extra equation by solving for one of
the fluxes in terms of the others.  This is not possible, however, in
the real string theory, where fluxes are quantized.  The mathematical
problem of determining which fluxes are associated with a vacuum with
this further constraint becomes one of solving a system of equations
over the integers.  Geometrically, we are considering the intersection
of the cone $C$ discussed in \S2.3\ with the hypersurface $W = 0$.
How many, if any, integral points lie on this hypersurface is a
problem which cannot be solved using a simple continuous approximation
to the fluxes.  But the continuous approximation is a necessary step
in deriving the statistical theory of \refs{\AD,\DD}.  Therefore, we
do not expect this theory to be easily generalized to capture
properties of the special subspace of $W=0$ vacua.

Instead, we must use more number-theoretic methods, as outlined in the
previous subsection.  The most basic question is, what fraction of the
solutions have $W=0$?  Generalizing the results of \S2.4, one finds
that a subtlety arises: it is easy to see that the equations $W =
D_\phi W = 0$ together are equivalent to the axio-dilaton-independent
relations $f \cdot \Pi = h \cdot \Pi = 0$.  Consequently instead of
\nvphi\ one finds
\equation\nvphiw{
N_{\rm vacua} (L; \phi, z; W=0) \sim{L^{b_3-(\eta + {\cal D})/2}\over H (\phi)^{\eta - {\cal D}} }\,,
}
where there are more powers of $1/\sqrt{L}$ due to the $W=0$ equation,
but actually {\it fewer} powers of $1/H$, and
thus
\equation\nvtw{
N_{\rm vacua}(L; z; W=0) \sim \sum_{H=1}^{H_{\rm max}} H^{2 {\cal
D}-\eta} L^{b_3 - (\eta + {\cal D})/2} \,.
}
We will look for $W=0$ vacua and determine how many of these vacua
there are in several examples.  In the concluding section, we will
speculate about the answer in more general models, informed by our
present results.

\subsec{Vacua with discrete symmetry}

The other special kind of flux vacuum we are interested in is one with
an enhanced discrete symmetry.  In this subsection we discuss these
vacua, beginning by describing precisely what is meant by the
statement that a vacuum possesses
an enhanced symmetry.

Each choice of fluxes defines not just a vacuum value for the moduli
fields, but a complete low-energy theory for these fields, with the
fluxes appearing as coupling constants.  A proper symmetry of this
low-energy theory must be a transformation of the fields, but not the
couplings, which are not treated as dynamical quantities.

Hence, when we speak of ``a vacuum with an enhanced symmetry" this is shorthand for a set of fluxes for which the low-energy theory of the moduli, including the superpotential, is symmetric under a transformation of the moduli fields alone, with the fluxes remaining inert.  Once a symmetry is established in the low-energy Lagrangian, one may ask whether the vacuum values of the moduli fields spontaneously break it.

We now discuss the framework within which all the symmetries we
discuss emerge.  The full string theory has a number of symmetries
acting both on the fields and on the fluxes as well.  Let us consider
a group $G$ of such symmetries; a subgroup of the modular group $G
\subset {\cal G}$ will be our primary example.  In general
$G$-transformations cannot be considered low-energy symmetries since
they act on the fluxes.  One can, however, associate with $G$ the set
of transformations $H$ that act on moduli just as $G$ does, but do not
change the fluxes.  In general $H$ will not be a symmetry of the
system.  It is natural to ask, however, whether there are
circumstances under which $H$ is indeed a good symmetry.

Since the kinetic terms for $\phi, z^a$ are independent of the fluxes, $G$-invariance implies they are invariant under $H$ as well.  Hence the question of the validity of $H$ reduces to a question about the potential $V_{f,h}(\phi,z^a)$ associated with the fluxes $f$ and $h$.  Let $f,h,\phi,z^a$ be mapped to $f',h',\phi',{z^a}'$ under $G$ and to $f,h,\phi',{z^a}'$ under $H$. $H$-invariance requires
\eqn\potinvar{
V_{f,h}(\phi', {z^a}') = V_{f,h}(\phi, z^a) \,.
}
The assumption of $G$-invariance, however, means $V_{f',h'}(\phi', {z^a}') = V_{f,h}(\phi, z^a)$, which implies the condition \potinvar\ is equivalent to
\eqn\potinvartwo{
V_{f',h'}(\phi, z^a) = V_{f,h}(\phi, z^a) \,.
}
That is, the condition \potinvar\ that the potential is invariant under the $H$-transformation, is equivalent to a condition that the potential is equal as a function of the moduli to another potential generated by $G$-image fluxes.

One condition for two potentials to be equal, of course, is that they have the same minimum.  As discussed in the previous section, the $DW=0$ vacuum is the unique absolute minimum of the no-scale potential; hence fluxes $f,h$ and their images $f',h'$ can only satisfy \potinvartwo\ if they lead to the same solutions to \origfflat.  But by acting with $G$ on \origfflat\ we know that the vacuum solution $\phi_0, z_0^a$ with fluxes $f,h$ is mapped to a vacuum solution ${\phi_0}', {z_0^a}'$ for the fluxes $f',h'$.  Thus we see that the potential can only be $H$-invariant if
$\phi_0 = {\phi_0}', z_0^a = {z_0^a}'$, that is, if its vacuum sits at a {\it fixed point} of $G$ on the moduli space.

All our examples of discrete symmetries will be of this type: if the fluxes are such that the vacuum value of the moduli lies at a fixed point of $G$, than the low-energy theory may be invariant under the corresponding transformation $H$ of the moduli alone.  For example, for $G \subset {\cal G}$ there are several candidate fixed points.  Fixed points of $SL(2,Z)$ preserving $G = {\bf Z}_2$ and $G = {\bf Z}_3$ occur on the dilaton moduli space; the Landau-Ginzburg point in the
one-parameter Calabi-Yau models $M_k$ we consider in \S5\ is a fixed point
stabilizing a $G = {\bf Z}_k \subset \Gamma$.  We shall find a number of vacua sitting at these fixed points that have an associated $H$ symmetry, though we shall also learn that this does not always happen; the vacuum can sit at the fixed point of $G$ without the $H$ symmetry existing.

These vacua can be counted using the techniques of \S2.4.  In particular, since they occur for fixed values of the moduli, this analysis is well-suited.  For the cases where they sit at a particular point in the dilaton moduli space, instead of summing over the heights to obtain \nvt, one instead evaluates \nvphi\ at the relevant value of $H(\phi_0)$.  We shall have several examples of counting vacua with enhanced symmetries.

One should note that the argument given above implicitly assumes that
the vacuum solution $\phi_0, z_0^a$ is unique.  If there is a flat
direction at $DW=0$ for fixed fluxes, it is possible to satisfy
\potinvartwo\ with $(\phi_0, z_0^a) \neq ({\phi_0}', {z_0^a}')$, as
long as both points lie along the flat direction. In the usual case
that the vacua sit at fixed points of $G$, the preserved symmetries
arising in these examples are never spontaneously broken, but for the
case of a flat direction they may be.

Besides the modular group ${\cal G}$, we shall mention one other source of potential symmetries, namely the complex conjugation transformation ${\cal C}$.  Given a solution $f,h,\phi_0, {z^a_0}$ to \fflat, one may take the complex conjugate to generate a new solution, provided one can write
\eqn\cconj{
\overline{\Pi}(\bar{z}^a) = U \cdot \Pi(\pm \bar{z}^a) \,,
}
for some matrix $U$ and choice of sign for $\pm \bar{z}^a$.  One then has a second solution to \fflat:
\eqn\cconjsoln{
f' = f \cdot U \,, h' = - h \cdot U \,, \phi' = - \bar\phi \,, {z^a}' = \pm \bar{z}^a \,,
}
where we arranged the signs on $h'$ and $\phi'$ to ensure that Im $\phi' > 0$.   The transformation \cconjsoln\ is an antiholomorphic involution of the space, and sends $W(\phi, z^i) \to \overline{W}(\bar\phi,\bar{z}^i)$.   Such a complex conjugation transformation will appear for all of our examples.  Note that unlike the modular group ${\cal G}$, ${\cal C}$ relates inequivalent vacua, and one should not mod out by it.  

Symmetries of the low-energy theory can also descend from ${\cal C}$, as we demonstrate in \S3.5.2, \S4.3.4 and \S5.2; we will find that in general these transformations of the moduli alone act as combinations of $W \to \overline{W}$ and a K\"ahler transformation.  As is familar from the analogous situation in the heterotic theory \StromW,  such a transformation is only a symmetry of the low-energy action when combined with spacetime parity.  In the heterotic case, chiral matter was conjugated under an antiholomorphic transformation, meaning the symmetry corresponded to spacetime $CP$; we expect a similar identification here, although we do not consider matter explicitly.

We shall find in some cases a relation between the $W=0$ vacua and
discrete symmetries: we shall in some cases (but not always) be able to
``explain" $W=0$ such vacua in terms of a preserved discrete
R-symmetry under which $W(\phi_0) \to - W(\phi_0)$. We will comment on these symmetries when they arise.

We have argued that symmetries of the complete theory acting on fluxes
and moduli may restrict to symmetries of the moduli alone when the
fluxes produce a fixed point vacuum.  It is possible to imagine other
symmetries of the low-energy theories of the moduli that are not
restrictions of this type, but we have no such examples.  Certainly
transformations associated to the modular group and to complex conjugation
are by far the most
natural, and we will accordingly focus on them.

\newsec{Flux vacua in a rigid Calabi-Yau}

The simplest model in which to study flux vacua is a rigid Calabi-Yau
with no complex structure moduli.  This model was analyzed in
\refs{\AD,\DD} and the scaling \crude\ was reproduced.
We review this model in detail here, adding comments on the
distribution of vacua as well as analysis of vacua with $W = 0$ and
with discrete symmetries.  Most questions about this model can be
answered fairly easily analytically, so that it is a good trial
laboratory for methods which can be generalized to study more
complicated models.  We analyze the number and distribution of vacua
in this model using the methods described in \S2.4, giving a simple
example of the application of these techniques.  For comparison, we
also review the analysis of \refs{\AD,\DD} using the continuous
approximation method.  While this model admits no $W = 0$ vacua, it
does admit vacua with discrete symmetries.  We find several classes of
such symmetric vacua, with ${\bf Z}_2$ and complex conjugation symmetries.  We enumerate the vacua in each class.  In the case of complex conjugation, the enumeration of vacua with this symmetry requires a number-theoretic analysis going beyond the continuous methods of \AD.

\subsec{Superpotential and equations of motion}

For a rigid Calabi-Yau, there are no complex structure moduli, so
$b_3 = 2$ and a  symplectic basis for $H_3(M)$ consists of
a single $A$ cycle and a single $B$ cycle.  We take the periods of the
holomorphic three-form $\Omega$ to be
\eqn\rigidper{\int_{B}\Omega = 1,~~\int_{A} \Omega = i.}
The resulting flux superpotential is
\eqn\fluxpotr{W = A \phi + B \,, \quad \quad
A \equiv - h_1 - i h_2 \,, \quad B \equiv f_1 + i f_2.
}
The D3-brane charge associated with these fluxes is
\eqn\dthreec{N_{\rm flux} = f_1 h_2 - h_1 f_2 \,,}
and the equation for a vacuum in this model has the form
\equation\rigidvacuum{D_\phi W = A \bar{\phi} + B = 0.}
Solving for $\phi$ given the fluxes, one finds
\eqn\tauis{\bar \phi = - {B \over A}\,.}
The requirement that ${\rm Im}\ \phi > 0$ directly
implies $N_{\rm flux} > 0$.

\subsec{Generic vacua}

For the rigid model the entire modular group is ${\cal G} =
SL(2,Z)_\phi$.  To avoid overcounting $SL(2,Z)$ copies of vacua, we
wish to gauge-fix ${\cal G}$.  We can fix the gauge either by choosing a
fundamental region for $\phi$ or by making a canonical choice for
the fluxes.  We will use both of these approaches to gauge fixing for
the various models studied throughout the paper, so we review both
here.

The first approach to gauge fixing is to choose  $\phi$ in the usual
fundamental region of  $SL(2,Z)$,
\equation\phifixing{\phi \in{\cal F}_{\rm D} =\{z:
-1/2 < {\rm Re}\; z \le 1/2, | z | \ge 1,
| z | \neq 1 \; {\rm for\ Re}\; z < 0\}.}
This does not entirely fix $SL(2,Z)$; one must remember that fluxes with opposite
 sign (which sit at the same point in ${\cal F}_{\rm D}$) are $SL(2,Z)$-equivalent
  and hence there is an overall weighting of $1/2$ to properly count vacua.  In addition,
there are two points on ${\cal F}_{\rm D}$ that stabilize a subgroup of $SL(2,Z)$,
and hence one must further mod out by a factor $1/2$ for vacua at the ${\bf Z}_2$ fixed point $\phi = i$, and by $1/3$ for vacua at the ${\bf Z}_3$ fixed point $\phi = \exp(i \pi/3)$.  (These fixed points have the potential to support discrete symmetries, as we will discuss in \S3.5.)
As an example of correct counting of flux vacua using these weights, at $N_{{\rm flux}}= 1$ there are four flux combinations leading to solutions of \rigidvacuum,
$(A,B) = (1,i),(i,-1),(-1,-i),(-i,1)$.
Together these represent a single physical vacuum at $\phi = i$.

The other way to fix $SL(2,Z)$ is to constrain the fluxes to take a
canonical form.  For example, we can require (as in
\AD) that \eqn\fluxcon{h_1 = 0, \; \; \; ~0 \leq f_2 < h_2~.}
The tadpole condition $N_{\rm flux} \leq L$ then gives $f_1 h_2 \leq L$.

It is easier to count the total number of flux vacua using this
second representation, in which each flux combination satisfying the
constraints represents a single vacuum.
The total number of inequivalent vacua with $ N_{\rm flux}\leq L$ in the
fluxes is then \AD
\eqn\totalnum{N_{\rm vacua}(L) =
\sum_{m = 1}^{L}
\sum_{k\vert m} k = \sum_{m = 1}^{L}  \sigma (m)
= \sum_{k  = 1}^{L}  k\lfloor {L \over k}\rfloor
\sim  {\pi^2 \over 12} L^2~.}
where $\sigma (m)$ is the sum of the divisors of $m$ and $\lfloor x \rfloor$
denotes the greatest integer $\leq x$.

\subsec{Distribution of vacua}

It was shown in \AD\ that when $L$ is sufficiently large that the
fluxes can be considered as sampled uniformly from the continuum, the
density of flux vacua on the moduli space of $\phi$ goes as \fullres\
$L^2/({\rm Im} \; \phi)^2$.  It was furthermore noted in \DD\ that the
distribution of vacua at fixed $L$ features ``voids" containing no
vacua except for an accumulation at their centers; for example, the
points $\phi = ki, k \in {\bf Z}$ are the centers of the largest
voids.  We now describe these results using three different
approaches.  While aside from some small new observations regarding
the void structure the results we describe here are not new, and
essentially recapitulate what was found in \refs{\AD,\DD}, this simple model
provides an excellent framework in which to compare the different
methods we use throughout the paper in more complicated examples.

\break
\noindent{\it Scaling analysis}
\medskip

First, let us consider the scaling analysis discussed in \S2.4.  We
can directly apply the arguments leading up to \nvt.  In this case our
field extension is ${\cal F} = {\bf Q}[i]$, the complex rational
numbers, with degree ${\cal D} = 2$.  We
have $h_{2, 1} = 0$, $b_3 = 2$, so the 4 fluxes satisfy $\eta = 2$
equations.  Given a value $\phi = (n + im)/k$ for the axio-dilaton, which has height $H (\phi) = {\rm max} (m,n,k)$, we have from \nvphi
\equation\nvphirigid{
N_{\rm vacua}(L,\phi) \sim{L\over H (\phi)^2} \,.
}
This gives an estimate for the number of vacua at a fixed axio-dilaton
$\phi$, as a function of the height of $\phi$.  To get the total
number of vacua, we must sum over $\phi$.  There are order $H^3$
values of $\phi$ with height $\leq H$, and therefore order $H^2$ with
height of order $H$.
This gives the estimate
from \nvt
\equation\rigid{
N_{\rm vacua}(L) \sim \sum_{H = 1}^{L}  H^2
{L\over H (\phi)^2}\sim L^2 \,.
}
This estimate gives us information about the scaling of the
total number of vacua as a function of $L$, and furthermore tells us
something about the distribution of values of $\phi$.  This
distribution is expressed here in terms of the height function $H$,
however, rather than the more familiar quantity $ {\rm Im}~\phi$.
Unlike $ {\rm Im}~\phi$, which is a smooth continuous function on
the axio-dilaton moduli space, the height function is a
number-theoretic object which only is defined at special values of
$\phi$ and which cannot be continued to a continuous function.
To compute the distribution on moduli space as a function of ${\rm
Im} ~\phi$, we need a slightly more detailed analysis. This can be done fairly easily as follows.

For ${\rm Im}~\phi = m/k\gg 1$, in the fundamental domain the height of $(n +
im)/k$ is just $m$.  The number of values of $\phi$ of height $H$ with
imaginary part $m/k > y$ then goes as $(H/y)^2$, so the density of $\phi$
at height $H$ goes as $H^2/({\rm Im}~\phi)^3$.
Each term in our linear equations is now of order $H \sqrt{L/{\rm Im}~\phi}$,
since $B/A$ is of order ${\rm Im}~\phi$, so the expected number of vacua at fixed
${\rm Im}~\phi$ goes as
\equation\rigidy{
N_{\rm vacua} \sim \sum_{H = 1}^{L}  {H^2\over ({\rm Im}~\phi)^3}
{L \, {\rm Im}~\phi \over H (\phi)^2}\sim {L^2\over ({\rm Im}~\phi)^2} \,.
}
Thus, we see that the methods described in \S2.4 allow us to compute
both the total number of vacua and their distribution, up to overall
constants.  To determine these constants, a more careful analysis must
be done in which an explicit gauge-fixing is carried out.  We now
compare two more detailed methods of analysis using the two different
gauge fixings discussed above.

\medskip
\noindent{\it Continuous analysis}
\medskip

For comparison, we review the method used in \AD\ to analyze this model
by approximating the fluxes as continuous variables.
Consider the gauge fixing $\phi \in{\cal F}_{\rm D}$.
The number of complex numbers $z = z_1
+ iz_2$ with $z_1, z_2 \in {\bf Z}$ of magnitude $| z |$ goes as $2 \pi |z| d |z|$.  For large $|z|$, the
phases of these numbers are uniformly distributed on the circle.
Thus, we can estimate the number of pairs $A, B$ \fluxpotr\ giving rise to any
particular $\phi$ in the fundamental domain as follows.
Taking $A = |A| e^{i \theta}, B = |B| e^{i ( \theta + \psi)}
= e^{i \theta} \beta$ we have
\equation\ab{L \ge  |A||B| \sin \psi, \;\;\;\;\; {\rm Im}~\phi = \left| {B\over A}\right| \sin \psi \,,}
so $|A|\le \sqrt{L/{\rm Im}~\phi}$.
Thus, the distribution of vacua scales as
\equation\phiestimate{
N_{\rm vacua} (L; \phi) \sim {1\over 2}
\int_0^{\sqrt{L/{\rm Im}~\phi}} 2 \pi |A| \; d|A| \;
\int  d^2 \beta \; \delta^2 (\phi -{1\over |A|}\beta)
={\pi\over 4}{L^2\over ({\rm Im}~\phi)^2} \,,
}
where the overall factor of 1/2 arises
due to the
$S = -1$ symmetry mentioned above.
The integral over moduli space gives $\pi^2 L^2/12$, in agreement with
the  computation \totalnum.

We can also understand the voids easily from this point of view.  For
any $A$ and $y \in {\bf Z}$ we can choose $B = iyA$ as long as $|A|^2 <
L/y$.  Thus, we expect on the order of $\pi L/2y$ vacua at each integral
point $\phi = yi$.  These solutions appear for any $A$.  The nearest
other vacua must have a $B$ which differs by at least one, and which
is thus separated by at least a distance of $\sqrt{y/L}$.  This simple
analysis thus gives us an easy way of deriving the density of
vacua on moduli space as well as the void structure.

\medskip
\noindent{\it Discrete analysis}
\medskip

We now review a more discrete approach to the analysis, using the
second approach to gauge fixing, where we fix the fluxes as in
\fluxcon.  This approach was discussed in \AD, but is closer in spirit
to the height-based analysis above than the continuous approximation
approach.  With the gauge fixing \fluxcon,
the moduli $\phi$ associated with the given fluxes fit a very
simple pattern, lying on the lattice of points
\equation\lattice{\phi ={f_2 + if_1\over h_2}\,,}
with $0 \leq f_2 < h_2$, $f_1 h_2 < L$.
By using the SL(2,Z) symmetry $\phi \rightarrow \phi -1$, we can shift
the set of allowed fluxes to lie in the band $-1/2 < {\rm Re}\; \phi
\le 1/2$.  While the total number of resulting flux solutions in this
gauge fixing is easy to compute and is given by \totalnum, the
distribution of fluxes is less obvious.  The fluxes just described can
be thought of as a set of ``layers'' of $h_2 \lfloor L/h_2 \rfloor$ vacua,
each with density $h_2^2$, arranged uniformly on the region $1/h_2 \leq
{\rm Im} \; \phi \leq\lfloor L/h_2 \rfloor/h_2$.  This would seem to give
rise to a distribution in ${\rm Im}~\phi$
scaling as
\equation\wrongscaling{\sum_{h_2 = 1}^{\sqrt{L/{\rm Im}~\phi}} h_2^2 \sim \left(
{L\over {\rm Im}~\phi} \right)^{3/2} \,.}
Note, however, that in this gauge fixing most of the vacua lie in the
region $|\phi | < 1$.  These points must be moved back into the
fundamental domain with an $SL(2,Z) $ transformation in order to compute
the real density.  As the vacua in the flux gauge fixing are most
dense in the region with small $ {\rm Im} \; \phi$, these vacua
lie in many different modular regions.  If the distributions of vacua
on the different modular regions are essentially uncorrelated, the
image on the fundamental region should give  rise to a distribution
which has a functional form invariant under modular
transformations---this is presumably one way of explaining the $1/({\rm Im}~\phi)^2$
dependence of the distribution in this gauge fixing, but this
explanation is certainly less transparent than in the direct gauge
fixing of $\phi$ in the fundamental region.

We can also consider the distribution of voids in the flux gauge
fixing.  For the layers described above, the appearance of the voids
is quite transparent: each layer contains a point at each integral
point $\phi = y i$, and adjacent points on that layer are separated by
a distance $1/h_2$ with $y h_2^2 < L$, so that the size of the hole is
clearly again $\sqrt{{\rm Im}~\phi/L}$.  The situation is complicated by the
images of the vacua in the region $| \phi | < 1$.  Note, however that
the Jacobian of the map from this region into the fundamental domain
is greater than one, and therefore increases the spacing of the vacua.

\medskip
\noindent{\it Comments on exact analysis vs. asymptotics}
\medskip

In the preceding discussion we have given analytic estimates for the
rate of growth of the total number of vacua and of the distribution of
the vacua on moduli space.  These analytic estimates are based on two
types of analysis.  In the continuous analysis approach of \AD, the
estimates are good only when we have a system of equations which is
not overconstrained, and when the fluxes are large enough that their
integral discreteness can be ignored and we can use a continuous
approximation.  In the discrete type of analysis, the continuous limit
is based on asymptotic formulae for number-theoretic properties of the
integers, which can be encoded in functions like $\sigma (k)$.
In this model, we found that both approximation methods lead to the
same asymptotic formula for the number of vacua, \phiestimate.
An important question is how large the fluxes must be
for these approximations to be valid.  Some discussion of this
issue was given in
\refs{\AD,\DD}.  For the rigid CY model discussed here, it is easy to put the exact
discrete equations on a computer and demonstrate that the
approximations are already valid at quite small values of the fluxes.

It is important to
note, however, that in general the discreteness of the fluxes will
strongly affect any equations involving quantities which are not
summed over all integers below some bound $L$, but rather involve specific
properties of a particular integer $N$.
For example, as discussed above the total number of flux vacua in the
rigid CY model at a particular value of $N_{{\rm flux}}$ is given by
\equation\fixedrigid{N_{\rm vacua} (N_{\rm flux}) = \sigma (N_{\rm
flux}) =
\sum_{k | N_{\rm flux}} k \,. }
This function of $N_{\rm flux}$ is not  smooth and depends
on the number theoretic properties of the argument. Both this
function and its fluctuations are of order $N_{\rm flux}$ -- this
is clearly illustrated in the figure below.  On the
other hand, the total number of vacua with $N_{\rm flux} \leq L$ is
much better behaved--this function, as described above, scales as $L^2$,
with fluctuations which decrease in relative scale as $L \rightarrow
\infty$.  Thus, in comparing precise analytic or
numerical calculations to
asymptotic estimates, we will be much better off when we compare to
quantities like the total number of vacua with $N_{\rm flux} \leq L$,
involving inequalities, rather than quantities like $N_{\rm vacua}
(N_{\rm flux})$, which are much more dependent on factorization
properties of their arguments.
\ifig\figa{Plot of $N_{\rm vacua}(N_{\rm flux})$ up to $N_{\rm flux}=1000$.
}
{\epsfxsize=0.8\hsize\epsfbox{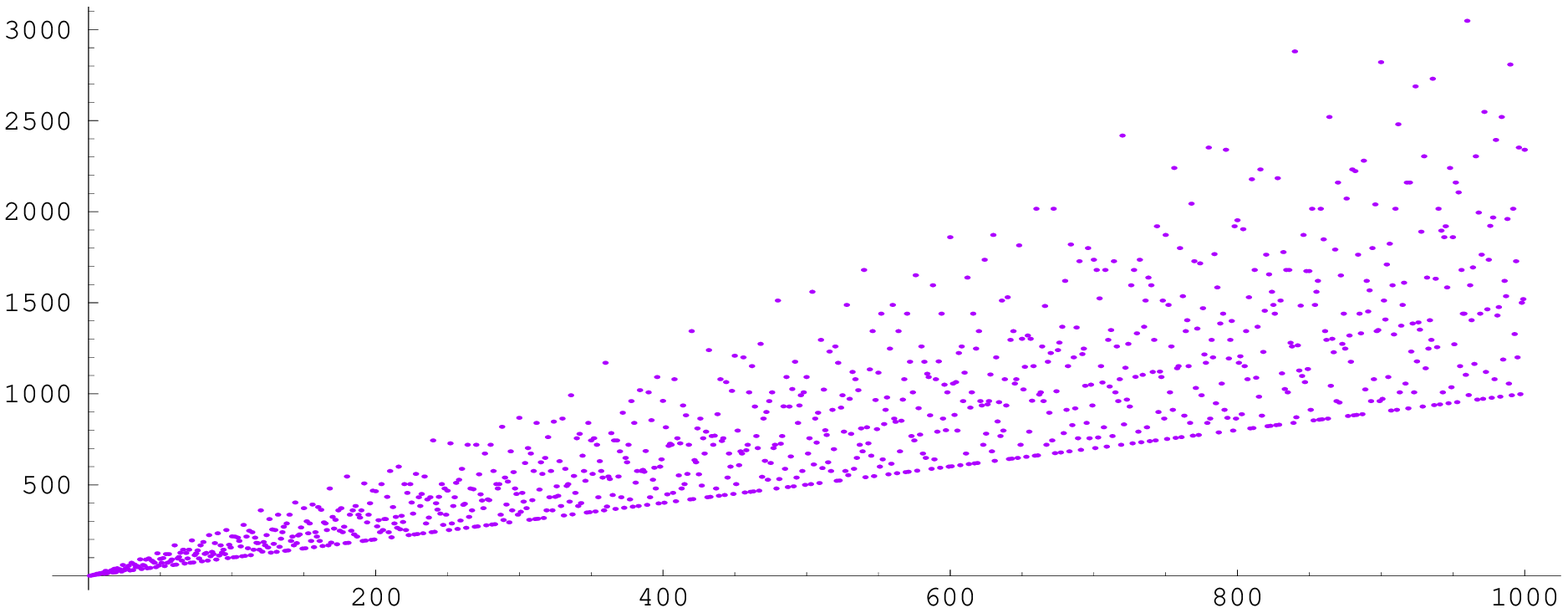}}

\subsec{Vacua with $W = 0$}

In this simple model it is easy to show that the system does not admit
any $W =0$ vacua.  $W=0$ implies, using \fluxpotr, that $A\phi+ B =
0$, which according to the analysis of $\S2.4$ should constitute
${\cal D}({\bf Q}[i]) = 2$ additional constraints on the four fluxes,
which naively should lead to solutions.  It is easy to see, however,
that combined with $DW = 0$, these particular constraints force
$N_{\rm flux} = 0$, leaving no physically acceptable flux vacua.  This
is an example of a subtlety in the counting of constraints, where
naively the number of solutions should be nontrivial but the special
form of the constraints removes them all.  One can further note that
satisfying $W = DW = 0$ requires Im $\phi=0$, which is not an
admissible point in the moduli space.

\subsec{Discrete symmetries}

Following the arguments of \S2.6, we look for enhanced symmetries
in theories whose vacua sit at fixed points in the axio-dilaton
moduli space.  For subgroups of $SL(2,Z)$, there are fixed points
preserving ${\bf Z}_2$ and ${\bf Z}_3$; in the fundamental domain these are
located at $\phi = i$ and $\phi = \exp( \pi i / 3)$ respectively.
Fixed points of complex conjugation ${\cal C}$, which in this case
just flips the sign of the RR scalar, lie on the imaginary $\phi$
axis.  We consider each of these types of discrete symmetries in
turn.

\medskip
\noindent{\it 3.5.1 ${\bf Z}_2 \subset SL(2,Z)$}
\medskip

For the rigid CY model, flux vacua exist precisely at the ${\bf Z}_2$ fixed
point $\phi = i$.  This point is invariant under the transformation $S
\equiv \pmatrix{\;\, 0 \;\;\; 1 \cr -1 \; \; 0}$.  As $S^2 = -1$, the
so-called ${\bf Z}_2$ is really a ${\bf Z}_4$ when action on the fluxes is
included.  As discussed above, therefore, the vacua at $\phi = i$ will
come in quartets, each set of four standing for one inequivalent
vacuum.

The point $\phi = i$ is an element of ${\bf Q}[i]$ of height
$1$.  From the scaling analysis of \S2.4 we expect to have on the
order of $L$ vacua at this point.
More specifically, fluxes compatible with $\phi   = i$ are of the form
\eqn\taui{h_1 = - f_2, ~~h_2 = f_1 \,, \quad N_{\rm flux} = f_1^2 + f_2^2~.}
Dividing out by ${\bf Z}_4$,  we find that
the number of vacua with $N_{\rm flux} \leq L$ at $\phi=i$ is
then
\eqn\numis{N_{\rm vacua}(L,\phi=i) \sim \pi{L\over 4}~.}
The ratio of vacua at the ${\bf Z}_2$ point to generic vacua is thus
asymptotically $3 /(\pi L)$.

We next ask whether the low-energy theory inherits the fixed point symmetry.  As mentioned in the previous section, this would mean the potential is invariant under the restriction of the fixed point symmetry transformation to $\phi$ alone, with fluxes invariant.  The $SL(2,Z)$ transformation $-1$ is trivial acting on $\phi$, and hence we are left asking about $H  = {\bf Z}_2$ on $\phi$ generated by $S$:
\eqn\rigidztwo{
\phi \to - {1 \over \phi} \,.
}
It is easy to see that \taui\ is equivalent
to $B = i A$.  Consequently for these fluxes the superpotential is simply
\eqn\rigidwform{
W(\phi) = (\phi + i) A \,.
}
It is then straightforward that the transformation \rigidztwo\ gives
\eqn\rigidw{
W(\phi) \to {i \over \phi} W(\phi) \,.
}
Hence up to a phase, $W(\phi)$ transforms in the canonical way \kahlertrans\
under $SL(2,Z)$, implying that the low-energy theory is indeed invariant under
${\bf Z}_2$ for any fluxes leading to a vacuum at $\phi = i$.

This is our first example of a preserved discrete symmetry: all vacua sitting at the fixed point ``inherit" a ${\bf Z}_2$ symmetry acting on the moduli alone.  As we shall see, for more complicated examples, this will not always be the case.

There is an alternate derivation of \rigidw\ that will generalize to more complicated models.  We note that another way of expressing the constraint \taui\ is $h = f \cdot S$,
where the matrix $S$ is acting now acting on the space of periods, {\it i.e.}\ it rotates $f_1$ into $f_2$ rather than $f_1$ into $h_1$.  One can then show that under \rigidztwo,
\eqn\rigidflux{
(f - \phi h) \rightarrow {1 \over \phi} ( f- \phi h) \cdot S \,,
}
Using $S \Pi = i \Pi$, \rigidw\ then follows.  This kind of argument will be useful later when the periods are more complicated, but are known to be eigenvectors of certain transformations, namely the modular transformations $\Gamma$.

Note in \rigidw\ that because in the vacuum $\phi_0 = i$, we have
$W(\phi_0) \to  W(\phi_0)$.  Hence even though this
${\bf Z}_2$ acts on the superpotential  it does not constrain $W$ to vanish in the vacuum;
as noted in the previous subsection, there are no such $W=0$ vacua in this model.

As discussed previously, the ${\bf Z}_2$ point
is surrounded by a ``void'', whose size shrinks as $1/\sqrt{L}$ as
$L$ is increased.  So at reasonably small values of $L$, there are no
vacua nearby with an only slightly broken ${\bf Z}_2$.

There is also a fixed point on dilaton moduli space preserving ${\bf Z}_3$,
but it is easy to show that there are no vacua precisely at the ${\bf Z}_3$
point, simply because $\bar \phi$ in \tauis\ has rational real and
imaginary parts.  In the language of \S2.4, this value of $\phi$ does
not lie in the field extension ${\cal F}$ over the rationals ${\bf Q}$
defined by the periods.
We shall find later a model with one complex
structure modulus where the ${\bf Z}_3$ point on the $\phi$-plane becomes
accessible.

If, however, one is interested in only approximate
symmetries, one may consider vacua within some small neighborhood of
size $\epsilon$ around the ${\bf Z}_3$ point; the density of vacua there
will scale like $\epsilon^2 L^2$.  For reasonable values $L$ and a
small symmetry breaking scale $\epsilon$, the ${\bf Z}_2$ point captures
many more vacua than the ${\bf Z}_3$ point.  Note that as soon as $\epsilon >
1 /\sqrt{L}$, one starts including vacua from outside the
void around the ${\bf Z}_2$ point.  So at any fixed $\epsilon$, for very
large $L$ there will be approximately as many ${\bf Z}_2$ vacua as ${\bf Z}_3$
vacua.  For smaller values of $L$ so $\epsilon < 1 /\sqrt{L}$ (but still $L \gg 1$, where the continuous approximation to the fluxes
should apply), the ${\bf Z}_2$ vacua
are more numerous.

\medskip
\noindent{\it 3.5.2 Complex conjugation ${\bf Z}_2$}
\medskip

Finally, we consider the ${\bf Z}_2$ associated to complex conjugation
${\cal C}$.  For the rigid CY model \cconj\ is satisfied somewhat
trivially with $U = \sigma_3$, the third Pauli matrix, so one can map
one vacuum to another (inequivalent) one via
\eqn\rigidcconj{
{f_1}' = f_1 \,, {f_2}' = - f_2 \,, {h_1}' = - h_1 \,, {h_2}' = h_2 \,, \quad \phi' = - \bar\phi \,.
}
The fixed points on the $\phi$-plane are of course at $\phi = i
 \beta$, which are vacua for fluxes
\eqn\cconjflux{
f_1 = \beta h_2 \,,  f_2 = - \beta h_1 \,
}
with $\beta \in {\bf Q}$ a rational number.
The superpotential determined by \cconjflux\ is then
\eqn\rigidcconjw{
W(\phi) =  A( \phi + i \beta) \,. }
One can then verify that under $\phi \to - \bar\phi$ with the fluxes inert, $V(\phi)$ is invariant.  In fact, this moduli-only transformation sends
\eqn\rigidconjtrans{
W \to - {A \over A^*} \, \overline{W} \,.
}
Thus we find that it acts as a combination of an antiholomorphic transformation and a K\"ahler transformation \kahlertrans, and hence when combined with spacetime parity it is a symmetry of the low-energy theory.  Note that when $\beta = 1$, \taui\ holds and the additional ${\bf Z}_2$ \rigidztwo\ is also present.

We can estimate the number of vacua with the complex conjugation ${\bf Z}_2$
symmetry by restricting the sum over $\phi$ at height $H$ in \rigid\ to
$\phi$ with vanishing real part.  The number of such $\phi$
goes as $H$ instead of $H^2$, so the total number of vacua on the
imaginary axis should go roughly as $L \log L$.

Let us now carry out a more careful analysis to check this rough
asymptotic estimate.  We are now looking at an overconstrained
system, where extra conditions are imposed which geometrically
constrain us to a hypersurface in the cone $C$ over the space of
fluxes.  Thus, the continuous methods of \AD\ cannot be applied in this
case and we must resort to a more discrete method of analysis.  This
is the first example we encounter in this paper where more
number-theoretic methods are required.  While the following analysis
may seem somewhat particular to this problem, we will find
parallel questions arising later in the paper, for example in the
discussion of $W = 0$ vacua on the torus in section \S4.3.

We can characterize the vacuum counting problem as a problem of
counting integers subject to a certain set of constraints.
We can write fluxes leading to ${\rm Re}~\phi =0$ as
\equation\ccpieces{
A = k (p + iq)\,, \quad \quad B = il (p + iq) \,,
}
for $k, l, p, q$ integers.  To avoid overcounting, we can choose $p +
iq$ to be {\sl primitive} in the sense that $p, q$ are relatively
prime.  This gives a 1-1 correspondence between solutions of the
vacuum equations with ${\rm Re}~\phi = 0$ and $N_{\rm flux} \leq L$, and integers $k, l, p, q$ satisfying the conditions that ${\rm gcd} (p, q) = 1$ and
\equation\ccconstraint{
kl (p^2 + q^2) \leq L \,.
}
We can write this as a sum
\equation\ccsum{
N_{\rm vacua}(L; {\rm Re}~\phi = 0) =
{1\over 2}
\sum_{(p, q) = 1}^{p^2 + q^2 \leq L}  \;
\sum_{n = 1}^{\lfloor{L \over p^2 + q^2}\rfloor}  d (n)
}
where $d (n)$ is the number of divisors of $n$ and counts the factors $kl$.  The 1/2
arises because we want to fix $SL(2,Z)$ by taking $\beta = l/k \geq 1$.
We can approximate the sum over $p, q$ by an integral, where just as
in
\nasymptotic\ we include a factor of $\zeta (2) = \pi^2/6$ in the
denominator to remove integers with common factors, and where we use
the use the fact that asymptotically
\equation\dsum{
\sum_{n = 1}^{N}  d (n) \sim N ~{\rm log}~ N\,,
}
which comes from the fact that asymptotically the number of divisors
of $n$ goes as $d (n) \sim {\rm log}~ n$.
We then have
\equation\cc{
N_{\rm vacua}(L; {\rm Re}~\phi = 0) =
\int_1^{\sqrt{L}}d\rho\;{\pi \rho\over 4 \zeta (2)} {L\over \rho^2} {\rm log}~ {L\over
\rho^2}\sim{3\over 8 \pi} L ({\rm log}~ L)^2 \,.
}
This gives an estimate for the number of vacua with the discrete ${\bf Z}_2$
complex conjugation symmetry.  We have confirmed this result
numerically by direct computation.
Note that the naive asymptotic analysis
above missed an extra factor of ${\rm log}~ L$ coming from the
number-theoretic aspects of the problem.  We have found that the
fraction of vacua with this particular discrete symmetry goes as
\equation\cratio{
{N_{\rm vacua}(L; {\rm Re}~\phi = 0) \over N_{\rm vacua}(L)} \sim{9\over 2 \pi^3}
{({\rm log}~ L)^2\over L} \,.
}

\newsec{Flux vacua on $T^6 = (T^2)^3$}

The torus is the simplest possible manifold on which to compactify
string theory, so it is naturally one of the first places to study
problems associated with flux compactification \refs{\drs,\KST,\Freypol,\TT}.
We now turn to a study of vacua on a simple class of toroidal
compactifications, namely those which can be described as a direct
product $(T^2)^3$ of 3 tori with equal modular parameter $\tau$.  This
example, while more complicated than the rigid CY, still has the
advantage that the superpotential is of finite order in the moduli.
We solve the $DW = 0$ equations for general fluxes on the symmetric
torus, and show that the periods always take values in a field
extension of degree at most 6 over the rationals ${\bf Q}$.  We study
$W = 0$ vacua and vacua with ${\bf Z}_2$, ${\bf Z}_3$, and complex
conjugation symmetry.  In our analysis we encounter examples of
logarithmic $L$-scaling associated with a special cut through moduli
space, nontrivial $W=0$ vacua, and interesting number-theoretic
structures.  We find in particular that the number of $W = 0$ vacua
can be expressed in terms of class numbers, as anticipated by Moore
\mooreleshouches, and that although vacua with discrete symmetries lie
at special loci in the moduli space, they are reasonably plentiful.

We parametrize the torus as follows (our conventions are as in
\KST).  For completeness we will set up the system with a general
$T^6$, then specialize to the more restricted (and hence more tractable)
models of interest.  Let $x^i, y^i$ for $i=1,2,3$ be three pairs
of periodic coordinates $x^i \equiv x^i + 1,~
y^i \equiv y^i + 1$.
Choose three holomorphic 1-forms $dz^i = dx^i + \tau^{ij} dy^j$.
The complex structure is completely specified by the period matrix $\tau^{ij}$.
Choose the orientation
\eqn\orient{\int ~dx^1 \wedge dx^2 \wedge dx^3 \wedge dy^1 \wedge dy^2 \wedge dy^3 ~=~1~.}
Then a convenient basis for $H^3(T^6,{\bf Z})$ is given by
\eqn\cohbasis{\eqalign{&\alpha^0 = dx^1 \wedge dx^2 \wedge dx^3 \, ,\cr
&\alpha_{ij} = {1\over 2}\epsilon_{ilm} dx^l \wedge dx^m \wedge dy^j \, ,\cr
&\beta^{ij} = - {1\over 2}\epsilon_{jlm} dy^l \wedge dy^m \wedge dx^i \, ,\cr
&\beta^0 = dy^1 \wedge dy^2 \wedge dy^3~.}}
This basis satisfies the property
\eqn\ortho{\int ~\alpha_I \wedge \beta^J = \delta_I^J~.}
The holomorphic three-form can be expressed in terms of the holomorphic one-forms as
\eqn\omegais{\Omega = dz^1 \wedge dz^2 \wedge dz^3 \,,}
with the expansion
\eqn\thenom{\Omega = \alpha^0 + \alpha_{ij}\tau^{ij} - \beta^{ij}({\rm cof}\tau)_{ij} + \beta^0
({\rm det}\tau)\,,}
where
\eqn\cofis{({\rm cof}\tau)_{ij} \equiv ({\rm det}\tau)\tau^{-1,T} = {1\over 2}\epsilon_{ikm}\epsilon_{jpq}
\tau^{kp}\tau^{mq}~.}
The modular group ${\cal G}$ for a general $T^6$ flux compactification would include the $SL(6,Z)$ group which
acts on the toroidal moduli and the $SL(2,Z)$ S-duality group.  For the case of interest
to us, where the $T^6$ is a product of three identical two-tori with complex structure
parameter $\tau$, this will simplify to ${\cal G} = SL(2,Z)_{\phi} \times SL(2,Z)_{\tau}$.

\subsec{Superpotential and equations for flux vacua}

General fluxes can be expanded (with $(2\pi)^2 \alpha^\prime = 1$) as
\eqn\fis{\eqalign{&F_3 = a^0 \alpha^0 + a^{ij} \alpha_{ij} + b_{ij} \beta^{ij} + b_0 \beta^0\,, \cr
&H_3 = c^0 \alpha^0 + c^{ij} \alpha_{ij} + d_{ij} \beta^{ij} + d_0 \beta^0 \,.}}
Here $a^0,\ldots d_0$ are all integer parameters.
The Gukov-Vafa-Witten superpotential \gvw\ can be expanded using \thenom\ and \ortho\ as
\eqn\torusw{W = (a^0 - \phi c^0) ~{\rm det}\tau - (a^{ij} - \phi c^{ij})~({\rm cof}\tau)_{ij}
- (b_{ij} - \phi d_{ij})~\tau^{ij} - (b_0 - \phi d_0)~.}
At this point, we restrict our discussion to the symmetric cases which will be of interest
to us.  The most general fluxes we will consider will take the form
\eqn\forus{a^{ij} = a \delta^{ij}, ~b_{ij} = b \delta_{ij},~ c^{ij} = c \delta^{ij}, ~d_{ij} = d \delta_{ij}~.}
That is, all flux
matrices will be diagonal and proportional to the identity matrix.
In addition, we allow general values of the integers $a^0, b_0, c^0, d_0$.

The most general discrete solutions of the $DW=0$ equations with the
fluxes \forus\ take the form $\tau^{ij} = \tau \delta^{ij}$ -- that
is, the six-torus becomes a $(T^2)^3$ where all two-tori have the same
complex structure (as described for the $W=0$ case in \KST).  In fact,
six-tori of this form have an enhanced symmetry $S_3$ which permutes
the three two-tori; the fluxes \forus\ are the most general fluxes
consistent with this enhanced symmetry.  Moduli which would break the
symmetry are by definition charged, and so as in \GKTT\ we can argue
that if we ignore the charged moduli and find solutions for $\tau$ on
the symmetric locus, these solutions will lift to solutions of the
full theory including the charged moduli.

For the symmetric torus $(T^2)^3$ with fluxes \fis, the superpotential \torusw\ further simplifies to
\eqn\reducedw{
W(\phi, \tau)  = P_1(\tau) -\phi P_2(\tau)\, ,
}
where $P_1(\tau)$ and $P_2(\tau)$ are cubic
polynomials in the complex structure parameter $\tau$ defined by
\eqn\pa{\eqalign{
&P_1 (\tau)\equiv a^0 \tau^3- 3a\tau^2
- 3b \tau - b_0\,, \cr
&P_2 (\tau) \equiv c^0 \tau^3- 3c\tau^2
- 3d \tau - d_0\,.}}
The K\"ahler potential for the theory truncated to $\tau$ and $\phi$ is given, using \thenom\ and \kahler, by
\eqn\kis{{\cal K} =  - 3 ~{\rm log}(-i (\tau - \overline \tau)) -{\rm log}(-i(\phi - \overline\phi)) \,,}
and the D3-brane charge contribution for the symmetric torus is
\equation\tadpoles{ N_{\rm flux} = b_0 c^0-a^0 d_0 + 3 (bc-ad) \,.}
After some simplification, the vacuum equations  take the form
\equation\vacuumas{\eqalign{
P_1 (\tau) -\bar{\phi} P_2 (\tau) &= 0\,,\cr
P_1 (\tau) -\phi P_2 (\tau) &={1\over 3}
(\tau -\bar{\tau}) \left(P_1' (\tau) -\phi P_2' (\tau) \right)\,.
}}
We can solve for the dilaton $\phi$ using one of these equations (when
$P_2 (\tau) \neq 0$), and
then eliminate it from the other equation, leaving a single equation
which is quintic in $\tau, \bar{\tau}$:
\equation\quintic{
\eqalign{
Q(\tau, \bar{\tau}) = &
P_2 (\bar{\tau}) \left( 3 a \tau^2 + 6b \tau + 3b_0
                  -\bar{\tau} (3a^0 \tau^2 -6a \tau -3b) \right) \cr
&-P_1 (\bar{\tau}) \left(3c \tau^2 + 6d \tau + 3d_0
                  -\bar{\tau} (3c^0 \tau^2 -6c \tau -3d) \right) = 0 \,.
}}
While rather complicated,
this equation can be solved exactly.  The complete solution is
described in subsection \S4.3.
Before analyzing the complete system, however, we find it instructive
to first consider a simpler model in which we only turn on a subset of
the fluxes.  In this simplified problem, which we analyze in
subsection \S4.2, the solution of the equations
is more transparent, and already leads to interesting lessons about
torus vacua.

\subsec{Vacua on special symmetric $(T^2)^3$}

We can even further simplify the full symmetric
torus problem by restricting the
fluxes to have the special form
\eqn\specialflux{
d = -a  \,, \quad b = c = a^0 = d_0  =0 \,,
\quad b_0, c^0\ {\rm arbitrary}.
}
This reduced set of fluxes was
discussed in  \S8.1 of \KST, where restrictions were also placed on
the allowed moduli.  Here, we analyze all possible solutions to the equations of motion \vacuumas\ given this set of fluxes, with no restriction on moduli.

The equations of motion in this case are fairly
simple, but already lead to interesting phenomena including residual
flat directions, nontrivial $W = 0$ solutions, and an asymptotic
number of vacua with anomalous scaling.  It will turn out that the set of all {\it isolated} vacua (those not part of a flat direction) coincides with the ones with restricted moduli considered in \KST, and it is these that we will count.

With the restriction \specialflux\ the tadpole condition \tadpoles\ becomes
\eqn\tadpoc{b_0 c^0 + 3a^2 \leq L~.}

\break
\noindent
{\it 4.2.1 Vacuum equations}
\medskip

Consider the quintic \quintic.  Imposing the flux restriction
\specialflux, and writing $\tau = x + iy$, we get two quintic
equations in the
real variables $x, y$.  These equations have factors of $y^2, y$
respectively.  Since we are interested in solutions with $y \neq 0$,
we drop these factors and get the cubic and quartic equations
\equation\simpletwo{
\eqalign{
2a c^0\;y^2 \; x &  \;=  \;\left[-2a c^0 x^2+ 3a^2 -b_0 c^0 \right] \;x
\;\equiv \; p_2 (x) x\, ,\cr
a c^0\;y^4 &  \;=  \;ac^0 x^4 + (b_0 c^0-3a^2) x^2 + a b_0
\;\equiv \; p_4 (x)\,.\cr
}}
The first equation clearly has a solution when $x = 0$.  In this case
the second equation is satisfied when either $a = 0$ or
\equation\simpley{
y^4 \; = \; {b_0\over c^0} \,.
}
When $x \neq 0$, we can factor out an $x$ from the first equation, and
combine the two equations to get
\equation\simpleextra{
\eqalign{
0 &  \; = \; p_2 (x)^2 -4a c^0 \;p_4 (x)\cr
& \; =  \; 9a^4-10a^2 b_0 c^0 + b_0^2 (c^0)^2 \,.
}
}
Summarizing, we see that there are several continuous families of
solutions, and one discrete family of solutions with the restricted
class of fluxes we are considering.  The continuous families of
solutions arise when either $a = 0$, in which case we have an
arbitrary imaginary value for $\tau$, or when \simpleextra\ is
satisfied, in which case the real part of $\tau$ is arbitrary, and the
imaginary part is determined by $2a c^0y^2 = p_2 (x)$ (of course, if
$a = 0$ or $c^0 = 0$, $\tau$ is completely unconstrained).

The appearance of these continuous families is not surprising.  A certain
number of fluxes must be turned on to stabilize all moduli, and in
these cases we simply have too few nonzero fluxes to stabilize $\tau$
completely.  Similar flat directions for flux vacua have been
discussed in e.g. \refs{\drs,\KST,\Freypol,\TT}, mostly in situations
where the fluxes preserve extended supersymmetry.

The vacua we are particularly interested in are the isolated vacua,
which we have found lie at the points
\eqn\soln{\tau = i \left({b_0 \over c^0}\right)^{1/4}~,}
where $b_0 c^0 > 0$.
{}From \vacuumas, we have
\equation\simplephi{
\phi = {P_1 (\bar{\tau})\over P_2 (\bar{\tau})}
={-3a \bar{\tau}^2 -b_0 \over
c^0 \bar{\tau}^3-3a \bar{\tau}} = \tau \,.
}
Precisely these vacua were considered in \S8.1 of \KST.

\bigskip
\noindent
{\it 4.2.2 Counting vacua}
\medskip

In enumerating vacua we have to make sure we are not double counting,
so we must understand the action of $SL(2,Z)_\phi \times SL(2,Z)_\tau$
on the fluxes; the latter can be determined using $dz = dx + \tau dy$
along with the definitions \cohbasis\ and \fis.  The $SL(2,Z)_\phi$
transformation $-1$ can be fixed by assuming $b_0, c^0 > 0$.  One
finds that although more general transformations of each $SL(2,Z)$
factor individually violate \specialflux, the combination S-duality
\eqn\specialtrans{
\phi \to - 1/\phi, ~\tau \to -1/\tau
}
preserves it with the induced action on the fluxes
\eqn\specialsym{
b_0 \leftrightarrow c^0 \,, \quad \quad a \; {\rm fixed}\,.
}
One can deal with this redundancy by counting only vacua with $b_0 \geq c^0$, which restricts \soln\ to the usual fundamental domain.

To determine the scaling of the number of vacua in this family,
we can use a continuous argument in which we ignore the discreteness
of the fluxes, as the set of allowed
fluxes forms a cone in the full 3-dimensional flux space we are
considering.  The boundaries of the cone are the conditions $0 < c^0
\leq b_0$, and we want to count points in this cone with $b_0 c^0 +
3a^2 \leq L$.
In the continuous approximation, the number of allowed fluxes
should be
\eqn\nint{N_{{\rm vacua}}(L) \sim {1 \over 2} \int_{1}^{L} db_0 \int_{1}^{L/b_0} dc^0
~2 \sqrt{{L-b_0 c^0}\over {3}}~,} where the integration over $a$ was
just the integral of 1 from $- \sqrt{(L - b_0 c^0)/3}$ to $\sqrt{(L-
b_0 c^0)/3}$, and we have dealt with the redundancy \specialsym\ in
this approximation simply by dividing by 2.  The bounds on $b_0$,
$c^0$ are determined by \tadpoc\ and the requirement that \soln\ give
only non-singular solutions, which rules out either vanishing.
Performing the integration yields
\eqn\ntot{N_{\rm vacua}(L) \sim {2\over {3\sqrt{3}}} L^{3/2}\, {\rm log}\, L \,.}
The logarithmic correction to power-law scaling for the number of flux
vacua we find here is notable.  Unlike the logarithmic term we found
in \S3.5.2, this logarithm is an example of the phenomenon
mentioned in \S2.3 in which the cone $C$ containing the region of
interest in moduli space intersects with the vanishing locus $Z$ of
the quadratic form $N_{\rm flux}$ on the sphere $S^{2b_3 -1}$, and
goes to infinity.  In this case, while the total volume of the cone
for the full torus is finite, we have taken a slice through this cone
of infinite area.  The divergence of this infinite area is cut off by
the lattice as $\int 1/x \rightarrow \sum_{i = 1}^{ N} 1/i \sim {\rm log}
~N$.  Because this divergence comes from a region where the cone approaches the vanishing locus $Z$, in fact the continuous approximation is not
really valid in this case and we should check the overall coefficient
by performing a discrete calculation.  It turns out, however, that in
this case imposing the lower cutoff on the fluxes at 1 precisely gives
the right coefficient, so the answer above is correct.

\medskip
\noindent
{\it 4.2.3 $W=0$ vacua}

For the vacua \soln\ , \simplephi\ the superpotential takes the form
\eqn\supsss{W(\phi=\tau=i(b_0/c_0)^{1/4})
=2\sqrt{{b_0\over c^0}}(3a-\sqrt{b_0 c^0})\, . }
$W=0$ precisely if $b_0 c^0 = 9a^2$.  This means $N_{\rm flux} = 12a^2$.
The number of $W = 0$ vacua with $N_{\rm flux} \leq L$ is then given
by\foot{Thanks to N.\ Elkies for explaining to us how to derive the
constant factor here.}
\eqn\nvac{N_{\rm vacua}(L; W=0) =
\sum_{a = 1}^{\lfloor\sqrt{{L\over 12}}\rfloor}  d (9a^2)
 \sim {3\over 2 \pi^2}\sqrt{L\over 12}({\rm log}(L))^2
+{\cal O}(\sqrt{L} \,{\rm log}\; L).}
Therefore, we see that the fraction of $W=0$ vacua in this simple
toy model is
\eqn\ratio{{N_{\rm vacua}(L; W=0)\over N_{{\rm vacua}}(L)} \sim {9\over
8 \pi^2}\,{\log L\over L}~.}

\medskip
\noindent
{\it 4.2.4 Discrete symmetries}

Two enhanced symmetry points of ${\cal G} = SL(2,Z)_\phi \times
SL(2,Z)_\tau$ could in principle arise for solutions with $\tau = \phi
= i$ and $\tau=\phi=\exp(2\pi i/3)$, but only the former is consistent
with \soln; this is the fixed point of the ${\bf Z}_2$ symmetry action
discussed previously, arising for fluxes $b_0 = c^0$.

As before, we can ask whether this transformation of the moduli and
fluxes descends to a symmetry only acting on the moduli when the
vacuum sits at the fixed point.  Since the combined transformation is
always a symmetry, this is equivalent to asking whether transforming
the fluxes only is a symmetry, as in \potinvartwo.  It is obvious from
\specialsym\ that at $b_0 = c^0$ this is indeed the case.  Checking
explicitly, one finds that 
\eqn\specialw{
W(\phi, \tau) \rightarrow W(-{ 1\over \phi}, -{ 1 \over \tau}) = {1 \over \phi \tau^3} W(\phi, \tau) \,.
}
Checking the action on the K\"ahler potential \kis\ as well,
we find that this transformation is indeed a K\"ahler
transformation on the low-energy theory, and hence a symmetry.

To count the enhanced symmetry points, we should count the integer
choices of $b_0, a$ satisfying
\eqn\nowsat{b_0^2 + 3a^2 \leq L~.}
The continuous approximation is valid here
\eqn\numvace{N_{\rm vacua}(L,\tau=\phi=i)
\sim 2 \int_{0}^{\sqrt{L}} db_0 ~\sqrt{(L-b_0^2)/3}\sim {\pi\over 2\sqrt{3}}\,L~.}
The ratio of
vacua with enhanced discrete symmetries to generic vacua therefore scales
like
\eqn\ratiot{{N_{\rm vacua}(L,\tau=\phi=i)\over N_{\rm vacua}(L)}
\sim {3\pi\over 4}\,{1\over {\sqrt{L}  \, \log  L }}~.}

\subsec{Symmetric $(T^2)^3$}

Now we consider the more general symmetric torus vacua characterized
by vacuum equations \vacuumas, \quintic\ with arbitrary flux parameters
$a^0, \ldots, d_0$.  As for
the special symmetric torus with restricted fluxes considered in the
previous subsection, we can again solve the equations completely,
although the details are somewhat more complicated.  We have used the
exact solution to numerically generate a large number of sample vacua,
which we can compare with a prediction of Denef and Douglas
\DD.
We can also give a simple algebraic characterization of  $W=0$ vacua
in the general case, which allows us to compute the asymptotic
fraction of flux vacua in this category.  We also discuss discrete
symmetry points in this model.

\medskip
\noindent
{\it 4.3.1  Vacuum equations}

Using again the parameterization $\tau = x + iy$, we can write the
quintic \quintic\ as a pair of coupled equations in $x, y$ of the form
\equation\simpletwo{
\eqalign{
q_1 (x) y^2 &  \;=  q_3 (x) \,,
\cr
q_0 (x) y^4 & \;= q_4 (x)\,,\cr }} where $q_i (x)$ are respectively
quartic, cubic, linear, and constant functions of $x$ for $i = 4, 3,
1, 0$ with coefficients which are polynomials in the flux parameters
$a^0, \ldots, d_0$ with integer coefficients.  These functions are
given explicitly in the Appendix.  Combining these equations, we have
\equation\sextic{
q_1 (x)^2 q_4 (x) = q_0 (x) q_3 (x)^2 \,.
}
Naively this looks like it should be a 6th order equation in $x$, but
remarkably the terms of order 4, 5, and 6 all cancel, giving us a
cubic equation for $x$,
\equation\cubic{
\alpha_3x^3 + \alpha_2x^2 + \alpha_1 x + \alpha^0 = 0 \,.
}
The coefficients in  this cubic equation are presented explicitly
in the Appendix.
Cubic polynomials can be solved explicitly in closed form, so for
any set of fluxes we can compute the values of $x$ satisfying \cubic.
When either $q_1 (x) \neq 0$ or $q_0 (x) \neq 0$ we can then solve for
$y$.  As we found in the previous subsection, certain flux
combinations do not fix the modular parameter $\tau$ uniquely.
For example, \cubic\ vanishes identically for the flux combination
\equation\vanishing{
a^0 =0, \;a = 8, \; b = 4,  \; b_0 =-2,  \;
c^0 =-1, \; c = 7, \; d = 3, \; d_0 = -2\,,
}
and thus $x$ is undetermined for these fluxes, indicating the presence
of a flat direction in the complex structure moduli space.  We are
primarily interested in vacua where all complex structure moduli and
the axio-dilaton are fixed, so we do not bother trying to classify the
fluxes with this property.

In general, we see that $x$ belongs to a field extension ${\cal F}_x$
of degree at most $3$ over ${\bf Q}$.  When $q_1 (x) \neq 0$, $y$
obeys a quadratic equation with coefficients in ${\cal F}_x$.  The
resulting $\tau = x + iy$ lives in a field extension ${\cal F}$ of
degree at most 6 over ${\bf Q}$.  For any particular values of the
flux parameters $a^0, \ldots, d_0$ we can solve the equations
explicitly and give algebraic expressions for the possible allowed
values of $\tau$.

\medskip
\noindent
{\it 4.3.2  Counting vacua}

In \DD, Denef and Douglas used the approach of \AD\ to estimate the
asymptotic number of vacua with $N_{\rm flux} \leq L$ on the symmetric
torus as $L \rightarrow \infty$, including a proper treatment of the
$SL(2,Z)_\phi \times SL(2,Z)_\tau$ gauge fixing.  They found that the
total number of vacua should go as $L^4$, and that the distribution on
the complex structure modulus $\tau$ should be proportional to
$1/({\rm Im} \; \tau)^2$.  We have used the exact solution described
in the previous subsection to perform an independent test of the
number and distribution of vacua on the symmetric torus.  We used a
Monte Carlo algorithm to generate a large number of solutions, gauge
fixing $SL(2,Z)_\phi$ with conditions on the fluxes, and gauge fixing
$SL(2,Z)_\tau$ by only keeping solutions with $\tau$ in the
fundamental domain.  Our numerical results are compatible with the
results of \DD, both for the total number of vacua and the
distribution on moduli space.  We found  the $L^4$ scaling for $L$
in the range $20 \leq L \leq 100$,
where we have $N_{\rm vacua} \sim C L^4$
with $C \approx 0.13$.
We have also confirmed numerically that when $N_{\rm
flux} > 20$, the distribution of vacua is compatible with $1/({\rm Im}
\; \tau)^2$.   For values of $L < 20$, the number of vacua  shrinks more
rapidly than $L^4$ as $L$ decreases, indicating that below this value
the number of vacua is suppressed due to lattice effects.  At small
values of $L$ the existing vacua are dominated by small values of
${\rm Im} \;
\tau$, indicating that the region of moduli space with large ${\rm Im}
\; \tau$ is near a void in the lattice structure.

\ifig\nstau{Distribution
of generic torus vacua in complex structure fundamental domain for $L=18$.}
{\epsfxsize=0.8\hsize\epsfbox{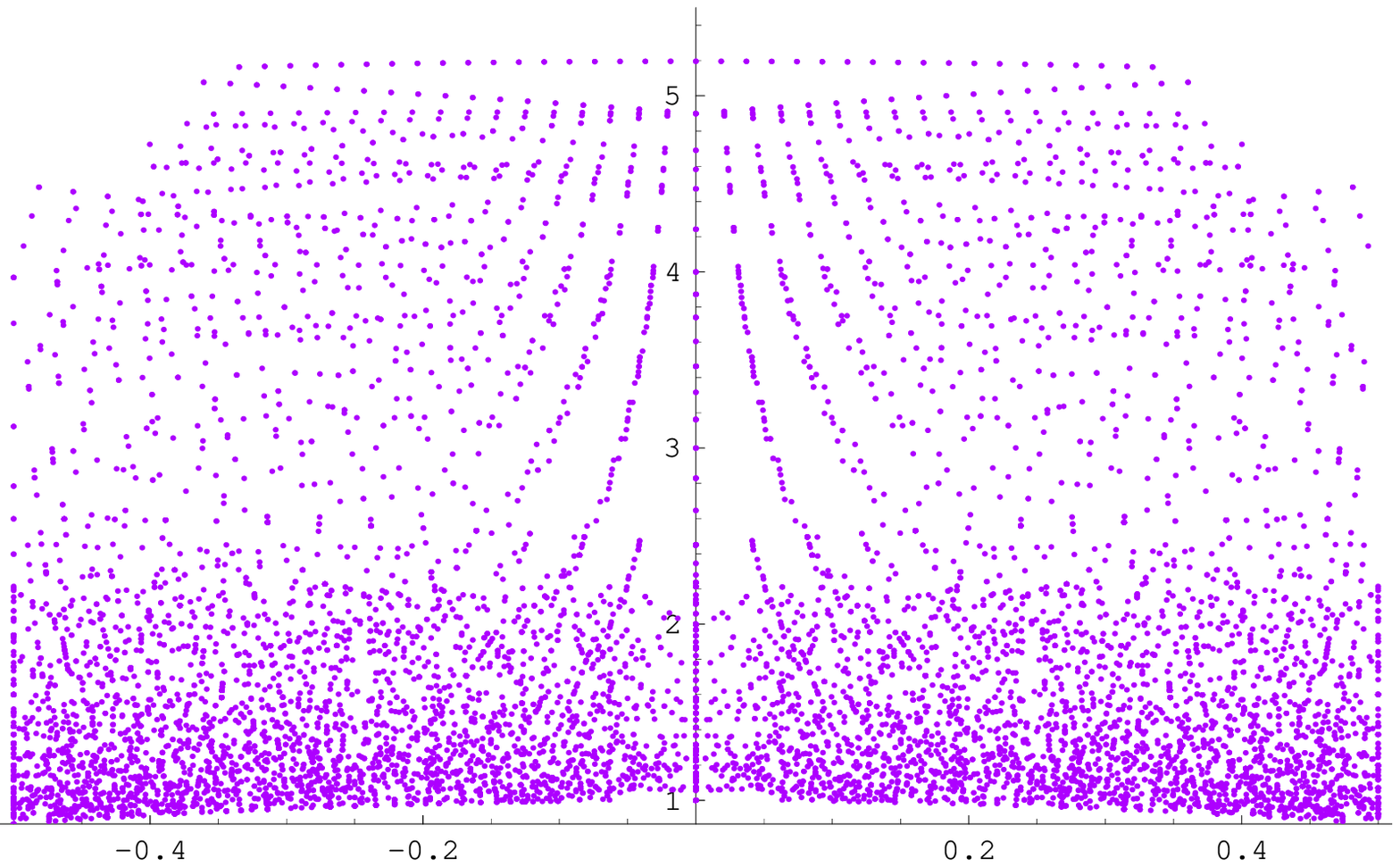}}

We have graphed the distribution of vacua in the $\tau$ fundamental
region for small values of $N_{\rm flux} \leq 18$ in \nstau.
As in the rigid CY model, for which a graph appears in \DD, there is quite
a bit of structure to the points.  In particular, something like the
void found in \DD\ appears in the vicinity of the imaginary axis.  This
phenomenon may be related to the large number of vacua which arise along this axis, including those described in \S4.2, many of which have
a discrete symmetry under $\tau \rightarrow -\bar{\tau}$, as we
discuss in \S4.3.4.
This void structure may also be related to the fact that the
imaginary axis is one place admitting one-parameter families of flux
vacua, as discussed in \S4.2.  Since we are now considering vacua
with periods which generally are associated with a field extension
over ${\bf Q}$ of degree up to 6, the allowed values of $\tau$ have
much greater freedom.  It would be nice to develop a better
understanding of the structure of this family of vacua.

Some comments on the details of our numerical investigation may be of
help in future efforts of this type.  We gauge-fixed the
$SL(2,Z)_\tau$ symmetry by choosing $a^0 = 0$, $0 \leq c < a$.  This
gauge-fixing is not complete, as it does not deal with situations
where $c^0 = 0$ or $ a = 0 $.  We did some further gauge fixing to
deal with most of the residual gauge symmetry, but these cases should
not contribute at leading order in $L$.  For each value of $N_{\rm flux} < L$ we
then chose random fluxes, selecting 5 of the nonzero fluxes in a range
$-L \leq b, d, \ldots \leq L$, and determining the remaining two
($c^0, b_0$) by choosing all possible factorizations of the value
$b_0c^0$ compatible with the other fluxes and \tadpoles. (This step increases the effectiveness of the algorithm
by a factor of $L^2$ since we essentially test $L^2$ values of the
fluxes at once.)

It is worth commenting on the range in which we
tested the fluxes.  While a gauge fixing of $SL(2,Z)_\phi$ by choosing
$\phi$ in the fundamental domain should give fluxes of order
$\sqrt{L}$, gauge fixing in flux space generically gives fluxes of
order $L$.  This is easy to check in the rigid CY model, where the fluxes
are uniformly distributed in a range $0
\leq H \leq L$.  For the torus model we are considering here,
empirical tests at a range of values of $L$, including a systematic
check for small $L< 9$ and a Monte Carlo check for $L < 100$ found no
vacua with individual fluxes $> N_{\rm flux}$, while vacua do arise
with individual fluxes which saturate the upper bound.

\medskip
\noindent
{\it 4.3.3 $W=0$ vacua}

Now let us consider the problem of finding vacua with $W = 0$ on the
symmetric torus.
It follows from \reducedw, \pa, and \vacuumas\  that to find
$W=0$ vacua, the modular parameter $\tau$ of the tori must solve
the two cubic equations
\eqn\pone{P_{1}(\tau) = a^0 \tau^3 - 3a \tau^2 - 3b \tau - b_0 = 0 \,,}
\eqn\ptwo{P_{2}(\tau) = c^0 \tau^3 - 3c \tau^2 - 3d \tau - d_0 = 0 \,.}
{}From the second equation of \vacuumas, we then have the dilaton
\equation\phipp{
\phi = {P_1'\over P_2'} \,.
}
The contribution to the D3-brane charge is again \tadpoles.

In order to be physical, $\tau$ must have a nonzero imaginary part.
But since the coefficients of \pone\ and \ptwo\ are real (and in fact
integral), complex roots must come in conjugate pairs.  Therefore, \pone\ and
\ptwo\ must share a common quadratic factor.
In fact as argued in \KST, they must factorize over the integers so that
\eqn\ponefac{P_1(\tau) = (f\tau + g) P(\tau)\,,}
\eqn\ptwofac{P_2(\tau) = k P(\tau)\,,}
where
\eqn\Pis{P(\tau) = l\tau^2 + m\tau + n \,,}
with $f, g, k,l,m,n \in {\bf Z}$.
Here we have used the  $SL(2,Z)_\phi$ symmetry to set $c^0=0$ and $0 \leq -b_0 < -d_0$, making $P_2(\tau)=0$ a quadratic equation.

This system of equations is closely parallel to that encountered in
\S3.5.2.  There we had a pair of fluxes in ${\bf Z}[i]$ with a common
factor.  We can use a very similar analysis to describe the number of
$W = 0$ vacua in this case.  Before going through the detailed
analysis, however, we can use the method of heights to give an
estimate of the scaling we expect.  In this case, the height is simply
the scale of the integer coefficients in the various polynomials.  As
mentioned in \S2.4, the height of a product of polynomials goes as the
product of the heights of the factors (this statement is transparent
for polynomials with integer coefficients).  We wish to know what fraction
of the pairs of cubic polynomials \pone, \ptwo\ share a common
quadratic factor.  The number of quadratics with coefficients of
height $q$ goes as $q^2$.  We multiply this quadratic by two linear
functions, each of height $\leq
\sqrt{L}/q$, as the coefficients in the cubics should be of height
$\sqrt{L}$.  The number of such linear functions is $L/q^2$.  Thus, we
expect the total fraction of cubics with coefficients of order
$\sqrt{L}$ which admit a common quadratic factor to go as
\equation\asymptoticw{
{N_{\rm vacua}(L; W=0)\over N_{\rm vacua}(L)} \sim{1\over L^4}
\sum_1^{ \sqrt{L}} q^2 \left( {L\over q^2} \right)^2 \sim
{1\over L^2} \sum_{ q = 1}^{\sqrt{L}} {1\over q^2} \,.  } Furthermore,
we expect the total contribution from quadratics of height $q$ to go
as $1/q^2$.

Note that while we have used the height here as a property of the
common quadratic polynomial factor $P$, there is a close connection
between this analysis and the algebraic properties of the modular
parameter $\tau$ where $P (\tau) = 0$.  The set of $\tau$ which
satisfy quadratic equations with integer coefficients are the
quadratic imaginary integers, which lie in complex extensions of
degree 2 over ${\bf Q}$.  By summing over all polynomials of all
heights, we are  equivalently performing a sum over all complex
degree 2 extensions of ${\bf Q}$ and over all $\tau$ which lie in each
of these extensions.  As we see in the following analysis, the
contributions to the sum naturally group themselves in terms of the
field extension, which is determined by the discriminant  of the
quadratic $P (\tau)$.

Now let us go through the analysis in more detail, following the same
approach as in \S3.5.2.
As there, we fix the redundancy in the parameterization
\ponefac, \ptwofac, \Pis\ by choosing ${\rm gcd} (l, m, n) = 1$;
$P (\tau)$ is thus a {\sl primitive} quadratic over the integers.
We furthermore fix the sign ambiguity by setting $l > 0$.
{}From the coefficients in \pone\ and \ptwo, we see that there are
further mod 3 constraints on the integers, so that
\eqn\threes{kl \equiv km \equiv gl+fm \equiv gm+fn \equiv 0 \; ({\rm
mod \; 3}) \,.}
In terms of the parameters $f, \ldots, n$, the tadpole condition
becomes
\eqn\tadpo{ 0 < (fk) (4ln-m^2) = 3N_{\rm  flux}\leq 3L~.}

We still need to fix the redundancy associated with
$SL(2,Z)_\tau$.  In this case,
it is convenient to do this by simply solving $P(\tau)=0$
and requiring that one of the roots lie in the fundamental
domain.
{}From
\equation\quadratic{
\tau ={-m \pm i \sqrt{4ln-m^2}\over 2l }\,,
}
we see that the necessary conditions are
\eqn\fundcon{
-l \leq m < l \leq n \,,}
with $m \leq 0$ when $l = n$.
These conditions guarantee that the discriminant is negative, or that
\equation\discriminant{
D(l, m, n) = 4ln-m^2 > 0 \,.
}
In addition, the previous $SL(2,Z)_\phi$ fixing requires
\eqn\kcons{0 \leq g < k ~.}
Note that the $SL(2,Z)_\tau$ fixed points of $\tau = i, \tau = \exp(\pi i /3)$
correspond to fluxes living in an $SL(2,Z)_\phi$ conjugacy class which
is invariant under the action of the relevant $SL(2,Z)_\tau$
generators.  For example, consider $P_1 = \tau^3 + \tau, P_2 = \tau^2
+ 1$.  Under the action of $S_\tau: \tau \rightarrow -1/\tau$, we have
$P_1 \rightarrow -P_2, P_2 \rightarrow P_1$, which can be put back in
canonical $SL(2,Z)_\phi$ form by $S_\phi$, which returns us to the
original polynomials.  The consequence of this is that we weight all
vacua, including those at the ${\bf Z}_2$ and ${\bf Z}_3$ points, with weight 1.

To compute the total number of vacua at fixed $N_{{\rm flux}}$, we
now simply sum over all combinations of integers satisfying $\tau
\in {\cal F}_{\rm D}$ (the fundamental domain), \threes, \tadpo,
and \kcons. Aside from the mod 3 constraints, this is a very
similar problem to that encountered in \S3.5.2. Let us first
consider a case where the mod 3 constraints are simple to
incorporate, namely when $D (l, m, n) \neq 0 ~({\rm mod }~3)$.  In
this case, the constraints \threes\ simply force the integers $f,
g, k$ to all be multiples of 3.  We write $f = 3\hat{f}, g = 3
\hat{g}, k = 3 \hat{k}$.  We now can write the sum for these
values of $D$ as (using $r = \hat{f} \hat{k}$) \equation\simplesum{
\sum_{l, m, n:
\tau \in{\cal F}, 3 | D} \; \; \; \sum_{r = 1}^{\lfloor{L\over 3
D}\rfloor} \; \; \sum_{\hat{k} | r} \; \hat{k} \,. } The factor of
$\hat{k}$ inside the summation includes all possible values of $g < k$
(including $g = 0$).  This summation can be written in terms of
standard number-theoretic functions.  The number of relatively prime
integers $l, m, n$ giving rise to a $\tau$ in the fundamental domain
with fixed $D$ is known as the class number $h (D)$.  For a discussion
of class numbers in the physics literature in a related context, see
\mooreaa.  The sum over $\hat{k}$ is just the sum over divisors of
$n$.  Thus, our summation in the case where $3 | D$ reduces to the
number-theoretic sum
\equation\ntone{ \sum_{D \in 3 {\bf Z}} h (D) \sum_{n =
1}^{\lfloor{L\over 3 D}\rfloor} \sigma (n) \,. } The mod 3
relations \threes\ only make a small change to this analysis.
Consider for example the case where $ 3 | D$ because $l \equiv m
\equiv 0$ (mod 3).  In this case the mod 3 constraints only force
$f = 3 \hat{f}$, while $g, k$ can be arbitrary.  The summation
above is now exactly the same except that instead of $3D$ in the
denominator of the upper bound for $n$, we just have $D$.  One can
check that in all cases where $D \equiv 0$ (mod 3), the correct
summation always has the factor of $D$ in place of $3D$.  Thus, we
can write our final formula for the number of $W = 0$ vacua on the
torus \equation\nttwo{ N_{\rm vacua} (L; W=0) = \sum_{D \in  {\bf
Z}} h (D) \sum_{n = 1}^{\lfloor{L\over t (D)}\rfloor} \sigma (n)
\,, } where \equation\tlmn{ t ( D) =\cases{D, & if $ D \equiv 0 \;
({\rm mod \; 3})$;\cr 3 D, & otherwise.\cr} } The fact that the
total number of vacua in this case can be expressed in terms of
the class number and other number-theoretic quantities was
anticipated in \mooreleshouches, and seems indicative of a deep
relationship between $W = 0$ vacua and arithmetic.

We can furthermore separate out the contributions to the total number of $W=0$ vacua \nttwo\ based on the value of the modulus $\tau$, given by the solution \quadratic\ to the primitive quadratic parameterized by relatively
prime integers $(l, m, n)$:
\equation\contribution{
N_{\rm vacua}(L; \tau = \tau_{l,m,n}; W=0) = \sum_{\hat{k}}^{L\over t (4ln-m^2)}
\hat{k} \;\lfloor{L\over \hat{k}\; t (4ln-m^2)} \rfloor
\sim{\pi^2\over 12} \; {L^2\over t (4ln-m^2)^2} \,.
}
For example, the number of vacua at $\tau = i$, associated with the
quadratic $P (\tau) = \tau^2 + 1$ and the integers $(l, m, n) = (1, 0,
1)$ is $\pi^2/1728 \;L^2\approx 0.00571 \; L^2$, while the number of vacua at
$\tau= \exp( \pi i /3)$ ($P = \tau^2 -\tau + 1$) goes as
$\pi^2/108 \; L^2 \approx 0.09139 \; L^2$.

{}From \contribution\ we see that, just as calculated using the height
argument, the contribution from
each primitive quadratic goes asymptotically as $L^2$, with a
coefficient which is of order
$1/q^4$.
The $q^2$ primitive quadratics at height $q$ combine to give a total
contribution at height $q$ of
$1/q^2$.
Using $\sum 1/n^2 = \pi^2/6$, a rough estimate for the total number of vacua is then given by $0.097 \;(\pi^2/6) \;L^2 \approx 0.159 L^2$.

To determine the precise asymptotic coefficient of $L^2$ in the total
number of vacua, we now simply need to carry out the sum
\equation\coefficientll{
N_{\rm flux}(L; W=0) =
{\pi^2 \over 12}
 \sum_{D} {h (D)\,L^2 \over t (D)^2} \;
=\;{\pi^2\over 12}
\sum_{l, m, n: \tau \in{\cal F}_{\rm D}}
{L^2 \over t (4l n-m^2)^2} \approx 0.143 \,L^2 \,.
}
This sum converges quite rapidly.  The contributions at fixed $q = n$
seem to go as ${{\rm log} ({\rm log}~ q) \over q^2}$,
indicating that more subtle
number theoretic considerations correct the heuristic argument above
by ${\rm log} ({\rm log}~ q)$ terms.  Note
also that the contributions at different
$q$ jump about with a periodicity of 6 due to the divisibility
properties of the integers.

\ifig\figIII{Log of the exact number of $W=0$ vacua with $N_{{\rm flux}} < L$
as a function of ${\rm log}~ L$ fit by predicted
curve ${\rm log}~(0.143)+ 2~{\rm log}~ L$ up to $L=5000$.}
{\epsfxsize=0.8\hsize\epsfbox{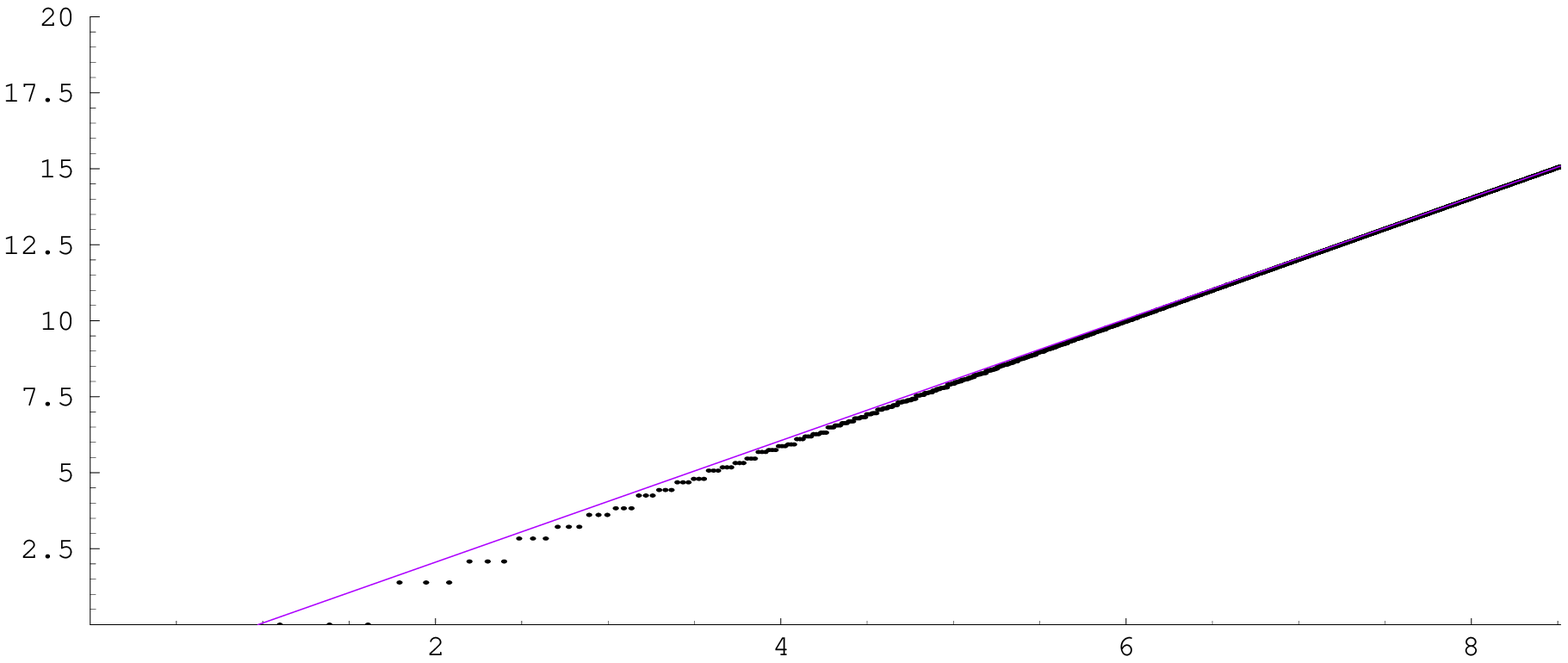}}

In \figIII\ we show
the exact number of $W=0$ vacua with $N_{{\rm flux}} < L$ and the curve
representing the asymptotic form at large $L$ in a log-log plot.
Note that while the sum \coefficientll\ converges rapidly in $q$,
the exact value of $N_{\rm flux}$ approaches the
asymptotic form much more slowly.

\ifig\figIa{Distribution
of $W=0$ vacua in complex structure fundamental domain for $L=5000$
for small values of Im $\tau$.}
{\epsfxsize=0.8\hsize\epsfbox{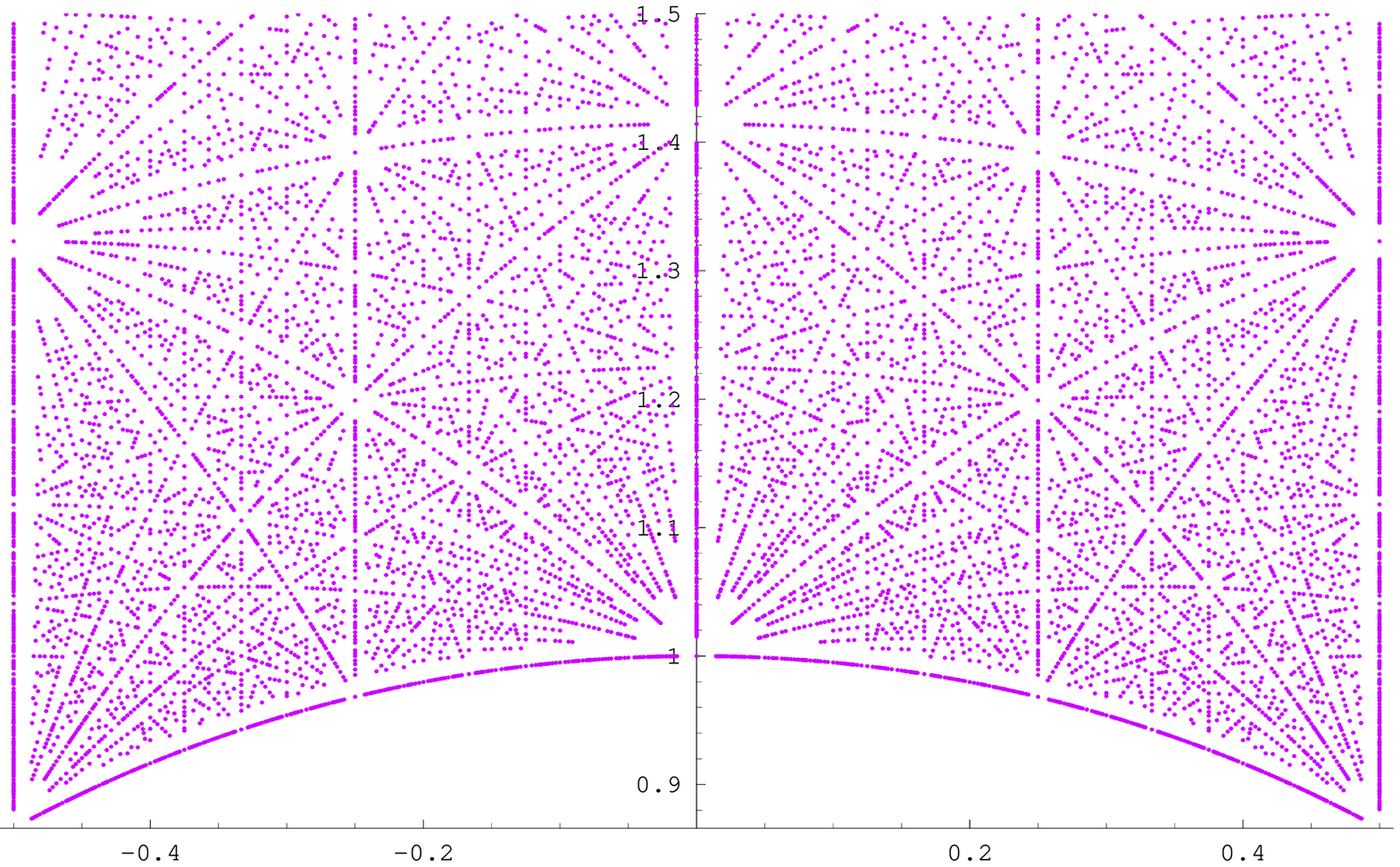}}

\ifig\figIb{Distribution
of $W=0$ vacua in complex structure fundamental domain for $L=2000$
for large values of Im $\tau$.}
{\epsfxsize=0.8\hsize\epsfbox{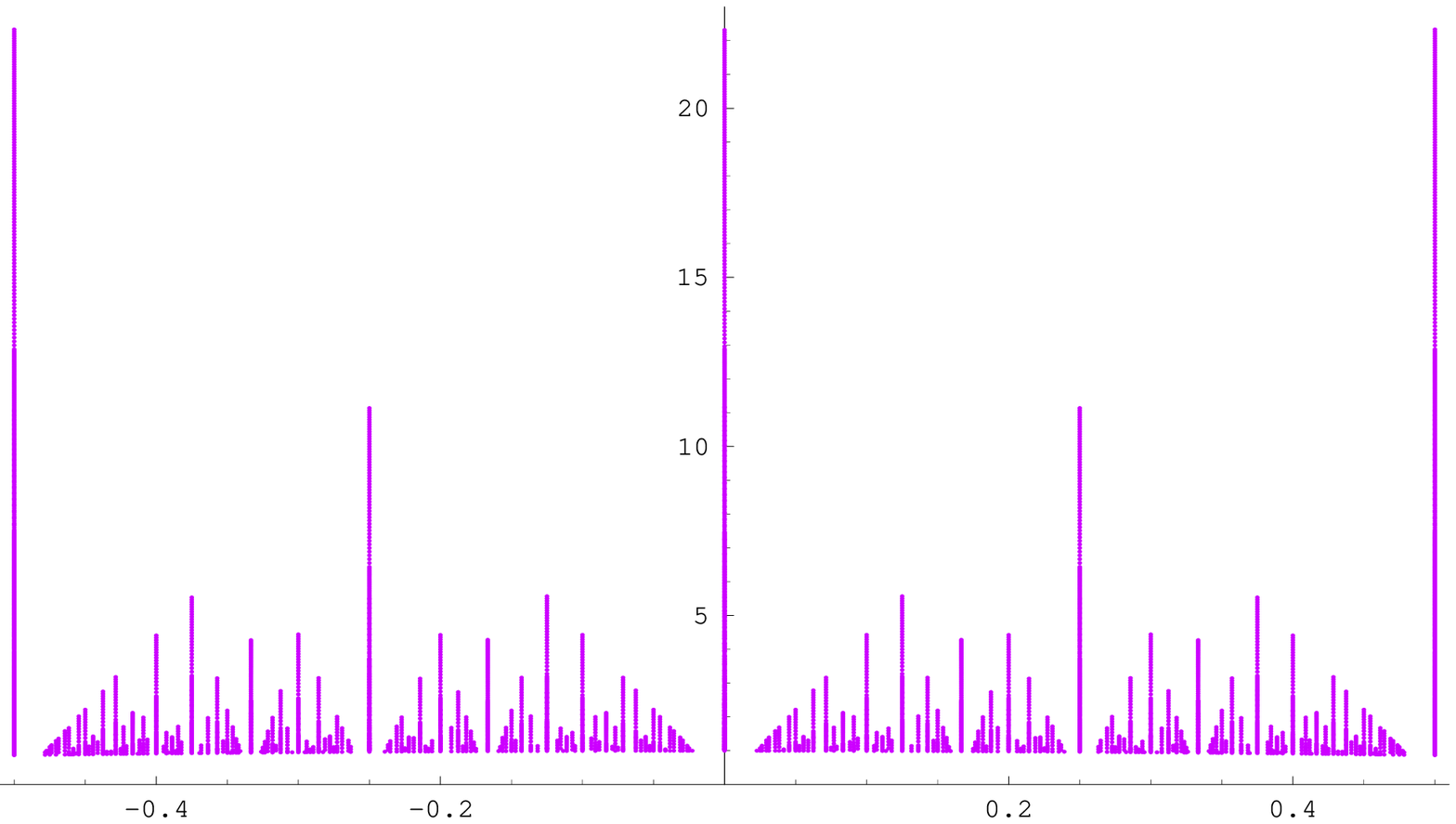}}

In \figIa\ and \figIb\ we illustrate the distribution of $W=0$ vacua
in the complex structure fundamental domain for small and large values
of ${\rm Im}~ \tau$ respectively.  The curves visible in \figIa\ can be interpreted as the loci of 
values of $\tau$ associated to vacua with a linear constraint imposed on $(l,m,n)$; using \quadratic, one sees that constraints not involving $n$ will fix $x = {\rm Re}~\tau$, while for $n= C_1 m + C_2 l$ one relates $x$ to $y = {\rm Im}~\tau$ as $y^2 + x^2 = -2C_1 x +C_2$.

The large scale peak structure in \figIb\  can be deduced from
\quadratic\ with the additional modular group fixing constraint \fundcon.
In particular, the heights of the maximal peaks are given by
\eqn\maximtau{{\rm max}({\rm Im}~\tau)\sim {\sqrt{L}\over 2}~
~~~~~{\rm at}~~~{\rm Re}~\tau=0,\pm 0.5~. }
The next peaks have height
${\rm Im}~\tau\sim\sqrt{L}/4$ at Re$~\tau=\pm 0.25$.
We also confirm numerically that the distribution
of vacua in the complex structure fundamental domain is
in accord with $1/({\rm Im}~ \tau)^2$.

\ifig\figIc{Void structure of distribution
of $W=0$ vacua in dilaton fundamental domain for $L=600$.}
{\epsfxsize=0.8\hsize\epsfbox{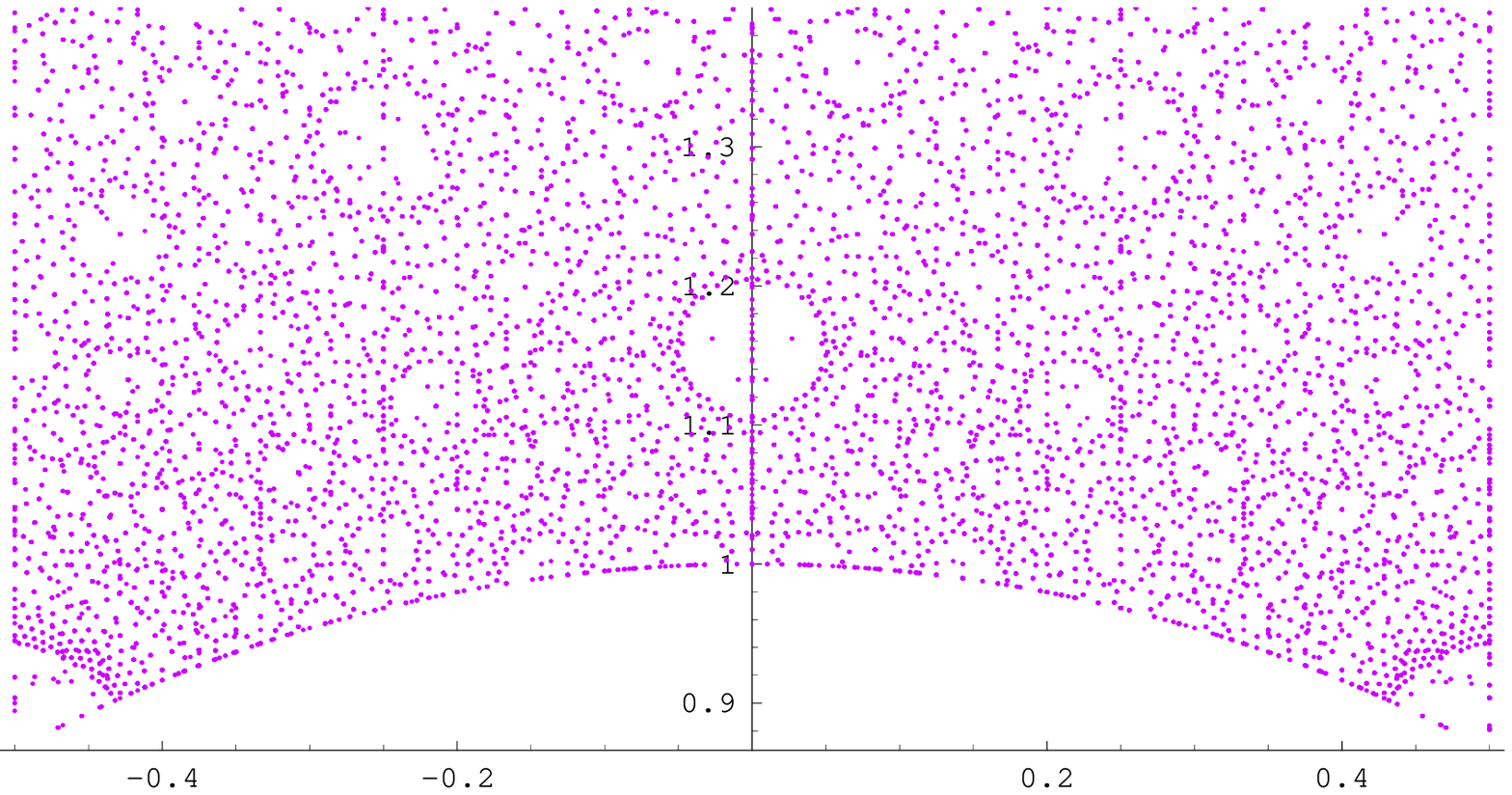}}

In \figIc\ we illustrate the distibution of $W=0$ vacua
in the axio-dilaton fundamental domain,
where the dilaton following \phipp\
is given by
\eqn\dilatonnnn{\phi={f\tau+g\over k}~,}
which we map to the fundamental domain. The void structure that appears
in this case has an analogous explanation to the voids which appeared
in the rigid CY case.

\medskip
\noindent
{\it 4.3.4 Discrete symmetries}

Two enhanced symmetry points arise for $(T^2)^3$, related to fixed points of the modular group ${\cal G} = SL(2,Z)_\phi \times SL(2,Z)_\tau$:
$\phi=\tau=i$ and $\phi=\tau=\exp(\pi i/3)$, with diagonal $SL(2,Z)$ transformations of the dilaton and complex structure.
We also have a complex conjugation symmetry which takes $\tau
\rightarrow -\bar{\tau}, \phi \rightarrow -\bar{\phi}$ similar to the
symmetry of the rigid CY discussed in section \S3.5.2.  We discuss
each of these discrete symmetries in turn.

\medskip
\noindent
{\it  ${\bf Z}_2 \subset SL(2,Z)$}

For the point $\phi=\tau=i$ there is a combined ${\bf Z}_2$ discrete symmetry
\eqn\identity{ \tau\rightarrow\tau'=-{1\over\tau},~~~~~~
\phi\rightarrow\phi'=-{1\over \phi}.
}
This symmetry arises for both $W=0$ and generic no-scale flux
solutions at the fixed point.  From
\vacuumas\ we determine the fluxes leading to vacua at this fixed point:
\equation\torusdiscretea{
a^0 = -d_0, \; \;
b_0 = c^0, \; \;
a =-d , \; \;
b = c\,.
}
The superpotential determined by fluxes \torusdiscretea\ then
transforms under \identity\ as:
\eqn\superpottr{W(\tau,\phi)
\rightarrow W(\tau',\phi')={1\over \tau^3\phi}W(\tau,\phi).}
We can count the number of vacua at $\phi = \tau = i$ easily by simply
imposing the corresponding linear conditions on the flux.  Since the
moduli have heights of order 1, we expect the number of vacua to scale
as $L^2$ with a coefficient of order 1 when we only impose the $DW =
0$ conditions \vacuumas, and as $L$ when we also impose the $W = 0$
condition.  We have
\equation\torusbounda{
N_{\rm flux} = 3a^2 + 3b^2 + (a^0)^2 + b_0^2 \leq L~,
}
so the total number of vacua is
\equation\torusnumbera{
N_{\rm vacua}(L; \phi = \tau = i)  \sim{\pi^2\over 12} L^2 \,.
}
Imposing the further constraint $W = 0$ gives us the further relation
on the fluxes
\equation\torusdiscreteb{
a^0 =- 3b , \; \;
b_0 = 3a\,,
}
leading to
\equation\torusboundb{
N_{\rm flux} = 12 (a^2 + b^2) \leq L~.
}
Then the total number of vacua with $W = 0$ and the discrete ${\bf Z}_2$ symmetry is
\equation\torusnumberb{
N_{\rm vacua}(L; \phi=\tau=i;W=0) \sim{\pi\over 24} L \,.
}

\medskip
\noindent
{\it  ${\bf Z}_3 \subset SL(2,Z)$}

For the point  $\phi=\tau=\exp(\pi i/3)$ there is a combined ${\bf Z}_3$ R-symmetry
\eqn\identityb{ \tau\rightarrow\tau'=-{1\over\tau-1},~~~~~~
\phi\rightarrow\phi'=-{1\over \phi-1}~.
}
In this case the symmetry arises only for $W=0$ solutions at the fixed point; the fluxes leading to these vacua are
\equation\torusdiscreteb{
\eqalign{
a^0 = -2b + d\,, \quad b_0 = c^0 = -b-d \,, \quad d_0 = b-2d \,, \quad
a = -b\,, \quad c = -d \,.
}}
The superpotential for such fluxes as a function of the complex
structure and dilaton transforms as
\eqn\superpottrb{W(\tau,\phi)
\rightarrow W(\tau',\phi')={1\over (\tau-1)^3(\phi-1)}W(\tau,\phi)~.}
In particular, at
the fixed point $W \to -\exp(\pi i/3) W$ which requires $W=0$.
We expect again that the number of vacua at this point will go as $L$
with a constant coefficient of order 1.
The value of $N_{\rm flux}$ given the fluxes \torusdiscreteb\ is
\equation\torusboundb{
N_{\rm flux} = 3 (b^2 -bd + d^2) \leq L.
}
Hence the total number of vacua with $W = 0$ and the discrete ${\bf Z}_3$ symmetry is
\equation\torusnumberb{
N_{\rm vacua}(L; \phi = \tau = \exp(\pi i/3); W=0) \sim{\pi\over 3 \sqrt{3}} L \,.
}

\break
\noindent
{\it  Complex conjugation ${\bf Z}_2$}

The torus possesses a complex conjugation symmetry ${\cal C}$ of the form \cconj, with $U = {\rm diag} (-1, 1, -1, 1)$; the full action is
\equation\toruscc{
a^0 \rightarrow -a^0\,, \quad
b  \rightarrow -b \,, \quad
c  \rightarrow  -c\,, \quad
d_0  \rightarrow -d_0\,, \quad
\tau  \rightarrow - \bar{\tau} \,, \quad
\phi  \rightarrow -\bar{\phi}\,.
}
For certain combinations of fluxes, the action of ${\cal C}$ takes the fluxes to themselves, up to a possible overall sign.  For such fluxes, the transformation $\tau  \rightarrow - \bar{\tau}$, $\phi  \rightarrow -\bar{\phi}$ acting on the moduli alone describes a symmetry of the low-energy theory when combined with four-dimensional parity; as for the complex conjugation symmetry discussed in section \S3.5.2, this is an antiholomorphic transformation combined with a possible K\"ahler transformation (if there is a sign).

The points in moduli space which are invariant under the action of
this symmetry are those for which both $\phi$ and $\tau$ are
pure imaginary, so we can write
\equation\expandpt{
\phi = i | \phi |, \;\;\;\;\; \;\;\;\;\;
\tau = i | \tau | \,.
}
At such points, the equations \vacuumas\ take the simplified form
\equation\ctorus{
\eqalign{
d_0 | \phi | + a^0 | \tau |^3 = 0 \,, \quad
b = c | \phi | \; | \tau | \,, \quad
b_0 = c^0 | \phi | \; | \tau |^3 \,, \quad
d | \phi | + a | \tau | = 0 \,.
}
}
{}From \simpletwo, where in this case $x = 0$, $y = | \tau |$, we see that
$\tau^4$ is always rational, and that sometimes $\tau^2$ is rational,
so that all solutions of \ctorus\ are associated with field
extensions of degree either 2 or 4.

Let us first take the case where
the field extension is of degree 4, so that $\tau = i (p/q)^{1/4}$,
with ${\rm gcd} (p, q) = 1$ and such that $p, q$ are not both perfect
squares.  In this case,  $| \tau |$ and $| \tau |^3$ are independent
over ${\bf Q}$.  For nonzero $\phi$ this means that the equations
\ctorus\ cannot all be satisfied nontrivially: either all
terms in the first two equations must vanish or all terms in the
second two equations must vanish.  This means we must set either $a^0 = b = c = d_0 = 0$ or $a = b_0 = c^0 = d = 0$, with the nonvanishing fluxes obey
two relations with $\phi$, $\tau$.
The nonzero fluxes in this case are precisely those which are taken to themselves up to a possible sign under the action of ${\cal C}$.  Thus, the vacua where $\phi$, $\tau$ are both pure
imaginary and live in the field extension of degree 4 over ${\bf Q}$
are precisely those where we expect a low-energy symmetry.

 We can use
a continuous flux argument to enumerate these vacua, as the
modular parameters are determined from the fluxes, so we expect order
$L^2$ vacua in this case.  In particular, in the situation where the
all terms in the first two equations vanish, we have $a^0 = b = c =
d_0 = 0$, leaving us with the fluxes $a,b_0, c^0, d$ satisfying the
last two equations in \ctorus.  This is almost identical to the vacuum
counting problem from \S4.2.2, except that now we no longer impose the
condition $d = -a$.  This gives us an extra factor of $\sqrt{L}$ above
\ntot.  As in that case, however, we expect additional ${\rm log} \;
L$ terms here also.

Note that the structure of the constraint
equations has simplified dramatically in this set of vacua.  In
general, for an extension of degree 4, we would expect 8 equations for
the 8 fluxes.  Because of our choice of fluxes and the structure of
the equations which results when we choose moduli of the particular
form $i(p/q)^{1/4}$, these equations have collapsed to only two
constraints relating the remaining fluxes and the modular parameters.
Applying the height-based arguments of \S2.4 in this situation to, for
example, the last two equations in \ctorus, we see that the last
equation suggests that if $\tau$ is defined by a quartic equation $q
\tau^4+ p = 0$ of height $H = {\rm max} \; (| q |, | p |)$, then
$\phi/\tau$ has height $\sqrt{L}$, so the third equation gives a
single constraint reducing the $L^2$ flux combinations by a factor of
$\sqrt{L} \cdot \sqrt{L}H$.  There are order $H$ $\tau$'s at height
$H$, each of which then contributes $ \sim L/H$ vacua, for an expected
total of $L^2$ vacua distributed fairly uniformly across the height of
$\tau$.

In the case where the field extension is of degree 2, the situation is
more complicated.  An argument based on heights suggests that
there are order $L^3$ such vacua.  These vacua, however, do not have the complex conjugation 
symmetry of the low-energy theory.

\newsec{Vacua in simple Calabi-Yau hypersurfaces}

We turn now to the last class of examples, the set of one-parameter
Calabi-Yau manifolds defined as hypersurfaces in weighted projective
space.  We analyze these vacua at and around a particular point in moduli space, the Landau-Ginzburg point.  We find that the structure of the number field in which the periods there take values plays a fundamental role in determining the nature of the solutions.  We focus on the two models in which vacua can arise at
the LG point, analyzing in each case the total number of vacua,
the number of vacua with $W = 0$, and the vacua with discrete
symmetries.  We find that in one case the generic arguments in \S2.4
are violated due to a convergence in the constraint equations, so that
more vacua arise than naively expected.  We find in general that
Landau-Ginzburg is a very special point in the moduli space, where the
number of vacua with $W = 0$ and with discrete symmetries can be of the same order as the total number of vacua, when nonzero.

The Calabi-Yau manifolds of interest, $M_k$ for $k=5,6,8,10$,
are defined by the equations
\eqn\cyeqs{\eqalign{&k=5:~\sum_{i=0}^{4} x_i^5 - 5 \psi \, x_0 x_1 x_2
x_3 x_4 = 0 \,, \quad x_i \in {{\bf P}^{4}} \,,\cr
&k=6:~\sum_{i=1}^{4}x_{i}^6 + 2 x_{0}^3 - 6 \psi\, x_0 x_1 x_2 x_3 x_4 = 0 \,, \quad
x_i \in {{\bf WP}^{4}}_{1,1,1,1,2} \,, \cr
&k=8:~\sum_{i=1}^{4} x_i^8 + 4 x_{0}^2 - 8 \psi \, x_0 x_1 x_2 x_3 x_4 = 0 \,, \quad
x_i \in {{\bf WP}^{4}}_{1,1,1,1,4}\,, \cr
&k=10:~\sum_{i=2}^{4} x_{i}^{10} + 2 x_{1}^5 + 5 x_{0}^2 - 10 \psi \, x_0 x_1 x_2 x_3 x_4 = 0\,, \quad x_i \in {{\bf WP}^{4}}_{1,1,1,2,5} \,.}}
The geometry of the $M_k$ spaces was
studied in \KT\ following the seminal work of Candelas et al.\ on
the quintic \cdgp\ (see also
\refs{\font, \morrison} for discussions of the one-parameter models), and
we follow their notation.  The mirrors $W_k$ of these
spaces have one complex structure modulus,
parametrized by $\psi$ in \cyeqs, and
are obtained by
dividing out by the appropriate discrete group ${\cal G}_k$ \GP.  In
the spirit of \GKTT, however, we work with the $M_k$ rather than their mirrors.
As long as we turn on only ${\cal G}_k$-invariant fluxes, those complex structure moduli that are not invariant under ${\cal G}_k$ can be set to zero, as they are in \cyeqs, leaving us with effectively one-parameter models.  This simplification is analogous to that made in the previous section to reduce to a model with a single $\tau$ modulus.

The moduli space of $\psi$ for each $M_k$ is quite similar: all four
have (mirror) Landau-Ginzburg points, conifold points, and
large-complex structure points at $\psi = 0, 1, \infty$ respectively.
For these one-parameter models, we focus on physics at and near the
Landau-Ginzburg (LG) point $\psi = 0$, which is a fixed point of the
modular group.  Because we work with a perturbative expansion of the periods around this
point, while we can solve perturbatively for solutions of the vacuum
equations $DW = 0$, we cannot identify points in the moduli space with
$W = 0$ or with discrete symmetries except precisely at the LG point,
since these characteristics cannot be definitively identified in
perturbation theory.

The period vector of each of the $M_k$ at $\psi = 0$ is associated
with a different field extension over ${\bf Q}$, and this algebraic
structure is vital to understanding the kinds of vacua that exist (or
fail to exist) there.  For each of the models, we ask whether vacua
exist at the LG point, whether $W=0$ vacua exist there, under what
circumstances discrete symmetries arise in the low-energy theory at
this point, and how many vacua with each of these properties there
are.  Despite the naive similarity of the models, we shall see that
$M_6$ and $M_8$ give rich sets of solutions on the LG locus, with
accompanying discrete symmetries and $W=0$ vacua in some cases, while
$M_5$ and $M_{10}$ do not.  In \S5.4, we also count vacua in a small
neighborhood around the LG point and the conifold point in $M_8$, to
check the predicted $L^{b_3}$ scaling.

Near the Landau-Ginzburg point $\psi = 0$ the periods admit an expansion \KT:
\eqn\wo{ w_{i}(\psi)= (2\pi i)^3 {1 \over k}
\sum_{n=1}^\infty {\exp({(k-1)\pi i \over k}n)\Gamma({n \over
k})\over\Gamma(n) \prod_{a=1}^4 \Gamma\left(1 - {n \nu_a \over k}\right)}
(\gamma\alpha^{i})^{n}\psi^{n-1} \,, }
valid for $|\psi|<1$.  Here
$\alpha$ is the $k^{th}$ root of unity $\alpha=\exp(2\pi i / k)$, $\gamma=k \prod_{a=0}^4 (\nu_{a})^{-\nu_a/k}$ and $\nu_a, a = 0,1,\ldots 4$ are the indices
characterizing the weighted projective space.  We have rescaled the
holomorphic three-form $\Omega$ by $1 / \psi$ relative to
\KT.

A convenient choice for the period vector is $w^{T}=(w_2(\psi),w_1(\psi),
w_0(\psi),w_{k-1}(\psi))$.  The periods in a symplectic basis $\Pi^{T}=({\cal
G}_{1},{\cal G}_{2},z^1,z^2)$ can be expressed in terms of this Picard-Fuchs
basis by means of the linear transformation $\Pi=m_k \cdot w$ where the
matrices $m_k$ have rational elements and are given in \KT.  The
symplectic basis period vector $\Pi$ then has the generic expansion
\eqn\perr{\Pi=
c_0 ~m_k \cdot \pmatrix{\alpha^2 \cr \alpha \cr 1 \cr \alpha^{k-1}}
 +c_1\psi~m_k \cdot \pmatrix{\alpha^4 \cr \alpha^2 \cr 1 \cr \alpha^{2(k-1)}}
 +{\cal O}(\psi^2)\equiv
 c_0  ~p_0
 +c_1 \psi ~p_1
 +{\cal O}(\psi^2) \,, }
where the $c_i$ are complex constants and the
$p_i$ are four-vectors with components that are rational linear combinations of powers of $\alpha$.

\subsec{Vacua at the Landau-Ginzburg point}

We turn on flux vectors $f \equiv (f_1, f_2, f_3, f_4)$ and $h \equiv (h_1, h_2, h_3, h_4)$.  The F-flatness conditions \fflat\ at $\psi = 0$ are then
\eqn\onevac{ (f - \bar\phi h) \cdot p_0 = 0 \,,
\quad \quad (f - \phi h) \cdot p_1 = 0 \,. }
In the cases $k=8$ and $k=10$ a subtlety arises: as is visible in \wo, $c_1 = 0$ in this case, and naively the $D_\psi W$ equation is trivial.  However, another consequence is that the inverse metric $g^{\psi \bar\psi}$ diverges, such that the vanishing of $g^{\psi \bar\psi} D_\psi W D_{\bar\psi} \overline{W}$ is nontrivial.  The equations \onevac\ should then be imposed with the vector $p_2$ substituted for $p_1$.

\medskip
\noindent
{\it 5.1.1 Vacuum counting and cyclotomic fields}

To solve the equations \onevac, we must consider the algebraic properties
of the vectors $p_i$.   The $p_i$ take values in rational combinations of
powers of the $k^{th}$ root of unity $\alpha$, which defines  the cyclotomic
 field ${\cal F}({\bf Q})_k$.  As we discussed in \S2.4, the degree of
 ${\cal F}_k$ is given by the Euler totient function ${\cal D}_k = \phi(k)$ \Euler.
 As is by now familiar, each equation in \onevac\ becomes ${\cal D}_k$ constraints
 on the integer fluxes when the moduli are specified.
Since there are 8 flux integers, we need $2\phi(k) < 8$ or $\phi(k) < 4$
if we are to generically expect nontrivial solutions; for $\phi(k)=4$ there
are as many equations as fluxes, but one would expect vanishing fluxes to
 provide the unique solution.
This is just the application of the logic that went into \nonsusy\ to this
 particular class of examples.

For the various one-parameter models, one finds $\phi(6) = 2$, but $\phi(5)=\phi(8)=\phi(10) = 4$.  Hence according to
the argument above, we expect nontrivial flux vacua at the Landau-Ginzburg
point on $M_6$, but not for the other spaces.  As we shall see in the next section,
there are indeed LG vacua in the $M_6$ model.  However, we shall find
that for certain restricted choices of the dilaton, there are LG vacua for $M_8$ as well.

These LG vacua at $M_8$ exist despite the counting of equations because
although \onevac\ produces $2 \phi(8) = 8$ conditions on the fluxes,
they are not all independent; for certain choices of the axio-dilaton
$\phi$, the equations degenerate into fewer conditions on the
fluxes. This was one of the loopholes we anticipated in \S2.4, that
can invalidate the expected scaling \nonsusy\ and yield larger numbers
of vacua.  Another example of this reduction in constraints appeared
for the complex conjugation symmetry on the torus discussed in \S4.3.4.

Returning to the logic of the analysis that led to \nonsusy, we can try to parametrize our expectations for this
nongeneric situation.
If for a given choice of the axio-dilaton VEV there are $m$ {\it relations} between
the $\eta = {\cal D}(b_{2,1}+1)$ equations on the integer flux parameters, as we shall find in the $k=8$ case, then one should modify the definition of $\kappa$ in \nonsusy: one finds an effective  $\kappa_{eff} = 2 b_3 -(\eta-m)$, and
the expected scaling of the number of solutions with $L$ is $N_{\rm vacua} \sim L^{(\kappa_{eff})/2}$.  For
the $k=8$ model we shall find, for special values of $\phi$, that
$m=4$.  For $M_5$ and $M_{10}$, this phenomenon does not occur
and there are no vacua at the LG point, as expected.

The type of analysis we carry out in this section should be applicable
for any Calabi-Yau which admits flux vacua where the periods have
ratios which are valued in an extension of the rationals of finite
degree.  While as we saw, for the torus all flux vacua have this
property, it may sound like an unusual condition for flux vacua on a
Calabi-Yau.  But in fact it has been shown to occur at the LG locus in
a wide variety of multiparameter models \BergKatz.
For example in one of the two-parametric CY models
there is a flat direction of $W=0$ vacua
constructed in \GKTT, with arithmetic periods---though in this case
the dilaton is not constrained due to the flat direction.
It may be that
this sort of arithmetic structure for the periods is in fact a
reasonably common occurrence for flux vacua.

\medskip
\noindent
{\it 5.1.2 The modular group}

Before studying the one-parameter models, we must also consider the issue of gauge redundancy.
The monodromy group $\Gamma$ of the complex structure moduli space has two generators: $A$, which generates phase rotations $\psi \to \alpha \psi$ with $\alpha = \exp (2 \pi i /k)$ around the LG point at $\psi = 0$, and $T$ which corresponds to the logarithmic monodromy ${\cal G}_2 \to {\cal G}_2 + z^2$ around the conifold singularity $\psi = 1$.  $A$ alone generates a ${\bf Z}_k \subset \Gamma$ subgroup, with an associated fixed point at $\psi = 0$; $T$ on the other hand is of infinite order.

In principle both $SL(2,Z)$ and
the modular group $\Gamma$
will generate an infinite number of copies of any fixed vacuum.   The former can be dealt with as it was for the previous examples, with
a constraint either on the fluxes or by choosing a fundamental region for the dilaton $\phi$.  We can try to fix $\Gamma$ similarly.  What
is a fundamental region for the action of $A$ and $T$?

$A$, which rotates around $\psi =0$ by a $k^{th}$ root of unity, obviously has as a fundamental domain a wedge of $2 \pi /k$ radians around this point.  For $T$, on the other hand, by choosing the periods \wo\ we have already implicitly chosen a fundamental domain.  $T$-transformations shift one between an infinite number of sheets defined around the conifold point $\psi =1$, where the periods have a logarithmic singularity.  By using the expansion \wo\ we have confined ourselves to a definite sheet.  Hence $T$-images of the
vacua we find will be solutions to the F-flatness equations with different periods, and we will not encounter them with our gauge choice
\wo.

Consequently, looking for vacua at or near $\psi = 0$ using \wo\ we find only the $k$ copies of a given vacuum generated by $A$; the $T$ redundancy is already implicitly fixed.  Numerical calculations with only $SL(2,Z)$ fixed confirm this by finding only a finite number of vacua at fixed $N_{\rm flux}$ at $\psi =0$, rather than the infinite number that would result if all $T$-images appeared.
In the examples we study, the effective modular group is thus $SL(2,Z) \times {\bf Z}_k$.

We now turn to a study of the one-parameter models in detail.  Since
only $M_6$ and $M_8$ have vacua at the LG point, we focus on those two
cases, enumerating generic LG vacua as well as those with discrete
symmetries and $W=0$.  In the case of $M_8$, we also report on a study
of the number of generic vacua in a small region around the LG point,
where solutions are found by using the expansion \wo\ and solving the
equations using perturbation theory in $\psi$.

\subsec{Landau-Ginzburg vacua for $M_6$}

Let us consider the case of $k=6$ first, the hypersurface in ${\bf WP}^4_{1,1,1,1,2}$. The root of unity is $\alpha
\equiv \exp({\pi i/ 3}) = (1 + i \sqrt{3})/2$.  Because $\alpha^2 = \alpha - 1$, the cyclotomic field ${\cal F}_6$ is just of degree two, with basis elements $\{ 1, \alpha\}$.  The
vectors $p_i$ can be written \eqn\perp{ p_0^{(6)} =m_6\cdot \pmatrix{\alpha-1 \cr \alpha \cr 1 \cr 1-\alpha}
=\pmatrix{1-\alpha \cr -1 \cr 3(2-\alpha) \cr \alpha-1} \,, \quad \quad p_1^{(6)} =m_6\cdot \pmatrix{-\alpha \cr
\alpha-1 \cr 1 \cr -\alpha} =\pmatrix{{1\over 3}(2-\alpha) \cr -1 \cr 3-\alpha \cr \alpha-2} \,. } Solutions of
\onevac\ then exist as long as $\phi$ is chosen in ${\cal F}_6$:
\eqn\sixtau{ \phi = t_1 + \alpha\,t_2 \,, }
with $t_1$, $t_2$ rational.
Because the field extension in this case is of degree 2, the equations
\onevac\   for a fixed value of $\phi$ impose 4 real relations on the fluxes $f_i, h_i$.  These relations can be conveniently expressed in terms of the monodromy matrix $A$  generating rotations by a root of unity around $\psi =0$:
\eqn\sixaperiods{
A \Pi(\psi) = \alpha \Pi(\alpha \psi) \,,
}
given explicitly\foot{The relation \sixaperiods\ differs from that given in \KT\ by the external factor $\alpha$ due to our different normalization for the periods \wo.} by
\eqn\asix{A =\pmatrix{1 & -1 & 0 & 1\cr
0 & 1 & 0 & -1\cr -3 & -3 & 1 & 3 \cr -6 & 4 & 1 & -3\cr}.}
Substituting \sixtau\ into \onevac, we see that an expression of the
form $f = t_1 h + t_2 h \cdot X$ holds as long as one can find an $X$
satisfying $X \cdot p_0 = \alpha^* p_0$, $X \cdot p_1 = \alpha p_1$.
Deriving $A \cdot p_0 = \alpha p_0$, $A \cdot p_1 = \alpha^2 p_1$ from
\perr\ and \sixaperiods, we obtain
\eqn\sixsoln{
f = t_1 h - t_2 h \cdot A^2 \,.
}
Notice that for fixed $t_i$, the $h_i$ must be chosen such that the $f_i$ are
integral.

The value of $N_{\rm flux}$, treating the degrees of freedom $h_i$ and
$t_i$ as independent, is
\eqn\nfluxsix{
N_{\rm flux}^{(6)} = - t_2\, h \cdot A^{2}\cdot\Sigma \cdot h \,,
}
where $\Sigma$ is defined below \kahler.
We see that $N_{\rm flux}$ is independent of $t_1$, indicating that the
magnitude of the D3-brane charge does not care about the value of the RR axion $C_0 = {\rm Re}\ \phi$; this is consistent with the absence of a $C_0 C_4 \wedge H_3 \wedge F_3$ coupling in the IIB supergravity action.

The fraction of vacua at the LG point $\psi=0$ can be
estimated for $M_6$ as follows.  We can use the height-based method of
analysis discussed in \S2.4 to estimate the number of
solutions to \sixsoln\ at fixed $\phi$.
Given the degree ${\cal D}_6=2$ and
the fact that there is one complex structure modulus, we have $\eta =
4$, reflecting the four real equations \sixsoln,
so that from \nonsusy\ we predict on the order of $L^2$ vacua for
any fixed $\phi$.
More precisely, using \nvphi\ we expect that for a given $\phi$ at
height $H(\phi)$,
\eqn\phih{N_{\rm vacua}(L; \phi,\psi=0) \sim {L^2 \over H^4}~.}
The factor of $1/H^4$ here just expresses the fact that roughly $1/H$
of the possible values for the $h_i$ are compatible with integral
values of the $f_i$ on the LHS of each of the four equations in
\sixsoln.  We can therefore estimate the total number of vacua by
summing over $\phi \in{\cal F}_6$ and using the fact that there are
${\cal O} (H^2)$ values of $\phi$ at a given $H$, so that
\eqn\totvac{N_{\rm vacua}(L; \psi=0) \sim L^2 \sum_{H=1}^{L} {1\over H^2}~.}
Note that the convergence of the sum over $H$ at large $L$ seems to indicate
that the freedom to choose the dilaton $t_1, t_2$ does not contribute
to the growth of the number of vacua; the $L^2$ scaling is consistent
with the choice of the four $h_i$ alone. As emphasized in \S2.4, this
estimate does not take into account non-generic features of the
constraint equations which may allow more or fewer solutions than the
generic argument would predict.  Indeed, as we see in the next
subsection, there are indeed ${\cal O} (L)$ solutions associated with
$\phi$ at height $L$ associated with special vacua where $W = 0$.  The
overall scaling of $L^2$ predicted by the generic height argument is still
correct, however it misses this special class of vacua which
contribute at large $H$ and which also total to contribute order $L^2$ vacua.

While we have argued that the number of vacua at the LG point goes as
$L^2$, we expect that the total number of vacua over all of moduli
space scales as \crude\ $L^{b_3} = L^4$.  Therefore, the ratio of LG
vacua to all vacua is \eqn\msixf{M_6 ~:~{N_{\rm vacua}(L; \psi=0)
\over N_{\rm vacua}(L)} \sim {1\over L^2}~.}

The same result can be also argued from consideration of the
restriction \nfluxsix\ which for allowed fixed dilaton value gives the
following scaling of the number of vacua
\eqn\phihhhhh{N_{\rm vacua}(L; \phi,\psi=0) \sim  {L^2 \over ({\rm Im}~\phi)^2}~,}
with an overall constant dependent on the arithmetic properties of the
numerator and denominator of $t_2$; this scaling is in accord with
the expected dilaton distribution in the fundamental
domain.

\medskip
\noindent
{\it 5.2.1 $W=0$ vacua}

Next we show that vacua with $W = 0$ exist at the LG point for $M_6$.  This equation leads to ${\cal D}_6 = 2$ additional constraints, and since the LG solutions \sixsoln\ still have four independent fluxes we generically expect solutions.  The equation $W=0$ becomes
\eqn\sixwzero{
(f - \phi h) \cdot p_0 = 0 \,,
}
which is solved by LG fluxes satisfying
\eqn\sixwzeroflux{
h = h \cdot A^3 \quad \rightarrow \quad  h = (-3 h_3+ h_4, 3 h_3, h_3, h_4)\,.
}
Together with \sixsoln\ this implies $f = f \cdot A^3$ as well.  Hence requiring $W=0$ leads to a codimension 2 subspace in the lattice of fluxes.

We can estimate the total number of $W=0$ vacua at the LG point using \nvphiw\ and \nvtw.  One finds  the total number of $W=0$ solutions will scale like
\eqn\totwo{N_{\rm vacua}(L; \psi=0; W=0) \sim \sum_{H=1}^L H^2 {L\over H^2} \sim L^2~.}
Note that the sum over heights increases the $L$ scaling by one power in this case, because of the modified contribution of the heights owing to the effective dilaton-independence of $W = D_\phi W =0$, as described in \S2.5.

Another way to present these vacua is as follows.  Instead of specifiying the fluxes $h_3, h_4$ and the dilaton parameters $t_1, t_2$, one may instead specify fluxes $f_3, f_4, h_3, h_4$.  The dilaton is then determined as:
\eqn\msixwdil{
\phi = {(\alpha-2) f_4 - 3 f_3 \over (\alpha -2) h_4 - 4 h_3} \,,
}
with the D3-brane charge
\eqn\Nfluxred{N_{\rm flux}=2(f_3 h_4-f_4 h_3) \,.}
We notice immediately the parallels with the generic vacua on the rigid CY studied in \S3: four flux parameters determining the dilaton, as in \tauis, and an analogous expression to \dthreec\ for $N_{\rm flux}$.

The counting of the vacua is hence identical to the rigid CY case.
Fixing SL$(2,Z)_\phi$ by choosing
\eqn\slzfix{f_4=0\,,~~~~~~~~~~0\leq h_3< f_3 \,,}
one obtains the scaling
\eqn\scal{N_{\rm vacua}(L; \psi=0; W=0)\sim {\pi^2\over 48}L^2 \,,}
in agreement with \totwo.

We turn now to studying the discrete symmetries of the model.  As we shall see, an R-symmetry may be associated with the $W=0$ vacua, ``explaining" the vanishing of the superpotential in the vacuum.

\medskip
\noindent
{\it 5.2.2 Discrete symmetries on $M_6$}

The LG point is a fixed point for ${\bf Z}_6 \subset \Gamma$, so one may
hope that this symmetry is preserved in the low-energy theory.
Additionally, the ${\bf Z}_2$ and ${\bf Z}_3$ points on the dilaton moduli space
are also potential sources of low-energy symmetry.  It is easy to see
that given the requirement \sixtau\ that the dilaton sits in ${\cal
F}_6$, the ${\bf Z}_2$ point is inaccessible, while vacua can sit on the ${\bf Z}_3$
point ($t_1 = 0, t_1 = 1$).  The accessibility of the ${\bf Z}_3$ point is
due to the sixth root of unity arising in both parts of the moduli
space.

We shall find in this case that the full fixed point symmetry ${\bf Z}_6$ is not generically a symmetry of the low-energy action for LG vacua.  However, there are cases where ${\bf Z}_2$, ${\bf Z}_3$ and even ${\bf Z}_6$ symmetries do appear; the latter two, however, also involve a nontrivial ${\bf Z}_3 \subset SL(2,Z)$.  We shall start with the smaller symmetry groups and work our way up.

\break
\noindent{\it ${\bf Z}_2$ symmetries}

For $W = 0$ vacua, we have thanks to \sixwzeroflux,
\eqn\wzerotrans{
W(\phi,\psi) = (f - \phi h) \cdot \Pi(\psi) = (f - \phi h) \cdot A^3 \cdot \Pi(\psi) = (f - \phi h) \cdot (\alpha^3 \Pi(\alpha^3 \psi)) \,,
}
where we have used \sixaperiods.  Since $\alpha^3 = -1$, this implies the superpotential is odd in $\psi$ for all values of the dilaton:
\eqn\wsixodd{
W(\phi, -\psi) = - W(\phi, \psi) \,.
}
Hence a ${\bf Z}_2$ R-symmetry arises for these vacua, and it is easy to see that this forces $W (\phi, \psi = 0) = 0$.
This is the R-symmetry ``responsible" for the vanishing of the vacuum superpotential -- in this example, this supports the
suggestion in \DGT\ that R-symmetries enforcing the $W=0$ condition may not be too uncommon in the space of flux vacua.
Clearly since $\psi \to \alpha^3 \psi$, we can think of this ${\bf Z}_2$ as a subset of the full ${\bf Z}_6$ acting only on moduli; the full ${\bf Z}_6$ is not in general a symmetry.

One may wonder whether analogous LG fluxes exist where the superpotential is even, and indeed this is the case, as fluxes satisfying
\eqn\wsixevenfluxes{
h = - h \cdot A^3 \quad \rightarrow  \quad h = (-3h_3 + 3 h_4, h_3, h_3, h_4) \,,
}
again together with the LG conditions \sixsoln\ imply
\eqn\wsixeven{
\quad W(\phi, -\psi) = W(\phi, \psi)~.
}
Hence for the fluxes \wsixevenfluxes\ we also have ${\bf Z}_2 \subset {\bf Z}_6$ preserved, but as a true symmetry rather than an R-symmetry.
No restriction is made on the value of the superpotential in this case and indeed it does not vanish.  One can show instead that the vacua obeying \wsixevenfluxes\ satisfy
\eqn\ddw{
D_\phi D_\psi W = 0 \,.
}
Flux vacua satisfying $DW=0$ are imaginary self-dual and contain only $(2,1)$ and $(0,3)$ fluxes; $W=0$ is the condition for the vanishing of the $(0,3)$ part.  Similarly, \ddw\ is the condition that the $(2,1)$ flux vanishes.\foot{It is known that ``attractor points of rank 2" can always be shown to host pure $(0,3)$ flux vacua; the LG point on $M_6$ is such a point \refs{\mooreaa,\mooreleshouches}.}

The counting of the $D_\phi D_\psi W =0$ vacua is entirely analogous to the $W=0$ case.  Following the analysis of the previous subsection, we can either use the method of heights as in \totwo, or we can note that given the expression for $N_{\rm flux}$ for these vacua,
\eqn\Nfluxred{N_{\rm flux}=2(f_4 h_3-f_3 h_4)\,, }
we again obtain the counting
\eqn\scal{N_{\rm vacua}(L; \psi=0; D_\phi D_\psi W = 0)\sim {\pi^2\over 48}L^2 \,.}
Another perspective on these results is as follows.  The expansion \wo\ for the periods indicates that the superpotential can be written in the form
\eqn\wsixexp{
W(\phi,\psi) = (f - \phi h) \cdot \left[ p_0 F_0(\psi^6) + p_1 \psi F_1(\psi^6) + p_1^* \psi^3 F_3(\psi^6) + p_0^* \psi^4 F_4(\psi^6) \right] \,,
}
where we have used $p_3 = p_1^*, p_4 = p_0^*$ and $c_{2 + 6n} = c_{5 + 6n} = 0$ for all $n$, and the $F_i$ are some ${\bf Z}_6$-invariant functions.  As we have discussed, $W = D_\phi W = 0$ imply $f \cdot p_0 = h \cdot p_0 = f \cdot p_0^* = h \cdot p_0^* = 0$, leaving $W(\phi,\psi)$ odd in $\psi$.  Similarly, the combination $D_\psi W = D_\phi D_\psi W =0$  implies $f \cdot p_1 = h \cdot p_1 = f \cdot p_1^* = h \cdot p_1^* = 0$, and consequently $W(\phi,\psi)$ for these vacua is even.

\medskip
\noindent{\it ${\bf Z}_3$ symmetries}

Let us now study the possibility of the ${\bf Z}_3$ on the dilaton moduli space being preserved in the low-energy action.  In order for fluxes to permit $\phi = \alpha$ as a vacuum, we must restrict the LG solutions \sixsoln\ to
\eqn\zthreesoln{
f = - h \cdot A^2 \,.
}
${\bf Z}_3$ is generated by the transformation $\phi \to (\phi - 1)/\phi$.  Is this a symmetry of the theory for the fluxes \zthreesoln?

To see how this might work, recall the preservation of ${\bf Z}_2$ in the rigid case \S3.5.1.  There the fluxes transformed such that $f - \phi h  \to (1 /\phi) (f - \phi h) \cdot S$, where $S$ was a matrix of which $\Pi$ was an eigenvector.  Similarly, it is possible to show that for fluxes \zthreesoln, one finds under $\phi \to (\phi - 1)/\phi$:
\eqn\zthreetrans{
(f - \phi h ) \to - {1 \over \phi} (f - \phi h) A^4 \,.
}
Given \sixaperiods, we find that the superpotential transforms as
\eqn\zthreew{
W(\phi,\psi) \to {\alpha \over \phi} W(\phi,\psi) \,,
}
provided we make the {\it combined} transformation
\eqn\msixzthree{
\phi \to (\phi - 1)/\phi \,, \quad \psi \to \alpha^2 \psi \,.}
Since \zthreew\ indeed has the form $W \to W/(c \phi + d)$ up to the phase, this leads to a K\"ahler transformation \kahlertrans\ and hence a symmetry.  Notice also that (just like the rigid ${\bf Z}_2$ case) when $\phi$ takes its vacuum value, here $\phi = \alpha$, we have $W( \phi = \alpha) \to W(\phi = \alpha)$.

Hence we find that for any LG solutions at the ${\bf Z}_3$ fixed point there is indeed a preserved ${\bf Z}_3$, but it is the diagonal between ${\bf Z}_3 \subset SL(2,Z)$ and ${\bf Z}_3 \subset {\bf Z}_6$: the preserved symmetry of the low-energy action is a combination of both dilaton and complex structure modulus transformations.  Without the latter, the former does not lead to a symmetry.

The counting of such vacua can proceed as follows.  Since we require that the dilaton sit at a fixed height, the number of vacua
is simply determined by the counting of \nonsusy\ with $H(\phi = \alpha) = 1$, with the result
\eqn\nvaczt{N_{\rm vacua}(L; \psi=0, \phi = \alpha) \sim L^2~.}
To make this more precise, note that for fixed dilaton $\phi = \alpha$, the $f$ fluxes are uniquely
determined from $h$ flux by \sixsoln\ with $t_1 = 0$ and $t_2 = 1$, and the number of vacua are then determined by positive quadratic form \nfluxsix\  again with $t_2 = 0$. This gives
\eqn\nvaaaa{
N_{\rm vacua}(L; \psi=0, \phi = \alpha)\sim {\pi^2\over 4}L^2~.
}
So we see that discrete symmetries are generic at the LG point, and are down by at most a power $1/L^2$ compared to all
vacua in this model.  The fact that an order one fraction of all LG vacua sit at $\phi = \alpha$ can be understood from the preference of the these vacua \phih\ to sit at small heights; since $\phi = \alpha$ has $H = 1$, it is much more common among all LG vacua than some other dilaton with a larger height.

\medskip
\noindent{\it ${\bf Z}_6$ symmetries}

Now we at last consider vacua where the full ${\bf Z}_6$ of $\psi \to \alpha \psi$ can appear as a low-energy symmetry.
The last example indicates that we may well need to combine this with some $SL(2,Z)$ transformation on $\phi$.
Under $\psi \to \alpha \psi$ we have
\eqn\psitrans{
W \to \alpha^* (f - \phi h) \cdot A \cdot \Pi(\psi) \,.
}
Without an $SL(2,Z)$ transform on $\phi$, we would require $f$ and $h$ both to separately be left eigenvectors of $A$; however, the eigenvectors of $A$ are necessarily complex, and hence this is not possible.

We find that with an $SL(2,Z)$ transform on $\phi$ and certain choices of fluxes, a ${\bf Z}_6$ symmetry can arise.  In addition to the transformation
\eqn\zsixtrans{
\psi \to \alpha \psi \,, \quad \phi \to {1 \over 1 - \phi} \,,
}
we must choose the fluxes so that both \zthreew\ as well as either \sixwzeroflux\
or \wsixevenfluxes\ (in addition to the LG condition \sixsoln) is satisfied.  That is, ${\bf Z}_6$ is a good symmetry precisely when {\it both} ${\bf Z}_3$ and one of the ${\bf Z}_2$s is present: one needs the dilaton at $\phi = \alpha$ as well as either $W=0$ or $D_\phi D_\psi W  =0$ vacua.

For the case of \sixwzeroflux, the $W=0$ vacua with the ${\bf Z}_2$ R-symmetry, we find
\eqn\zsixodd{
W(\phi,\psi) \to {\alpha^2 \over 1 - \phi} W (\phi,\psi) \,,
}
while for the $D_\phi D_\psi W = 0$ vacua with non-R ${\bf Z}_2$ \wsixevenfluxes\ we have
\eqn\zsixeven{
W(\phi,\psi) \to {\alpha^* \over 1 - \phi} W (\phi,\psi) \,.
}
These again have the proper $SL(2,Z)$ form $W \to W/(c \phi + d)$ for a K\"ahler transformation \kahlertrans, and hence are symmetries of the theory.  Moreover, it is simple to check that \zsixodd\ and \zsixeven\ square to the ${\bf Z}_3$ transform \zthreew, while they cube to the appropriate ${\bf Z}_2$ transformations
\wsixodd, \wsixeven.  Hence, when both the ${\bf Z}_3$ and a ${\bf Z}_2$ symmetry are present simultaneously, they are subsumed into a full ${\bf Z}_6$.

We finally count these two types of ${\bf Z}_6$ vacua.
{}From \nvphiw, the scaling for $W=0$ ${\bf Z}_6$ vacua at the fixed height $H=1$ is simply given by
\eqn\nwozt{N_{\rm vacua}(L; \psi=0, \phi=\alpha ;W=0) \sim L~,}
and one can convince oneself that since $D_\psi W = D_\phi D_\psi W =0$ also become effectively dilaton-indepenent equations, the result for the $D_\phi D_\psi W =0$ ${\bf Z}_6$ vacua is the same.
More precisely, one can note that the positive definite quadratic form \nfluxsix\ for $W=0$ solutions associated to \sixwzeroflux\ reduces to
\eqn\nfluxsixwzero{N_{\rm flux}=2(3 h_3^2+3h_3 h_4+h_4^2)\,,}
while for the $D_\phi D_\psi W =0$  solutions associated with \wsixevenfluxes\ it reduces to
\eqn\nfluxsixwzero{N_{\rm flux}=2(h_3^2+h_3 h_4+h_4^2)~.}
Consequently, in both cases the scaling becomes
\eqn\nwozt{N_{\rm vacua}(L; \psi=0, \phi=\alpha ;W=0) = N_{\rm vacua}(L; \psi=0, \phi=\alpha ;D_\phi D_\psi W=0) \sim {\pi\over 2\sqrt{3}} L~.}
Hence although the ${\bf Z}_2$ symmetries as well as the ${\bf Z}_3$ symmetry were generic within the set of LG vacua, all scaling as $L^2$, we find that the intersection between the ${\bf Z}_3$ and one of the ${\bf Z}_2$ sets, which possesses a ${\bf Z}_6$ symmetry, are supressed by $1/L$.

\medskip
\noindent
{\it Complex conjugation ${\cal C}$}

Before ending our discussion of the symmetries of $M_6$, we mention also the complex conjugation transformation ${\cal C}$; our discussion is general and actually applies to all the $M_k$.  Using \wo\ one can demonstrate that the relation \cconj\ indeed holds with $U_k = m_k \cdot V \cdot m_k^{-1}$, with the matrices $m_k$ given in \KT\ and
\eqn\ldef{
V \equiv \pmatrix{0 \;\;\;0 \;\;\;0 \;\;\;1 \cr 0 \;\;\;0 \;\;\;1 \;\;\;0 \cr 0 \;\;\;1 \;\;\;0 \;\;\;0 \cr 1 \;\;\;0 \;\;\;0 \;\;\;0} \,.
}
This means that ${\cal C}$ indeed exists as a transformation carrying vacua to other (inequivalent) vacua: note that $N_{\rm flux}$ is preserved.  

A low-energy symmetry acting on the moduli alone will be present for fluxes invariant under ${\cal C}$, that is $f = f \cdot U_k$ and $h = - h \cdot U_k$.  These vacua are present in all four models.  For $M_6$, the conditions become
\eqn\msixconj{
f_1 = - { 9\over 2} f_3 \,, \quad \quad f_2 = -2 f_4 \,, \quad \quad h_2 = h_3  =0 \,.
}
Not all of these vacua sit at the LG point; those that do additionally satisfy
\eqn\msixconjLG{
f_3 = t_2 (2 h_4 - h_1) \,, \quad \quad f_4 = t_2 (h_1 - {5 \over 2} h_4) \,,
}
and sit on the imaginary dilaton axis, $\phi = i \sqrt{3} t_2 /2$.  In the set of LG vacua, we may think of those with the complex conjugation symmetry as those that also satisfy $h_2 = h_3 = 0, t_1 = - t_2/2$.  The overall scaling will go like $L$, but we should expect logarithmic enhancement along the lines of \S3.5.2.  We leave the question of the distribution of non-LG vacua with complex conjugation to the future.

\subsec{Landau-Ginzburg vacua for $M_8$}

The root of unity for the $k=6$ case was particularly ``nice" because its real part is rational, leading to the
small degree of the extension.  The other cases are not so simple, and all three have extensions of degree four.
The $k=5$ and $k=10$ cases both can be spanned with the bases $\{ 1, \sqrt{5}, i \sqrt{5 + \sqrt{5}}, i \sqrt{5}
\sqrt{5 + \sqrt{5}} \}$, or alternately powers of the roots of unity such as $\{1, \alpha, \alpha^2, \alpha^4\}$.
In general one may allow the dilaton $\phi = t_0 + t_1 \alpha + t_2 \alpha^2 + t_4 \alpha^4$ with rational $t_i$.
Attempting to solve \onevac, one indeed finds eight equations, and no nonzero solution for the
fluxes.  So as we mentioned earlier,
\eqn\mfe{M_5, ~M_{10}~:~N_{\rm vacua}(L,\psi=0) = 0~.}

For the $k=8$ case, the root of unity is $\alpha = \exp(\pi i/4) = (1 + i)
/ \sqrt{2}$ with $\alpha^2 = i$, and hence a convenient basis for this
degree-four extension is $\{1, i, \sqrt{2}, i \sqrt{2}\}$.  One considers
the dilaton ansatz $\phi = t_1 + i t_2 + \sqrt{2} t_3 + i \sqrt{2} t_4$,
and solving the F-flatness conditions \onevac\ with the vectors
\eqn\eightp{ p_0^{(8)} =m_8\cdot \pmatrix{i \cr \alpha \cr 1 \cr -i\alpha}
=\pmatrix{{1 \over 2} - {i \over 2} - {i \over \sqrt{2}} \cr -1 \cr 3 - i
+ \sqrt{2} - i \sqrt{2} \cr -1 + {1 \over \sqrt{2}} + {i \over \sqrt{2}}}
 \,, \quad p_2^{(8)} =m_8\cdot \pmatrix{-i \cr i \alpha \cr 1 \cr -\alpha}
=\pmatrix{{1 \over 2} + {i \over 2} - {i \over \sqrt{2}} \cr -1 \cr 3 + i - \sqrt{2} - i \sqrt{2} \cr -1 - {1
\over \sqrt{2}} + {i \over \sqrt{2}}} } one finds in general that the fluxes are forced to vanish.  However, for
the simplifying assumption $t_3 = t_4 = 0$ something special happens: the four equations in $D_\psi W =0$ become
redundant with those in $D_\phi W =0$.

One can see this as follows: the $\psi$ equation and the conjugate of the
$\phi$ equation are $(f - \phi h ) \cdot p_2 =0$ and $(f - \phi h) \cdot
p_0^* = 0$.  It is evident from \eightp\ that $p_2$ and $p_0^*$ are
identified under the exchange $\sqrt{2} \rightarrow - \sqrt{2}$, or equivalently $\alpha \to \alpha^5$; this is a transformation of the Galois group of ${\cal F}_8$.  Hence
the F-flatness equations will be carried into each other under the same
identification, provided $\phi$ is invariant under $\sqrt{2} \rightarrow -
\sqrt{2}$, which is precisely the requirement $t_3 = t_4 =0$ we found.
Consequently, when $\phi$ contains no $\sqrt{2}$ factors, the dilaton and
$\psi$ conditions each reduce to the same four equations.

The LG solution to these equations is: \eqn\eighttau{ \phi = t_1 + i t_2 \,, } with $t_1$, $t_2$ rational, and just as in the $k=6$ case,
\eqn\eightsoln{
f = t_1 h - t_2 h \cdot A^2 \,,
}
with $A$ this time given by \KT:
\eqn\aeight{A =\pmatrix{ 1 & -1 & 0 & 1\cr
0 & 1 & 0 & -1\cr
-2 & -2 & 1 & 2 \cr
-4 & 4 & 1 & -3\cr}\,,}
and also satisfying \sixaperiods, but with the eighth root of unity.
The expression for $N_{\rm flux}$ in terms of the $h_i$ and $t_i$
is again
\eqn\nfluxeight{
N_{\rm flux}^{(8)} = - t_2\, h \cdot A^{2}\cdot\Sigma \cdot h \,.
}
Notice that again the D3 charge is independent of the RR axion.

As in
$M_6$, one finds here that
\eqn\nvacme{N_{\rm vacua}(\psi=0) \sim L^2 \sum_{H=1}^L {1\over H^2}~.}
This is because the counting works exactly for this nongeneric ${\cal D}=4$ case as it
would in a generic ${\cal D}=2$ case with one complex structure modulus: the sum over dilaton heights yields a factor of
$H^2$ in \nvt\ due to the specialization $t_3 = t_4 = 0$, while there are only four independent linear
constraining the flux integers instead of the expected eight.

\medskip
\noindent{\it{5.3.1 W=0 vacua}}

Although this coincidence of the $DW=0$ equations leads to LG vacua where they might not have been expected, there is no similar coincidence for $W=0$ vacua.
Imposing the $W=0$ equation adds ${\cal D}_8 = 4$ new conditions on the flux integers.  Unlike the case for $D_\phi W =0 $ and $D_\psi W = 0$, these conditions are not redundant with those already imposed. As a result $W=0$ vacua must satisfy eight conditions on the eight fluxes, and the only solution is at vanishing flux integers.  Hence there are no nontrivial $W=0$ vacua in this model.

\medskip
\noindent{\it{5.3.2 Discrete symmetries}}

Consider now discrete symmetries.  In addition to the ${\bf Z}_8$ from the LG point, these models can access the ${\bf Z}_2$ point on the dilaton moduli space for $t_1 = 0, t_2 = 1$, so these are the fixed points at which to look for low-energy symmetries;  the ${\bf Z}_3$ point on the dilaton moduli space is forbidden.

The most obvious symmetry comes from a property of the periods: as remarked before, all odd terms in $\psi$ vanish in $\Pi(\psi)$ for the $M_8$ and $M_{10}$ models.  Thus a ${\bf Z}_2$ is trivially present under $\psi \to - \psi$ with $W$ invariant.  Although this symmetry has no action on the fluxes, it is nonetheless an element of the modular group, and in fact since $\alpha^4 = -1$, it is ${\bf Z}_2 \subset {\bf Z}_8 \subset \Gamma$; modular transformations that do act on the fluxes square to this ${\bf Z}_2$.  Hence, although it is a transformation of the moduli alone, it should be considered a gauge transformation of the theory, what we called in \S2.6 a symmetry $G$, not a global transformation $H$; it is actually an identification of the moduli space.

Let us look for true global symmetries.  We consider vacua at the dilaton ${\bf Z}_2$ fixed point, which must satisfy:
\eqn\ztwo{
f = - h \cdot A^2 \,.
}
In a manner analogous to that of the $k=6$ case, we can show that under $\phi \to - 1/\phi$,
\eqn\ztwoflux{
(f - \phi h) \to {1 \over \phi} (f - \phi h) \cdot A^2 \,.
}
Consequently we have the combined transformation:
\eqn\ztwotrans{
\phi \to -1/\phi \,, \quad \psi \to i \psi \,, \quad \rightarrow \quad W(\phi,\psi) \to { i \over \phi} W(\phi, \psi) \,,
}
which is a K\"ahler transformation.  Hence we find that the ${\bf Z}_4 \subset {\bf Z}_8$ combines with the ${\bf Z}_2$ of the dilaton to create what is naively a preserved ${\bf Z}_4$ symmetry in the low-energy theory.  However, its square is the gauged ${\bf Z}_2$ described above; as a global symmetry, consequently, it is just a ${\bf Z}_2$.

These vacua can be counted analogously to the $M_6$ ${\bf Z}_3$ vacua in the previous subsection; one takes  \nvacme\ and excludes the sum over heights, instead setting $H(\phi = i) = 1$:
\eqn\nvacztt{N_{\rm vacua}(L; \phi=i, \psi=0) \sim L^2~.}
To obtain the coefficient, one again notes that for fixed dilaton $t_1 = 0$, $t_2 =1$ the  $f$ fluxes are uniquely
determined from the $h$ fluxes by \ztwo\ and the number of vacua is determined by the
positive quadratic form \nfluxeight\ . This gives
\eqn\nvaaaa{
N_{\rm vacua}(\phi=i, \psi=0)\sim {\pi^2\over 4}L^2~.
}
As with the ${\bf Z}_3$ vacua in $M_6$, we find that these ${\bf Z}_2$ vacua are generic among the LG vacua, both going as $L^2$, because they sit at the lowest height for the dilaton.

It is interesting to observe that we do not find any fluxes with the full ${\bf Z}_8$ preserved.
A transformation $\psi \to \alpha \psi$ with $\phi$ inert would require $f$ and $h$ to be
eigenvectors of $A$, but these are complex.  The ${\bf Z}_4$ \ztwotrans\ is the most one can get
 from an $SL(2,Z)$ transformation by $S$, and other $SL(2,Z)$ transformations are less useful.

As mentioned in the discussion of $M_6$, there is a complex conjugation operation ${\cal C}$ defined for all the $M_k$, and for $M_8$ there is a symmetry acting on the moduli alone for fluxes
 $f = f \cdot U_8$ and $h = - h \cdot U_8$: 
\eqn\meightconj{
f_1 = - 3 f_3 \,, \quad \quad f_2 = -2 f_4 \,, \quad \quad h_2 = h_3  =0 \,,
}
with vacua at the LG point also satisfying
\eqn\meightconjLG{
f_3 = t_2 (2 h_4 - h_1) \,, \quad \quad f_4 = t_2 (h_1 - 3 h_4) \,,
}
sitting on the imaginary dilaton axis, $\phi = i t_2$; the counting is identical to the case for $M_6$.

\subsec{L-scaling of number of vacua near the LG and conifold points in $M_8$}

In \wo\ and \perr\ we have presented the periods in an expansion about
the LG point $\psi=0$. We can use these expansions to perform a
perturbative calculation of the vacua near the LG point.  Although
this has not been our primary focus in this paper, we give numerical
results for the $L$ scaling of the number of vacua in the vicinity of
both the LG and the conifold points (an expansion of the relevant
periods around the conifold can be found in \GKT), finding the
expected scaling $N_{\rm flux}(L) \sim L^{b_3} = L^4$ \crude.  Note
that a similar calculation around the LG and conifold points in \GKT\ suffered from
incomplete gauge fixing, which we correct in the present paper. This
changes the scaling, bringing it into accord with the expected
$L^{b_3}$.

In performing these Monte Carlo calculations,
we looked for vacua at a
variety of values $N_{\rm flux} \leq L$.  At a given $N_{\rm flux}
\leq L$, as in the analogous computations for the torus which were
described in Section \S4.3.2, we looked for
vacua with fluxes $\max(|f|,|h|)\leq k N_{\rm flux}$, where we used $k =
1$.  Unlike for the torus, there do seem to be a nonzero fraction of
vacua with fluxes outside this bound.  Empirical tests, however,
indicate that more than half of all fluxes at fixed $N_{\rm flux}$
satisfy the bound in most of the region we considered.  Thus, the
scaling of the number of vacua as a function of $L$ should not be
affected, though the multiplicative constant we found may be too small
by a factor of order 1.  We could have imposed a larger bound, using,
say $k = 2$, which would capture almost all relevant vacua.  This,
however, would have slowed the program by a significant factor.

In both cases in our Monte Carlo  computations we implement
the following dilaton SL$(2,Z)$ gauge fixing:
\eqn\gaugefix{f_2=0\, ,~~~~~~ 0\leq h_1<f_1\, ,~~~~~~ h_2\neq 0~.}
At special values of the fluxes, additional discrete gauge identifications arise, but these
do not change the $L$ scaling.

In the vicinity of the LG point, our numerical results give the
following expected number of vacua for $|\psi|^2<0.5$ and $N_{\rm
flux}\leq L$:
\eqn\vacnumlg{N_{\rm vacua}(L)= 0.02\, L^4.}
Note that the LG ${\bf Z}_8$ monodromy should be fixed in this case.
We display our results in the figure below.
Fixing the conifold monodromy by requiring arg$(1-\psi)\in [-\pi,\pi)$ one finds
from Monte Carlo data
the following expected number of vacua in the vicinity of the conifold point
for $|1-\psi|<0.01$ and $N_{\rm flux}\leq L$:
\eqn\vacnumlg{N_{\rm vacua}(L)= 0.2\, L^4.}
Our results for the expansion about the conifold point are
also shown below.

\ifig\figI{
The numerical results for the expected number of vacua in vicinity of LG point
with $N_{\rm flux}<L$
for $L\in(1,100)$ fit by the curve ${0.02\, L^4}$.}
{\epsfxsize=0.8\hsize\epsfbox{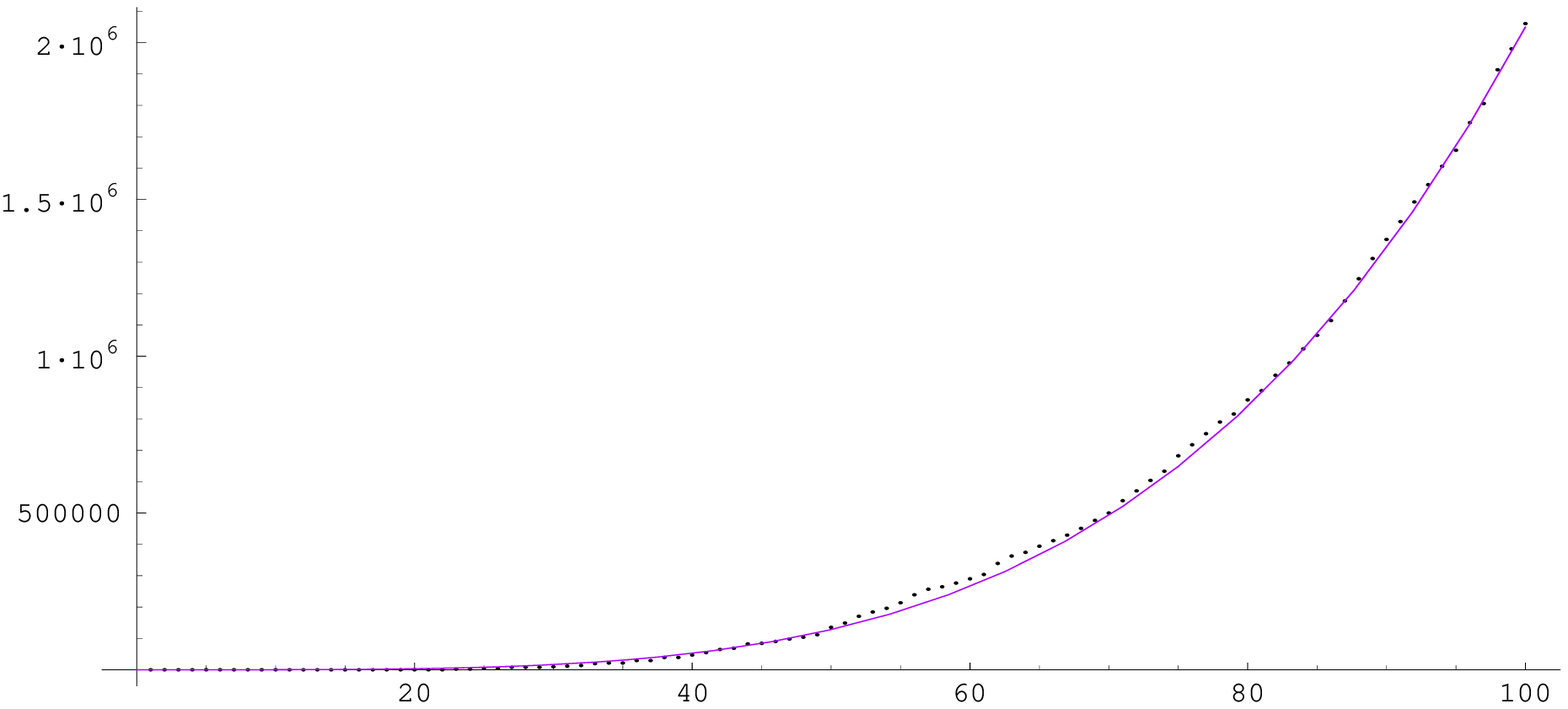}}

\ifig\figIIII{
The numerical results for the expected number of vacua in vicinity of conifold point
with $N_{\rm flux}<L$ for $L\in(1,100)$ fit by the curve ${0.2\, L^4}$.}
{\epsfxsize=0.8\hsize\epsfbox{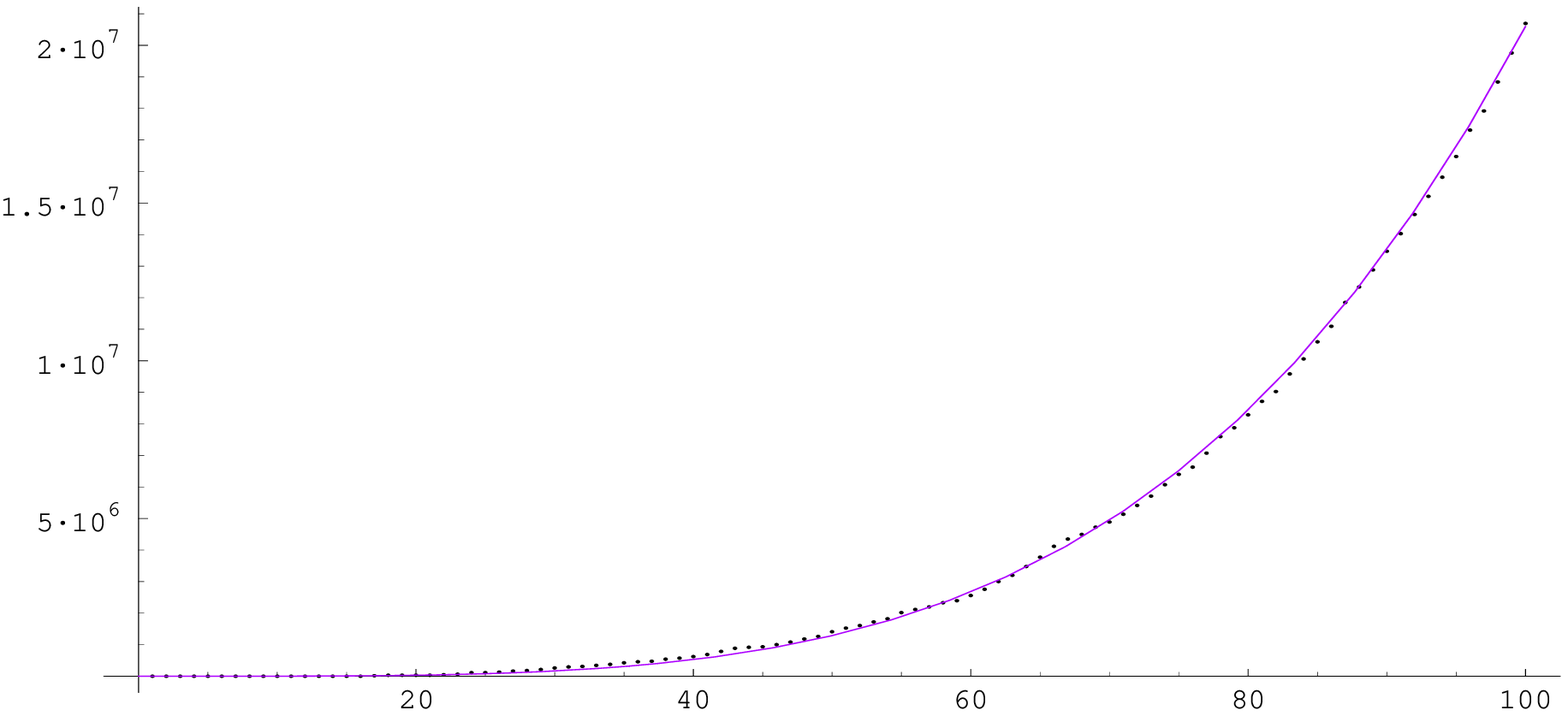}}

\newsec{Discussion}

In this paper, we have explored a variety of simple models for flux
vacua.  In particular, we have addressed the question of when flux
vacua preserve supersymmetry in the no-scale sense, that is, which
vacua have $W = 0$, as well as finding classes of flux vacua for
which the low-energy field theories have discrete symmetries.  Because
in general the approach of approximating the fluxes by continuous
parameters is not suited to solving the kind of overconstrained
problems we encounter in studying these two questions, we have
developed some simple approaches to addressing these types of
questions, using techniques motivated from number theory.

More specifically, we considered the models of the rigid Calabi-Yau,
the symmetric torus with a single modular parameter, and the
one-parameter Calabi-Yau models associated with hypersurfaces in
weighted projective space.  For the flux vacua we considered, the
periods (as well as their derivatives) live in a field ${\cal F}$ of
finite extension degree ${\cal D}$ over the rational numbers ${\bf
Q}$.  In general, the number of constraints on the fluxes is linear in
${\cal D}$, so that the form of the field extension plays an important
role in determining the number of allowed flux vacua.  We found an
exact solution for generic flux vacua on the one-parameter torus, with
the result that the moduli of all such flux vacua live in a
field extension of degree at most ${\cal D} = 6$ over the rationals.
On the one-parameter Calabi-Yaus, we considered vacua at the LG
point, where the periods take values in the cyclotomic fields arising
from extending the rationals by $k^{th}$ roots of unity, with degrees ${\cal D}=2$ and ${\cal D}=4$.

In the models we considered, we found that the number of vacua with
either $W = 0$ or with discrete symmetries is generally suppressed by
a small power of $L$ below the total number of vacua, although in some
cases we found anomalous scaling laws involving various powers of
${\rm log}~ L$.  These anomalous scalings arise from situations where
the continuous approximation breaks down because either the geometry
of the allowed volume in flux space includes a divergent region, or
because extra constraints force us to use number-theoretic methods of
analysis.

For most of the models we considered, there is a single
complex structure modulus and the total number of vacua goes as $L^4$.
We found that all $W = 0$ vacua on the symmetric torus have periods
associated with degree 2 extensions over the rationals, and that the
number of these vacua goes as $L^2$.  Families of vacua on the torus
with discrete symmetries also generally scale as $L^2$.  In the
one-parameter Calabi-Yau hypersurfaces that yielded vacua at the LG
point, the results were similar: the total number of vacua was $L^4$, and
in fact the number of vacua with $W=0$ or with discrete symmetries scaled with the ${\it same}$ power of $L$ as the total number of LG vacua, namely $L^2$.  In all these cases enhanced symmetry vacua are not suppressed by more than a power of $1/L^2$ relative to all vacua.

While the precise results we have derived here are insufficient to
make clear predictions for the scaling of $W = 0$ vacua and vacua with
discrete symmetries when the number of complex structure moduli is
large, we can gain substantial insight from the form of the equations
we have used to do the analysis.  In particular, as we have discussed
here, imposing the equation $W = 0$ imposes an additional
set of ${\cal D}$ equations on the fluxes at a point in moduli space
where the periods live in a field extension of finite degree ${\cal
D}$ over $ {\bf Q}$.  Since the number of additional constraints here
(relative to generic vacua at the same point in moduli space) is not
proportional to $b_3$, we expect that generically when the total
number of vacua (at the given point) grows as $L^{n}$, the number of
$W = 0$ vacua will be on the order of $L^{n-{\cal D}/2}$, and will not
be massively suppressed.

Given the significance in our analysis of the arithmetic structure of
the periods, an important question seems to be: in a Calabi-Yau with
large $b_3$, of the total of order $L^{b_3}$ flux vacua, what fraction
have periods which take values in fields of finite extension degree
${\cal D}$ over the rationals?  If these vacua with arithmetic periods
form a large fraction of the total set of flux vacua, then as long as
${\cal D}\ll b_3$, we expect that a significant fraction of the total
number of vacua will have $W = 0$.  Unfortunately it seems that
currently little is known about this question.  Based on the
appearance of an unexpectedly low extension degree ${\cal D} \leq 6$
for generic flux vacua on the torus, however, we are tempted to
conjecture that the number of vacua with finite ${\cal D}$ may be
quite large.  In fact, at this point we are unaware of any flux vacua
which can be proven to have transcendental periods.  A related
outstanding question is how to properly sum over families of vacua
associated with different field extensions; more advanced concepts in
algebraic number theory may be useful here.

If the above conjectures are correct, so that indeed there are
situations where $W = 0$ vacua are suppressed by a power of $L$ which
is independent of $b_3$, then $W=0$ vacua may be ``not too rare,'' in
a way in which the authors of \DGT\ have suggested may indicate that a
realistically small cosmological constant is not incompatible with
low-scale supersymmetry breaking.  Needless to say, a long and rather
tenuous chain of reasoning relates the mathematical problem we are
addressing here to the phenomenological question of the scale of
supersymmetry breaking.  Numerous issues need to be understood much
better before this connection can be convincingly justified.  In
particular, we need a better understanding of how a measure on the
space of vacua can be produced by cosmological dynamics, and how such
a measure can be incorporated into statistical methods of analysis
such as were used in \AD\ and in this paper.

Regarding discrete symmetries, based on the examples we have
considered it seems likely that there are two types of discrete
symmetries of flux vacua associated with Calabi-Yau manifolds with
large $b_3$.  Some symmetries will involve imposing on the order of
$b_3$ conditions on the fluxes, and should lead to a highly suppressed
multiplicity of vacua, while other discrete symmetries may only
require a fixed number of constraints independent of $b_3$, and may be
much more prevalent.  We analyzed symmetries in this paper which arise
from discrete subgroups of the modular group as well as symmetries
related to complex conjugation on the moduli space.  The prevalence of
these symmetries is related to the codimension of fixed point loci of
the relevant group on the moduli space.  Since the vacuum sitting at
the fixed point locus is a necessary, but not sufficient condition,
however, this codimension provides only an upper bound on the
frequency of these vacua; a more exact understanding of under
precisely which circumstances they arise would be of great interest
for model-building and cosmology.

There is a strong similarity between some of the questions we have
investigated here, particularly the study of $W = 0$ vacua, and the
study of ``rank two attractors'' associated with points on moduli
space admitting flux vacua with $G_3$ of type (0,3)
\refs{\mooreaa,\mooreleshouches}.  In contrast, $W=0$ vacua arise at
points in moduli space where $G_3$ is purely of type (2,1), while the
generic vacua satisfying the $DW = 0$ equations are a combination of
these two types.  The attractor conjectures in their most recent form
relate ``rank two attractors'' to Calabi-Yau varieties of CM type
(see, {\it e.g.}, \GV\ for an explanation of this and
related notions).  In problems with $h^{2,1}=1$ (as we have mostly
considered here) there is roughly a one-to-one correspondence between
those vacua with $G_3$ of type (0, 3) and those with $G_3$ of type (2,
1).  At large $b_3$, however, it should be much more common to find
points of the latter type than the former.  While the (0,3) condition
constrains $G_3$ to lie in a one-dimensional (complex) vector space,
the $W=0$ condition leaves open an $h^{2,1}$ dimensional subspace of
$H^3$, imposing as discussed above a number of additional constraints
independent of $b_3$.  Thus, we expect that although there are similar
issues involved in studying these two types of vacua, vacua with $W =
0$ are  likely to be much more prevalent than rank two attractors.  The
study of models with more than one complex structure parameter seems
necessary to make progress on this issue.

In this paper we have touched on potentially fascinating connections
between the study of $W=0$ vacua, vacua with discrete symmetries and number theory. For instance, we saw in \S4.3.3\ that the class numbers $h(D)$
play an important role in the counting of $W=0$ vacua on the torus,
as conjectured in \mooreleshouches.
More generally, we saw starting in \S2.4\ and running
through all of our examples
that simple ideas from the theory of heights
\dghs\ may prove to be of general use in exploring the taxonomy of
string vacua.  The stabilization of complex structure moduli through
fluxes leads to a remarkable interplay between continuous geometry and
discrete mathematics.  It seems possible that further investigations of
this connection will lead to a new and important role for number
theory as a mathematical tool for understanding string
compactifications and their phenomenology.

\bigskip\bigskip
\centerline{\bf{Acknowledgements}}
\medskip

We are grateful to N.~Arkani-Hamed, P.~Berglund, J.~de Jong, F.~Denef,
S.~Dimopoulos, M.~Dine, C.~Doran, M.~Douglas, G.~Dvali, N.~Elkies,
B.~Florea, S.~Gukov, B.~Kors, H.~Liu, J.~McGreevy, F.~Quevedo,
S.~Sethi, S.~Shenker, E.~Silverstein, L.~Susskind and S.~Thomas for
helpful discussions.  S.K.\ and W.T.\ thank the University of British
Columbia and the Pacific Institute for Mathematical Sciences, O.D.\
thanks Stanford University and MIT, and O.D., S.K.\ and W.T.\ thank
the Aspen Center for Physics for hospitality and support while this
work was in progress.  The work of O.D. was supported by NSF grant
PHY-0243680.  The work of A.G. was supported in part by INTAS
03-51-6346.  The work of S.K. was supported by a David and Lucile
Packard Foundation Fellowship for Science and Engineering, NSF grant
PHY-0097915, and the DOE under contract \#DE-AC02-76SF00515.  The work
of W.T.\ was supported by the DOE under contract
\#DE-FC02-94ER40818.
Any opinions, findings, and conclusions
or recommendations expressed in this material are those of the authors
and do not necessarily reflect the views of the above agencies.

\bigskip

\appendix{A}{Equations for exact solution of symmetric torus}

As discussed in \S4.3,
the equations of motion $DW = 0$ for the symmetric torus can be
combined into a pair of quintic equations for the real and imaginary
components of the modular parameter $\tau = x + iy$
\equation\simpletwo{
\eqalign{
q_1 (x) y^2 &  \;=  q_3 (x)\, , \cr
q_0 (x) y^4 &  \;=  q_4 (x)\, ,}}
where
\equation\polynomials{\eqalign{
q_0 (x) &  =  ac^0-a^0c\, ,\cr
q_1 (x) &  =  b c^0-  a^0 d + 2( a c^0 - a^0 c)x\, ,\cr
q_3 (x) &  =  b_0 c - a d_0 + (3 b c  - b_0 c^0 - 3 a d + a^0 d_0)x + (- 3 b c^0 +  3 a^0 d) x^2 + 2  (a^0 c - a c^0) x^3 \, ,\cr
q_4 (x) &  =  -b_0 d + b d_0  + 2 (- b_0 c + a d_0)x+ (- 3 b c + b_0 c^0 + 3 a d -  a^0 d_0) x^2\cr
& \;\;\;\;\; + 2 (b c^0 - a^0 d) x^3 + ( - a^0 c + a c^0) x^4\, .}}
These equations can be combined into a single cubic equation satisfied
by $x$
\equation\cubic{
\alpha_3x^3 + \alpha_2x^2 + \alpha_1 x + \alpha^0 = 0\, ,
}
where the coefficients are given by
\equation\coefficientsa{
\eqalign{
\alpha^0 & \; = \;a^0 b_0^2 c^3 - a b_0^2 c^2 c^0 - b^2 b_0 (c^0)^2 d +
2 a^0 b b_0 c^0 d^2 -
  (a^0)^2 b_0 d^3 - 2 a a^0 b_0 c^2 d_0
\cr & \;\;\;\;\;
+ 2 a^2 b_0 c c^0 d_0
+ b^3 (c^0)^2 d_0 -
  2 a^0 b^2 c^0 d d_0 + (a^0)^2 b d^2 d_0 + a^2 a^0 c d_0^2 - a^3 c^0 d_0^2\, ;}}
\equation\coefficientsb{
\eqalign{
\alpha_1 & \; = \; 6 a^0 b b_0 c^3 - 6 a b b_0 c^2 c^0 - 2 a^0 b_0^2 c^2 c^0 - 2 b^2 b_0 c (c^0)^2 +
  2 a b_0^2 c (c^0)^2 - 6 a a^0 b_0 c^2 d\cr
& \;\;\;\;\; + 6 a^2 b_0 c c^0 d+8 a^0 b b_0 c c^0 d -  4 a b b_0 (c^0)^2 d - 6 (a^0)^2 b_0 c d^2 + 4 a a^0 b_0 c^0 d^2\cr
& \;\;\;\;\;
  - 6 a a^0 b c^2 d_0 +  2 (a^0)^2 b_0 c^2 d_0   + 6 a^2 b c c^0 d_0 - 4 a^0 b^2 c c^0 d_0 + 6 a b^2 (c^0)^2 d_0\cr
& \;\;\;\;\;
  -2 a^2 b_0 (c^0)^2 d_0 + 6 a^2 a^0 c d d_0 + 4 (a^0)^2 b c d d_0  - 6 a^3 c^0 d d_0 -  8 a a^0 b c^0 d d_0\cr
& \;\;\;\;\;
  + 2 a (a^0)^2 d^2 d_0 - 2 a (a^0)^2 c d_0^2  + 2 a^2  a^0 c^0 d_0^2\, ;}}
\equation\coefficientsc{
\eqalign{
\alpha_2 & \; = \; 9 a^0 b^2 c^3 - 9 a b^2 c^2 c^0 - 4 a^0 b b_0 c^2 c^0 - 3 b^3 c (c^0)^2 +  4 a b b_0 c (c^0)^2 + a^0 b_0^2 c (c^0)^2 \cr
& \;\;\;\;\;  + b^2 b_0 (c^0)^3 - a b_0^2 (c^0)^3 -  18 a a^0 b c^2 d - 6 (a^0)^2 b_0 c^2 d + 18 a^2 b c c^0 d  + 6 a^0 b^2 c c^0 d\cr
& \;\;\;\;\;  +  16 a a^0 b_0 c c^0 d + 3 a b^2 (c^0)^2 d - 10 a^2 b_0 (c^0)^2 d - 2 a^0 b b_0 (c^0)^2 d +  9 a^2 a^0 c d^2 \cr
& \;\;\;\;\;  - 3 (a^0)^2 b c d^2- 9 a^3 c^0 d^2 - 6 a a^0 b c^0 d^2 +  (a^0)^2 b_0 c^0 d^2 + 3 a (a^0)^2 d^3 \cr
& \;\;\;\;\;  + 10 (a^0)^2 b c^2 d_0- 16 a a^0 b c c^0 d_0-  2 (a^0)^2 b_0 c c^0 d_0 + 6 a^2 b (c^0)^2 d_0 - a^0 b^2 (c^0)^2 d_0 \cr
& \;\;\;\;\;   + 4 a^2 a^0 c^0 d d_0 + 2 (a^0)^2 b c^0 d d_0 - (a^0)^3 d^2 d_0 + 2 a  a^0 b_0 (c^0)^2 d_0-  4 a (a^0)^2 c d d_0\cr
& \;\;\;\;\;  + (a^0)^3 c d_0^2 - a (a^0)^2 c^0 d_0^2\, ;}}
\equation\coefficientsd{
\eqalign{
\alpha_3 & \; = \;-4 (a^0)^2 b_0 c^3 - 6 a^0 b^2 c^2 c^0 +  8 a a^0 b_0 c^2 c^0 + 6 a b^2 c (c^0)^2 - 4 a^2 b_0 c (c^0)^2 \cr
& \;\;\;\;\; + 2 a^0 b b_0 c (c^0)^2+  2 b^3 (c^0)^3- 2 a b b_0 (c^0)^3 + 6 (a^0)^2 b c^2 d - 2 (a^0)^2 b_0 c c^0 d  \cr
& \;\;\;\;\; -  6 a^2 b (c^0)^2 d - 6 a^0 b^2 (c^0)^2 d+ 2 a a^0 b_0 (c^0)^2 d - 6 a  (a^0)^2 c d^2+  6 a^2 a^0 c^0 d^2 \cr
& \;\;\;\;\; + 6 (a^0)^2 b c^0 d^2 - 2 (a^0)^3 d^3 + 4 a (a^0)^2 c^2 d_0- 8 a^2 a^0 c c^0 d_0 - 2 (a^0)^2 b c c^0 d_0 \cr
& \;\;\;\;\; + 4 a^3 (c^0)^2 d_0 + 2 a a^0 b (c^0)^2 d_0 +  2 (a^0)^3 c d d_0 - 2 a (a^0)^2 c^0 d d_0\, .
}}

\listrefs
\end